\documentclass[runningheads]{llncs}
\usepackage{balance} 
\usepackage{amsmath,xspace}
\usepackage{etex}
\usepackage{centernot}
\usepackage{stmaryrd}
\usepackage{wrapfig}
\usepackage{amsfonts}
\usepackage{mathtools}
\usepackage{xspace}
\usepackage{color}

\usepackage{fixltx2e}
\usepackage{fancybox}
\usepackage{ifthen}
\usepackage{pgf}
\usepackage{tikz}
\usetikzlibrary{arrows,automata}
\usetikzlibrary{arrows.meta, positioning}
\usepackage{graphics}
\usepackage{soul}
\usepackage{accsupp} 
\usepackage{cancel}
\usepackage{textcomp}
\usepackage{calrsfs}
\usepackage{paralist}
\usepackage{enumitem}

\usepackage{changebar}

\usepackage[super]{nth}
\DeclareMathAlphabet{\pazocal}{OMS}{zplmf}{m}{n}
\newcommand{\mcal}[1]{\pazocal{#1}}
\newcommand{\bcal}[1]{\mathcal{#1}}

\newcommand{\rulename}[1]{\mbox{\textsc{#1}}}

\newcommand{\qt}[1]{``{#1}"}

\newcommand{\rTo}[1]{\xrightarrow{\ #1}}

\newlength{\arrow}
\newcounter{sqindex}

 \newcommand{\rom}[1]{ \textup{(\lowercase\expandafter{\romannumeral#1})}}

\usepackage[ruled]{algorithm2e} 

\SetAlFnt{\small}
\SetAlCapFnt{\small}
\SetAlCapNameFnt{\small}
\SetAlCapHSkip{0pt}
\IncMargin{-\parindent}

\newcommand \Until      {{\mathbin{\mcal{U}}\kern-.1em}}
\newcommand \Release     {{\mathbin{\mcal{R}}\kern-.1em}}

\newcommand \Since      {\mathbin{\mcal{S}\kern-.08em}}

\newcommand \g    {{\mathsf{{G}}\kern.08em}}
\newcommand \f    {{\mathsf{{F}}\kern.08em}}
\newcommand \UntilHat   {\mathbin{\LTLhat{\mcal{U}}\kern-.1em}}

\newcommand \SinceHat   {\mathbin{\LTLhat{\mcal{S}}\kern-.08em}}

\newcommand \ltl        {\textsc{ltl}\xspace}

\newcommand{\set}[1]{\{{#1}\}}
\newcommand{\conf}[1]{\langle{#1}\rangle}

\newcommand{\bcup}[3]{\bigcup_{#1}^{#2}{#3}}

\def\<#1>{\mathinner{\langle#1\rangle}}

\newcommand{\id}{i}

\newcommand{\Exp}[1]{2^{#1}}

\newcommand{\msf}[1]{\mathsf{#1}}

\newcommand{\typecvar}{{\scriptstyle@\msf{type}}}
\newcommand{\assigncvar}{{\scriptstyle@\msf{asgn}}}
\newcommand{\readycvar}{{\scriptstyle@\msf{rdy}}}
\newcommand{\lnkcvar}{{\scriptstyle@\msf{lnk}}}

\newcommand{\listen}{\rulename{ls}}

 \newcommand{\chanlabel}[2]{L^c_{#1}({#2})}
 \newcommand{\chank}[1]{L^c_k({#1})}

  \newcommand{\outlabel}[2]{L^o_{#1}({#2})}
 \newcommand{\outk}[1]{L^o_k({#1})}

\newcommand{\ch}{C}
\newcommand{\och}{Y}

\usepackage[disable]{todonotes} 
\usepackage{todonotes}

\usetikzlibrary{arrows.meta, positioning, shapes, backgrounds}
\usetikzlibrary{fit, backgrounds}

\usepackage{wrapfig}
\usepackage{framed}
\usepackage[inkscapeformat=pdf]{svg}
\usepackage{apxproof} 
\newtheoremrep{theorem}[theorem]{Theorem}[section]
\newtheoremrep{proposition}[theorem]{Proposition}[section]
\newtheoremrep{lemma}[theorem]{Lemma}[section]
\newtheoremrep{corollary}[theorem]{Corollary}[section]
\newtheoremrep{definition}[theorem]{Definition}[section]
\newtheoremrep{example}[theorem]{Example}[section]

\usepackage{algorithm2e}
\usepackage{svg}
\usepackage{xcolor}

\newcommand{\Init}[2]{\mathsf{Init}[#1 \| #2]}
\newcommand{\InitN}[2]{\mathsf{Init}[#1 \| \dots\|#2]}

\newcommand{\Deltaij}{(\Delta_i \cup \Delta_j)}
\newcommand{\Yij}{(Y_i \cup Y_j)}
\usepackage{lineno}
\usepackage{bbding}
\SetInd{1em}{1em} 
\begin{document}
%
\title{On Reconfigurable Bisimulation, with an Application to the Distributed Synthesis Problem}

%
\titlerunning{Reconfigurable Bisimulation and Distributed Synthesis}

%
 \author{
 	Yehia Abd
 	Alrahman
 	\and
 	Nir Piterman
 }

 \authorrunning{Y. Abd
 	Alrahman \& N. Piterman}

 \institute{University of Gothenburg \& Chalmers University of Technology, Gothenburg, Sweden\\
 	\email{\{yehia.abd.alrahman,nir.piterman\}@gu.se}
 	}

\maketitle              

\begin{abstract}
We address the problem of distributing a sequential system across asynchronous agents that jointly  recognize the same language. Existing approaches incur state-space blow-up due to fixed communication structures. We propose a solution based on \emph{reconfigurable communication}, where the communication structure is not fixed but instead computed and subsequently minimized. We rely on a novel notion of \emph{Parametric Reconfigurable Bisimulation}, which beyond classical state-space minimization, identifies the necessary communications for each agent while pruning the rest. We show how to compute this bisimulation, how to minimize communications per agent, and prove that the composition of the resulting agents remains bisimilar to the original sequential system.
As a key application, we enable \emph{teamwork synthesis}, a variant of distributed synthesis, from global specifications.
While distributed synthesis is undecidable, we show that teamwork synthesis is decidable.
We further validate our results with a detailed experimental evaluation.
\end{abstract}

\section{Introduction}\label{sec:intro}
\vspace{-5pt}

The problem of distributing a sequential system  into a set of asynchronous systems
recognizing the same language is
well-established~\cite{Zielonka87,DBLP:conf/concur/StefanescuEM03,MukundS97,KrishnaM13,GimbertMMW22}.
 This is a challenging problem, as
a feasible solution to the problem aspires to unlock the
\emph{synthesis} of distributed systems from declarative
specifications~\cite{PnueliR90,FinkbeinerS05}.
The latter problem is known to be generally undecidable, and is only
decidable for very restricted configurations and with a very high
complexity.
It is widely agreed that the undecidability is mainly due to
the partial knowledge of the asynchronous systems, or equivalently the lack of \qt{sufficient coordination}.
As the general problem is undecidable, approaches that try to tackle it
must compromise and find ways to bypass its undecidability.

A notable algorithmic approach is in the context of asynchronous automata.
In a landmark result, Zielonka demonstrated that the language
recognised by a deterministic automaton can be distributed across a set
of asynchronous automata \cite{Zielonka87}.
That is,
given a fixed communication mapping from events to automata and a regular  language that
respects the asynchrony of events, one can construct a
set of asynchronous automata, communicating over these events, and
jointly recognise the same language.
Here, asynchronicity of events means that only events that have the
same automata listening to them are synchronised. Events that do not
share automata listening to them are executed independently.

This result is significant because it enables the construction of asynchronous automata from a sequential automaton while preserving a fixed communication structure. However, the construction is expensive: the resulting asynchronous automata are doubly exponential in the number of local automata and exponential in the size of the sequential automaton~\cite{MukundS97}. For acyclic communication graphs with binary channels, this complexity can be reduced to quadratic in the size of the original automaton~\cite{KrishnaM13}; see also a recent extension to more general configurations~\cite{DBLP:conf/fossacs/LehautMP26}.
Despite these results, Zielonka's construction has important limitations. Deviations from the language of the sequential system may be detected only after they have occurred, with no mechanism for recovery. Moreover, the construction itself is intricate, and efforts to simplify it continue even after nearly three decades. Akshay et al.~\cite{akshay2013implementing} further showed that existing constructions of asynchronous automata inherently suffer from deadlocks and nondeterministic guessing, and that eliminating these limitations requires restricting the input language to an impractically narrow class.
More fundamentally, Zielonka's approach relies on the input language respecting the asynchrony of events, formalized by the \(I\)-closedness property. This requirement limits its applicability to general distributed synthesis because deciding whether a given regular language admits an \(I\)-closed sublanguage is itself undecidable~\cite{DBLP:conf/concur/StefanescuEM03}. Consequently, the general distributed synthesis problem remains undecidable.

In this paper, we leverage reconfigurable communication~\cite{AbdAlrahmanP21,AlrahmanMP22} to obtain a new decidability result for distribution, without constraining the input language. In reconfigurable settings, communication interfaces are
state-based.
Namely,
an agent decides in a given state whether to participate in (or
discard) specific events during execution. Thus, instead of restricting the input language to fit a fixed communication structure (\`a la Zielonka), we decide
algorithmically what events (and when) each agent needs to
participate in so that jointly with others they recognise the input language. In this way, the restriction is shifted from the input language to the communication model.

\noindent
{\bf Contributions:} after a high-level overview (Sect.~\ref{sec:new overview}), \rom{1} we introduce asynchronous reconfigurable transition systems (TS)(Sect.\ref{sec:model});
\rom{2} we propose a novel notion of \emph{
Reconfigurable Bisimulation} (Sect.\ref{sec:bisim}) that
prunes unnecessary interactions; 
\rom{3} we show how to compute this bisimulation
and that the minimisation produces TSs that are, at worst, as large as the sequential system, and whose composition is bisimilar to
the  sequential system (in Sect.\ref{sec:bisim});  \rom{4}  we
use this distribution (in Sect.\ref{sec:app}) in order to enable teamwork synthesis, a variant of distributed synthesis, from global specifications, and thus
bypass undecidability by our ability to reconfigure communications;
\rom{5} We demonstrate the effectiveness of our approach experimentally.
All proofs are included in Appendix~A.


\vspace{-6pt}
\section{Overview}
\vspace{-4pt}
\label{sec:new overview}
\vspace{-5pt}

We present our approach through a simple example to illustrate the intuition. 

\begin{figure}[bt]
\begin{center}
\begin{tikzpicture}[
    ->, >=stealth',
    auto,
    semithick,
    node distance=2cm,
    state/.style={circle, draw, minimum size=18pt, inner sep=1.5pt},
    num/.style={font=\scriptsize}
]

\node[state] (q0) at (0,2.2) {0};
\node[state] (q1) at (1,2.2) {1};
\node[state] (q2) at (2,2.2) {2};
\node[state] (q3) at (4,2.2) {3};
\node[state] (q4) at (2,3.2) {4};
\node[state] (q6) at (3,3.2) {6};
\node[state] (q7) at (4,3.2) {7};
\node[state] (q5) at (5,3.2) {5};

\node[coordinate] (init) at (0,3) {};
\draw[->] (init) -- (q0);


\draw (q0)  edge node {a} (q1);
\draw (q1)  edge node {f} (q2);
\draw (q2)  edge node {a} (q3);
\draw (q2)  edge node [left] {p} (q4);
\draw (q3)  edge node [pos=0.3,right] {p} (q5);
\draw (q4)  edge[bend left=60] node [pos=0.3,below] {a} (q5);
\draw (q4)  edge[bend right=30] node [pos=0.2,below] {d} (q0);
\draw (q5)  edge[bend right=40] node [pos=0.8,above] {f} (q6);
\draw (q5)  edge[bend right=110] node [pos=0.7,above] {d} (q1);
\draw (q6)  edge node [below] {a} (q7);
\draw (q6)  edge node [left] {d} (q2);
\draw (q7)  edge node {d} (q3);


\node[state] (q0dash) at (5,0) {0\textcolor{gray}{-}};
\node[state] (q0d) at (5,1) {0\textcolor{blue}{d}};
\node[state] (q1a) at (6.5,0) {1\textcolor{gray}{a}};
\node[state] (q1d) at (6.5,1) {1\textcolor{blue}{d}};
\node[state] (q2f) at (8,0) {2\textcolor{gray}{f}};
\node[state] (q2d) at (8,1) {2\textcolor{blue}{d}};
\node[state] (q3a) at (10,0) {3\textcolor{gray}{a}};
\node[state] (q3d) at (10,1) {3\textcolor{blue}{d}};
\node[state] (q4) at (8,2.2) {4\textcolor{gray}{p}};
\node[state] (q6) at (9,2.2) {6\textcolor{gray}{f}};
\node[state] (q7) at (10,2.2) {7\textcolor{gray}{a}};
\node[state] (q5p) at (11,2.2) {5\textcolor{gray}{p}};
\node[state] (q5a) at (11,3) {5\textcolor{gray}{a}};

\node[draw=red,dotted,fit=(q0dash) (q1a) (q2f) (q3a)] {};
\node[draw=red,dotted,fit=(q0d) (q1d) (q2d) (q3d)] {};
\node[draw=red,dotted,fit=(q4) (q6) (q7) (q5p) (q5a)] {};

\node[coordinate] (init) at (4.3,0) {};
\draw[->] (init) -- (q0dash);


\draw (q0dash)  edge node {a} (q1a);
\draw (q0d)  edge node [pos=0.3,below] {a} (q1a);
\draw (q1a)  edge node {f} (q2f);
\draw (q1d)  edge node [below,pos=0.3] {f} (q2f);
\draw (q2f)  edge node {a} (q3a);
\draw (q2d)  edge node {a} (q3a);
\draw (q2f)  edge[bend left=30] node [left] {p} (q4);
\draw (q2d)  edge[bend left=30] node [left] {p} (q4);
\draw (q3a)  edge[bend right=30] node [right] {p} (q5p);
\draw (q3d)  edge[bend right=30] node [right] {p} (q5p);
\draw (q4)  edge[bend left=60] node [pos=0.3,below] {a} (q5a);
\draw (q4)  edge[bend right=30] node [pos=0.2,above] {d} (q0d);
\draw (q5p)  edge[bend right=40] node [pos=0.5,above] {f} (q6);
\draw (q5a)  edge[bend right=40] node [pos=0.8,above] {f} (q6);
\draw (q5p)  edge[out=45,in=-225,looseness=2] node [above,pos=0.7] {d} (q1d);
\draw (q5a)  edge[bend right=110] node [pos=0.7,above] {d} (q1d);
\draw (q6)  edge node [below] {a} (q7);
\draw (q6)  edge node [left] {d} (q2d);
\draw (q7)  edge node {d} (q3d);

\node[state] (state1) at (1,0) {0\textcolor{gray}{-}};
\node[num]   [above=0pt of state1,yshift=-5pt,xshift=10pt]  {$\{p\}$};
\node[state] (state2) at (1,1) {0\textcolor{blue}{d}};
\node[num]   [above=0pt of state2,xshift=-10pt,yshift=-3pt]  {$\{p,a,f\}$};
\node[state] (state3) at (3,1) {4\textcolor{gray}{p}};
\node[num]   [above=0pt of state3,yshift=-3pt,xshift=6pt]  {$\{d\}$};
\node[coordinate] (init) at (0.3,0) {};
\draw[->] (init) -- (state1);

\draw (state1) edge[out=-10,in=-70] node [below] {p} (state3);
\draw (state2) edge[out=-10,in=-70] node [pos=0.4,below] {p} (state3);
\draw (state3) edge node [above] {d} (state2);
\draw (state2) edge[bend right=50] node [left] {a,f} (state1);

\end{tikzpicture}
\end{center}
\vspace*{-5mm}
\caption{\label{fig:fourarm} Centralised Controller (top left), controller with state labeled by actions (right), and the quotient system for arm 3 with states labeled by actions (bottom left).}
\vspace*{-4mm}
\end{figure}

Consider a machine involving four components: an arm (arm 1) putting items on an input tray, an arm (arm 2) taking items from the input tray passing them to a processing machine that puts the processed items on an output tray (machine), and, finally, an arm (arm 3) taking the items out from the output tray. 
For simplicity, each component and each tray can hold/process exactly one item. 
The system is specified using a sequential system in the style of discrete event systems \cite{ramadge89} in Figure~\ref{fig:fourarm} (top left).
Action $a$ corresponds to arm 1 putting an item in the in-tray (add).
Action $f$ corresponds to arm 2 taking the item from the in-tray (forward).
Action $p$ corresponds to the machine processing the item and delivering it to the out-tray (process).
Finally, action $d$ corresponds to arm 3 delivering the item from the out-tray (deliver).

We aim to distribute the control of the components to separate controllers. Accordingly, we compute a minimised reconfigurable communication structure for each controller, based on the component's actions and with respect to the language of the sequential system. Clearly, arm 1 should be in charge of $a$, arm 2 of $f$, the machine of $p$, and arm 3 of $d$.
Each arm may require additional information about other actions, which we aim to minimize as much as possible.

We rewrite  the system in Figure~\ref{fig:fourarm} (top left) equivalently so that every state is labeled by the action that led to it. See Figure~\ref{fig:fourarm} (right), and ignore red for now. The language of the original system is thus encoded in the state labels, allowing transitions to be safely pruned later if deemed unnecessary.  

We create four copies of the system, one for each controller, where the only difference is in distinguishing between `owned' actions (initiate) and `un-owned' actions (react), by removing the state labels corresponding to unowned actions. 
However, two states that are distinct remain distinct.
This corresponds to the greying out of $a$, $f$, and $p$ from arm 3's point of view in Figure~\ref{fig:fourarm}. 
For example, $5a$ and $5p$ are distinct but their label has been removed. 
Thus, we consider $\textcolor{blue}{d}$ as one label and the greyed $\textcolor{gray}{\_}$, $\textcolor{gray}{a}$, $\textcolor{gray}{f}$, and $\textcolor{gray}{p}$ as the empty label. 
Conceptually, we consider that each state, once reached, synchronizes exclusively on the actions enabled from that state and does not block any other communications. Thus, we consider it as a \emph{reconfigurable} transition system. Clearly, if we compose the four copies, where they synchronize on joint actions, the composition corresponds to the original, and all greyed labels get filled back.

We now analyze each copy in the context of the composition of the other three. 
We concentrate here on arm 3 in the context of the composition of arms 1 and 2 and the machine, referring to arm 3 as \emph{arm} and the rest as \emph{parameter}. Recall that the parameter has the same structure as the arm but with inverted labels, where arm 3 has the label $d$ (and others ignored) the parameter has the labels $a$, $f$, and $p$ (and $d$ ignored). 
In particular, each state appears both as an arm state and as a parameter state. We compute  \emph{reconfigurable bisimulation} as follows: for each parameter state~$\epsilon$, we partition the arm states into equivalence classes. Intuitively, this means that, in the composition of arm~3 with the parameter, computations starting from equivalent arm states under~$\epsilon$ are indistinguishable. However, different parameter states may induce different partitions over the arm states; we therefore compute a single bisimulation-consistent partition that maximizes agreement among them.
Consider the equivalence class $\set{0\textcolor{gray}{\_},1\textcolor{gray}{a},2\textcolor{gray}{f},3\textcolor{gray}{a}}$ of the partition depicted in red in Figure~\ref{fig:fourarm} (right), it must hold that all corresponding parameter states  in the composition $\set{0\textcolor{blue}{\_},1\textcolor{blue}{a},2\textcolor{blue}{f},3\textcolor{blue}{a}}$ agree on this equivalence; otherwise the equivalence class is inconsistent. 
In general, the partitions of different parameter states could be arbitrarily incomparable, and finding a unique partition that maximizes their agreement is reduced to a clique-partitioning problem-- an NP-hard problem~\cite{GJ79}; we therefore use a standard polynomial-time greedy algorithm in our tool.
We repeat the process by analyzing each component in the context of the other three as parameter.
 
Each component is quotiented with respect to the identified agreement classes, and its states are reconfigured to eliminate unnecessary react transitions.
Figure~\ref{fig:fourarm} (bottom left) shows the quotient system corresponding to arm~3, where each state is labeled by a representative of its equivalence class. The agreement class highlighted in red, $\set{0\textcolor{gray}{\_},1\textcolor{gray}{a},2\textcolor{gray}{f},3\textcolor{gray}{a}}$ is collapsed into a single state $0\textcolor{gray}{\_}$ in the quotient. This collapse removes all internal react transitions (i.e., the sequence $a,f,a$), while retaining only the external transition $p$. The state $0\textcolor{gray}{\_}$ is then reconfigured to listen exclusively to action $p$, indicated by the set $\set{p}$ above the state, and to ignore all other actions. Consequently, the quotient of arm~3 effectively disconnects from the composition until the parameter initiates on $p$.
Thus, reconfiguration induces a special composition semantics in which a system that is not listening to an action  does not hinder or block that action. Our approach guarantees that the (reconfigurable) composition of the four arms is bisimilar (in the classical sense) to the state-labeled copy of the transition system we started from, despite the removal of synchronizations. 



Unlike our approach, Zielonka’s distribution does not apply to non I-closed input languages, and thus requires prior problem-specific insight to manually adjust actions connectivity. Even for this simple system, one must permanently determine that arm~1 connects to $a$ and $f$, arm~2 connects to $a$, $f$, and $p$, the machine connects to $f$, $p$, and $d$, and arm~3 connects to $p$ and $d$, as follows:
\medskip

$\qquad\qquad\qquad\fbox{\mbox{arm 1}} \stackrel{a,f}{\longleftrightarrow}
\fbox{\mbox{arm 2}}  \stackrel{f,p}{\longleftrightarrow}
\fbox{\mbox{machine}}  \stackrel{p,d}{\longleftrightarrow}
\fbox{\mbox{arm 3}}
$
\medskip

Obviously, there are actions that are connected \emph{permanently} to more than two components, leaving no room to disconnect when information are not needed during execution.
In this example, the fixed connectivity does \emph{not} correspond to a tree \cite{KrishnaM13}, but it is \emph{tree-like}, as the connectivity of all non-binary actions is continuous along the tree (e.g., $p$ spans arm 2, the machine, and arm 3), and hence the quadratic distribution \cite{DBLP:conf/fossacs/LehautMP26} applies. However, this leads to almost fully synchronised systems due to static assignments of blocking \emph{non-binary} actions. 

We computed the full distribution using our approach and Zielonka’s. Our quotient systems for arm~1, arm~2, arm~3, and the machine have 3, 5, 3, and 5 states, respectively—much smaller than the state-labeled copy of the original system—whereas Zielonka’s construction yields larger systems (cf. Appendix~\ref{app:more overview}). Nonblocking reconfigurable actions can reduce connectivity, yielding smaller systems and lower communication overhead.
In Section~\ref{sec:eval}, we quantitatively evaluate our approach against the optimal Zielonka-type construction~\cite{GenestGMW10} on the highly connected system $\msf{Path}_n$, which exhibits an exponential lower bound for Zielonka-type methods. Appendix~\ref{sec:ts-zie} shows the constructions, Zielonka’s  unavoidable early violations, and lack of recovery. In contrast, our approach avoids violations and scales better in both system size and required connectivity.

\section{Reconfigurable Transition Systems (TS)}\label{sec:model}
\vspace{-8pt}
We assume a global set of channels $\ch$ that agents use for communication.
An agent can use a subset $\och$ of $\ch$ to (uniquely) initiate communication on.
Moreover, an agent can only \emph{react} to communications initiated by
others on $C\setminus \och$.
Initiate transitions are non-blocking/autonomous
(i.e., can happen either individually or jointly with other reacting
agents) while react transitions are blocking (can only happen jointly
with corresponding initiate transitions from another agent).

\begin{definition}[Transition System]\label{def:shadow} A
Transition System (TS)  $T$ is of the form
$\langle S,s_0, \och,O,\listen,L,
\Delta\rangle$, where:
\begin{compactitem}
\item $S$ is the set of states of $T$ and $s_0\in S$ is its
initial state.

\item $\och$ and $O$  are the  interface
of $T$, we sometimes denote it as $\langle \och,O\rangle$, where
\begin{compactitem}
\item $\och\subseteq \ch$ is the only set of channels that the agent can use to initiate communication. All other channels in $\ch\backslash \och$ can only be used to react to communication from other agents.
\item $O$ is an output (or actuation) alphabet.
\end{compactitem}
\item $\listen: S\rightarrow\Exp{\ch}$ is a channel listening
function defining (per state) the enabled channels that $T$ currently listens to (both initiate and react channels).

\item $L:
S\rightarrow\Exp{\och}\times\Exp{O}$
 is a labelling function. We use
$L$  to label states with recently used initiate channels from $\och$ and also
 the produced output $O$. We will use $\chank{s}\subseteq \och$ and
 $\outk{s}\subseteq O$ to denote the channel label of $s$ (and
 correspondingly the output label of $s$). Note that unlike
 $\listen$, we have that $\chank{s}\subseteq \och$, i.e., it cannot
 contain channel labels in $\ch\backslash \och$ (i.e., reacting channels).
That is, $\chank{s}$ and $\outk{s}$ are local contributions from $T$.

\item $\Delta\subseteq S\times \ch\times S$ is the
transition relation of $T$.
We write $\Delta(s,c)$ to denote $\{ s' ~|~ (s,c,s')\in \Delta\}$.
The transition relation satisfies the following:
%
%
\begin{compactitem}
\item
For every state $s\in S$ and every channel $c\in \ch$, we have that
$\Delta(s,c)\neq\emptyset$ iff $c \in
\listen(s)$.
That is, $\listen$ always includes exactly the \emph{enabled}
channels in $\Delta$, regardless whether they are initiate $(\och)$ or react $(\ch\backslash \och)$ channels.
\item For every state $s\in S$ and every ``initiate'' channel
$c \in \och$, if $\Delta(s,c)\neq \emptyset$ then
$c\in\chank{s'}$ for every $s'\in \Delta(s,c)$.
That is, after a communication on an ``initiate'' channel, the
channel is included in the channel label of target state, 
recording the input contribution of $T$
%
%
%
%
%
\end{compactitem}

\end{compactitem}
\end{definition}


A TS $T$ is \emph{communication-closed} iff its transition relation $\Delta$ only contains self-initiated transitions, i.e., $\Delta\subseteq S\times \och\times S$ ($T$ cannot react to external communication on $\ch\backslash \och$), and thus $T$ can take all transitions independently.
It is \emph{deterministic} if $|\Delta(s,c)|\leq 1$ for all $s$ and $c$.
A run of a communication-closed $T$ is the infinite sequence $r=s_0c_0s_1c_1s_2\dots$ such
that for all $\id\geq 0: (s_{\id},c_{\id},s_{\id+1})\in \Delta$ and $s_0$ is the
initial state.
An execution of $T$ is the projection of a run $r$ on \emph{state
labels}. That is, for a run  $r=s_0c_0s_1c_1s_2\dots$, there is an
execution $w \in (\Exp{\och}\times \Exp{O})^\omega$ induced by $r$ such that $w=L(s_0)L(s_1)L(s_2)\dots$.
We use $\mathcal{L}_T$ to denote the language of $T$, i.e., the set of all infinite
sequences in $(\Exp{\och}\times \Exp{O})^{\omega}$. Notice that $\mathcal{L}_T$ 
is defined through a projection onto state labels. 

\begin{definition}[Asynchronous Composition $\|$]\label{def:comp}
Given a set of two TSs $\set{T_k}_{k\in\set{1,2}}$  such that  ${T_k} = \langle S_k,s^0_k,\och_k,O_k,\listen_k,L_k,
\Delta_k\rangle$
, we define their composition
${T_1\|T_2} = \langle S,s^0,\och_{\|}, O_{\|},\rulename{ls},L,
\Delta\rangle$ as follows:

\begin{compactitem}
\item
  $S = S_1\times S_2$ and
   $s^0 = (s_1^0,s_2^0)$.

\item
	$\och_{\|}={\och_1\cup \och_2}$ and
	$O_{\|}={O_1\cup O_2}$.

\item $\Delta =$
%
%
%
%
%
%
%
\vspace*{-10pt}
\[
\left\{ \left(
\begin{array}{@{}c@{}}
\strut (s_1, s_2)\\[-0.5ex]
,c, \\[-0.5ex]
(s'_1, s'_2)
\end{array}
\right)
\middle|\;
\begin{array}{@{}l@{}}
c \in \och_1,\ (s_1, c, s'_1) \in \Delta_1\ \text{and}\\
\qquad 
\begin{cases}
\!(1)~ c \in \listen_2(s_2),\ (s_2, c, s'_2) \in \Delta_2 \mbox{ or}\\[-0.8ex]
\!(2)~ c \notin \listen_2(s_2),\ s'_2 = s_2
\end{cases}\\ 

\mbox{or }
c \in \och_2,\ (s_2, c, s'_2) \in \Delta_2\ \text{and}\\
\qquad 
\begin{cases}
\!(1)~ c \in \listen_1(s_1),\ (s_1, c, s'_1) \in \Delta_1 \mbox{ or}\\[-0.8ex]
\!(2)~ c \notin \listen_1(s_1),\ s'_1 = s_1
\end{cases} \\ 
\mbox{or }
c \notin (\och_1 \cup \och_2)\ \text{and}\\
\qquad \begin{cases}
\!(1)~ c \in \listen_i(s_i), (s_i,c, s_i') \in \Delta_i, \mbox{ for all}\ i\in\{1,2\} \hfill\mbox{ or}\\[-0.3ex]
\!(2)~ c \in \listen_1(s_1)\setminus \listen_2(s_2),\ (s_1, c, s'_1) \in \Delta_1,\ 
s'_2 = s_2 \hfill\mbox{ or}\\[-0.3ex]
\!(3)~ c \in \listen_2(s_2)\setminus \listen_1(s_1),(s_2, c, s'_2) \in \Delta_2, s'_1 = s_1
\end{cases}
\end{array}
\hspace{-4mm} \right\}
\]
\vspace*{-6pt}
\item
  $\listen((s_1, s_2)) = \listen({s_1})\cup\listen({s_2})$

\item $L((s_1, s_2)) =(\chanlabel{1}{s_1}\cup\chanlabel{2}{s_2},
{\outlabel{1}{s_1}\cup\outlabel{2}{s_2}})$
\end{compactitem}
\end{definition}
The transition relation $\Delta$ of the composition defines a specialized \emph{reconfigurable channelled multicast}. Intuitively, the composition ${T_1 \parallel T_2}$ initiates a communication on channel $c$ and supplies a transition 
$
((s_1, s_2), c, (s'_1, s'_2)) \in \Delta
$
if either ${T}_1$ or ${T}_2$ is able to initiate it. That is, either:
\begin{itemize}
  \item[] \qquad\( c \in \och_1 \) and \( (s_1, c, s'_1) \in \Delta_1 \)\quad or\quad  \( c \in \och_2 \) and \( (s_2, c, s'_2) \in \Delta_2 \)
\end{itemize}
In the former case, if ${T}_2$ is currently listening (i.e., \( c \in \listen_2(s_2) \)), then it must supply a matching react transition \( (s_2, c, s'_2) \in \Delta_2 \); otherwise, if \( c \notin \listen_2(s_2) \), then ${T}_2$ does not observe the communication and remains in place, i.e., \( s'_2 = s_2 \). The latter case is symmetric.

The composition ${T_1 \parallel T_2}$ may also react to an externally initiated communication on channel $c \in \ch\setminus (\och_1\cup \och_2)$, producing a transition
$
((s_1, s_2), c, (s'_1, s'_2)) \in \Delta
$
provided that neither ${T}_1$ nor ${T}_2$ is able to initiate it (i.e., \( c \notin \och_1 \) and \( c \notin \och_2 \)). In this case, the communication is initiated externally, and the response of each component depends on its current listening set. If both are listening on \( c \), then they must jointly supply matching react transitions \( (s_1, c, s'_1) \in \Delta_1 \) and \( (s_2, c, s'_2) \in \Delta_2 \). If only one is listening, then that component supplies a react transition, while the other remains unchanged.

Unlike Zielonka's multicast, a TS cannot listen to a channel in a state without also supplying a matching transition—i.e., an initiate (send) transition cannot be blocked by passive listeners. Moreover, the composition yields a valid transition system. When the union of the initiating channel sets satisfies \( \bigcup_k \och_k = \ch \), the resulting system is \emph{communication-closed}.

Given a TS $T =\langle S
,s_0,\och,O,\listen, L,\Delta \rangle$, a state $s\in S$, we use
the notations:
\begin{compactitem}
\item {\bf $s$ initiates on $c$, written $(s\rTo{c}_! s')$, iff}
$c\in \och$, $(s,c, s')\in\Delta$ and $c\in\chanlabel{}{s'}$,
i.e., $T$ is initiating a communication on $c$ from state $s$ to state $s'$.
\item {\bf $s$ reacts to $c$, written $(s\rTo{c}_? s')$,  iff}  $c\in (\ch\setminus \och)$,
$c\in\listen(s)$,
$(s,c, s')\in\Delta$, i.e., $T$ is in state $s$ that can react to a communication on $c$ from other agents.

\end{compactitem}

\begin{definition}[Strong Bisimulation~\cite{leifer2000deriving}]\label{def:sbisim}
 A strong bisimulation $\mathcal{R}\subseteq S\times S$  over the set
 of states of $S$ of a TS ${T}$ is a \emph{symmetric} relation such that whenever
$(s_1,s_2)\in\mathcal{R}$ and
for all $c\in \ch,$ we have that:
\begin{compactenum}
\item $L(s_1) = L(s_2)$,
\item $\ s_1\rTo{c}_! s'_1$  \quad implies\quad $\exists s'_2.$  $s_2\rTo{c}_! s'_2$ and $(s'_1,s'_2)\in\mathcal{R}$

\item $\ s_1\rTo{c}_? s'_1$\quad implies\quad $\exists s'_2.$  $s_2\ \rTo{c}_?\  s'_2\  \ \mbox{and}\ \ (s'_1,s'_2)\in\mathcal{R}$
\end{compactenum}
Two states $s_1$ and $s_2$ are strongly bisimilar, written $s_1\sim s_2$, {\bf iff} there exists a strong bisimulation $\mathcal{R}$ such that $(s_1,s_2)\in\mathcal{R}$. Moreover, two TSs $T_1,T_2$ are strongly bisimilar, written $T_1\sim T_2$, {\bf iff} their initial states are strongly bisimilar in the TS that is their disjoint union. 
\end{definition}
\begin{lemma}[$\sim$ preserves parallel composition~\cite{leifer2000deriving}]\label{lem:bsimclose} Consider two TSs $T_1,T_2$, we have that: $\quad T_1\sim T_2$\ implies\ $(T_1\| T)\sim(T_2\| T)$ for all TS $T$.
\end{lemma}
It is well known that strong bisimulation is finer than (trace-) language-equivalence~\cite{bloom1995bisimulation}, because the latter is insensitive to branching. Thus,
$ T_1\sim T_2$\ implies\ $\mathcal{L}_{T_1} = \mathcal{L}_{T_2} $ but not the converse.

%
\begin{lemmarep}\label{lem:com+assoc}
Given two TSs
${T}_1$ and ${T}_2$ we have that:
\begin{compactenum}
\item $\|$ is commutative: ${T}_1\|{T}_2\sim{T}_2\|{T}_1$;
\item $\|$ is associative: $({T}_1\|{T}_2) \| {T}_3\sim{T}_1\|({T}_2 \| {T}_3)$ for every  TS $T_3$;
\end{compactenum}
\end{lemmarep}
\begin{proof}
We prove each statement separately. In both statement, the proof proceeds
by case analysis on the joint transition.
\begin{description}
\item[\bf ($\|$ is commutative):] 

To prove that $({T}_1 \| {T}_2)\ \sim\ ({T}_2 \| {T}_1)$, it is sufficient to prove that their initial states are equivalent by the definition of $\sim$. Let us use the sets $S_1$ and $S_2$ to denote the set of states in $T_1$ and $T_2$ respectively. Moreover, we use $s^0_1$ and $s^0_2$ to denote the initial states of  $T_1$ and $T_2$ respectively. Thus, we need to show:
\[
(s^0_1,s^0_2)\sim (s^0_2,s^0_1)
\]

We construct a relation $\mcal{R}$ as follows:
\[
\mcal{R} = \left\{((s,r),(r,s))~\middle|~ s\in S_1,\text{ and } r\in S_2\right\}
\]
\medskip

Now, we must show:

\begin{enumerate}
    \item $((s^0_1,s^0_2),(s^0_2,s^0_1)) \in \mcal{R}$, and
    \item $\mcal{R}$ is a strong bisimulation.
\end{enumerate}

\paragraph{(1)} The initial state pair $((s^0_1,s^0_2),(s^0_2,s^0_1))$  is in $\mcal{R}$ by construction, since:
\[
(s^0_1\in S_1 \text{ and }  s^0_2\in S_2)
\]

\paragraph{(2)} To show that $\mcal{R}$ is a bisimulation, let $((s_1,s_2),(s_2,s_1)) \in \mcal{R}$ and verify the bisimulation conditions.

\begin{itemize}

\item \textbf{Label agreement:} Show that $L((s_1,s_2)) = L((s_2,s_1))$.\\

By composition in Def.~\ref{def:comp}, $L((s_1, s_2)) =(\chanlabel{1}{s_1}\cup\chanlabel{2}{s_2},
{\outlabel{1}{s_1}\cup\outlabel{2}{s_2}})$.\\
Moreover,  $L((s_2, s_1)) =(\chanlabel{2}{s_2}\cup\chanlabel{1}{s_1},
{\outlabel{2}{s_2}\cup\outlabel{1}{s_1}})$.\\
Thus:
\[
L((s_1,s_2)) = L((s_2,s_2))
\]

\item \textbf{Transition preservation:} Show that whenever $((s_1,s_2),(s_2,s_1)) \in \mcal{R}$, the following conditions (and their symmetrical ones) hold:

\begin{enumerate}
\item If $(s_1,s_2) \rTo{c}_! (s'_1,s'_2)$, then there exist $(s'_2, s'_1)$ such that
\[
(s_2,s_1) \rTo{c}_! (s'_2,s'_1) \quad \text{and} \quad ((s'_1,s'_2),(s'_2,s'_1)) \in \mcal{R}
\]
\begin{itemize}
\item By the definition of  $\rTo{c}_! $, we have that $((s_1,s_2),c, (s'_1,s'_2))\in \Delta_{12}$ ($\Delta_{12}$ is the transition relation of ${T}_1\|{T}_2$). Now, By Def.~\ref{def:comp} this can be derived in the following cases:
\begin{enumerate}
\item  $c\in \och_1$, $(s_1,c,s'_1)\in  \Delta_{1}$, $c\in\listen_{2}(s_2)$ and $(s_2,c,s'_2)\in  \Delta_{2}$:  Thus, it follows that $((s_2,s_1),c, (s'_2,s'_1))\in \Delta_{21}$ ($\Delta_{21}$ is the relation of ${T}_2\|{T}_1$) is also derivable by the second main case of $\Delta_{21}$, the subitem $(1)$; and $((s'_1,s'_2),(s'_2,s'_1)) \in \mcal{R}$ as required.

\item $c\in \och_1$, $(s_1,c,s'_1)\in  \Delta_{1}$ and $c\notin\listen_{2}(s_2)$ and ($s_2=s'_2$): It follows that $((s_2,s_1),c, (s'_2,s'_1))\in \Delta_{21}$ is derivable by the second main case of $\Delta_{21}$, the subitem $(2)$; and $((s'_1,s'_2),(s'_2,s'_1)) \in \mcal{R}$ as required.

\item  $c\in \och_2$, $(s_2,c,s'_2)\in  \Delta_{2}$ and $(s_1,c,s'_1)\in  \Delta_{1}$:  This is the symmetric case of case (1), where it is also derivable by the first main case of $\Delta_{21}$, the subitem $(1)$

\item $c\in \och_2$, $(s_2,c,s'_2)\in  \Delta_{2}$ and $c\notin\listen_{1}(s_1)$ and ($s_1=s'_1$): This is the symmetric case of case (2),  where it is also derivable by the first main case of $\Delta_{21}$, the subitem $(2)$

\end{enumerate}

\end{itemize}
\item If $(s_1,s_2) \rTo{c}_? (s'_1,s'_2)$, then there exist $(s'_2, s'_1)$ such that
\[
(s_2,s_1) \rTo{c}_? (s'_2,s'_1) \quad \text{and} \quad ((s'_1,s'_2),(s'_2,s'_1)) \in \mcal{R}
\]
\item[] By definition of $ \rTo{c}_?$, we have that $((s_1,s_2),c, (s'_1,s'_2))\in \Delta_{12}$ and $c\notin (Y_1\cup Y_2)$. Now by Def.~\ref{def:comp}, this can be derived in the following cases:
\begin{enumerate}
\item  $c \in (\listen_1(s_1) \cap \listen_2(s_2))$, $(s_1,c,s'_1)\in  \Delta_{1}$ and $(s_2,c,s'_2)\in  \Delta_{2}$:  It follows that $((s_2,s_1),c, (s'_2,s'_1))\in \Delta_{21}$ is derivable by last main case, subitem $(1)$; and $((s'_1,s'_2),(s'_2,s'_1)) \in \mcal{R}$ as required.

\item $c \in \listen_1(s_1)$, $(s_1,c,s'_1)\in  \Delta_{1}$ and $c\notin\listen_{2}(s_2)$ and ($s_2=s'_2$): It follows that $((s_2,s_1),c, (s'_2,s'_1))\in \Delta_{21}$ is derivable again by last main case, subitem $3$; and $((s'_1,s'_2),(s'_2,s'_1)) \in \mcal{R}$ as required.

\item $c \in \listen_2(s_2)$, $(s_2,c,s'_2)\in  \Delta_{2}$ and $c\notin\listen_{1}(s_1)$ and ($s_1=s'_1$): is derivable again by last main case, subitem $2$; and $((s'_1,s'_2),(s'_2,s'_1)) \in \mcal{R}$ as required.

\end{enumerate}
\end{enumerate}

\end{itemize}

\item[\bf ($\|$ is associative):] 
To prove that $({T}_1\|{T}_2) \| {T}_3\sim{T}_1\|({T}_2 \| {T}_3)$ for every  TS $T_3$, it is sufficient to prove that their initial states are equivalent by the definition of $\sim$. Let us use the sets $S_1, S_2$ and $S_3$ to denote the set of states in $T_1, T_2$ and $T_3$ respectively. Moreover, we use $s^0_1, s^0_2$ and $s^0_3$ to denote the initial states of  $T_1, T_2$ and $T_3$ respectively. Thus, we need to show:
\[
((s^0_1,s^0_2), s^0_3)\sim (s^0_1,(s^0_2, s^0_3))
\]

We construct a relation $\mcal{R}$ as follows:
\[
\mcal{R} = \left\{((s_1,s_2), s_3),(s_1,(s_2, s_3))~\middle|~ s_1\in S_1, s_2\in S_2\text{ and } s_3\in S_3\right\}
\]
\medskip

Now, we must show:

\begin{enumerate}
    \item $((s^0_1,s^0_2), s^0_3),(s^0_1,(s^0_2, s^0_3)) \in \mcal{R}$, and
    \item $\mcal{R}$ is a strong bisimulation.
\end{enumerate}

\paragraph{(1)} The initial state pair $((s_1,s_2), s_3),(s_1,(s_2, s_3))$  is in $\mcal{R}$ by construction, since:
\[
(s^0_1\in S_1, s^0_2\in S_2 \text{ and }  s^0_3\in S_3)
\]

\paragraph{(2)} To show that $\mcal{R}$ is a bisimulation, let $((s_1,s_2), s_3),(s_1,(s_2, s_3)) \in \mcal{R}$ and verify the bisimulation conditions.

\begin{itemize}

\item \textbf{Label agreement:} Show that $L(((s_1,s_2), s_3)) = L((s_1,(s_2, s_3)))$.\\

This is immediate by composition in Def.~\ref{def:comp}. 
\item \textbf{Transition preservation:} Show that whenever $((s_1,s_2), s_3),(s_1,(s_2, s_3)) \in \mcal{R}$, the following conditions (and their symmetrical ones) hold:

\begin{enumerate}
\item If $((s_1,s_2), s_3) \rTo{c}_! ((s'_1,s'_2), s'_3)$, then there exist $(s'_1,(s'_2, s'_3))$ such that
\[
(s_1,(s_2, s_3)) \rTo{c}_! (s'_1,(s'_2, s'_3)) \quad \text{and} \quad (((s'_1,s'_2), s'_3),(s'_1,(s'_2, s'_3))) \in \mcal{R}
\]
\begin{itemize}
\item By Def.~\ref{def:comp}, we have $(((s_1,s_2),s_3),c, ((s'_1,s'_2),s'_3))\in \Delta_{(12)3}$ in the following cases:
\begin{enumerate}
\item  $c\in (\och_1\cup \och_2)$, $((s_1,s_2),c,(s'_1,s'_2))\in \Delta_{12}$,  $c\in\listen_{3}(s_3)$ and $(s_3,c,s'_3)\in \Delta_3$:  As before, there are many cases depending on whether $c\in \och_1$ or $c\in \och_2$, $((s_1,s_2),c,(s'_1,s'_2))\in \Delta_{12}$, we only consider the case for $c\in \och_1$, $(s_1,c,s'_1)\in  \Delta_{1}$, $c\in\listen_{3}(s_3)$  and $(s_2,c,s'_2)\in  \Delta_{2}$; and other cases follow similarly. It follows that
$((s_1,(s_2,s_3)),$ $c, (s'_1,(s'_2,s'_3))\in \Delta_{1(23)}$ where $(s_1$ initiates and $((s_2,s_3)$ reacts such that $(s_2,c, s'_2)\in  \Delta_{2}$ and $(s_3,c, s'_3)\in \Delta_3$; and $(((s'_1,s'_2), s'_3),(s'_1,(s'_2, s'_3))) \in \mcal{R}$ as required.
\item $c\in (\och_1\cup \och_2)$, $((s_1,s_2),c,(s'_1,s'_2))\in \Delta_{12}$ and $c\notin\listen_{3}(s_3)$ and ($s_3=s'_3$): It follows that
$((s_1,(s_2,s_3)),c, (s'_1,(s'_2,s_3))\in \Delta_{1(23)}$ where $c\in \och_1$, $(s_1,c,s'_1)\in  \Delta_{1}$ and $((s_2,s_3),c, (s'_2,s_3))\in \Delta_{23}$ such that $c\in\listen_{2}(s_2)$ , $(s_2,c, s'_2)\in  \Delta_{2}$ and  $c\notin\listen_{3}(s_3)$ and ($s_3=s'_3$); and $(((s'_1,s'_2), s_3),(s'_1,(s'_2, s_3))) \in \mcal{R}$ as required.

\item  $c\in \och_3$ and $c\notin\listen_{12}(s_1,s_2)$:  This is the symmetric case of case (1);

\item $c\in \och_3$, $(s_3,c,s'_3)\in \Delta_3$ and $c\notin\listen_{12}(s_1,s_2)$ and ($(s_1,s_2)=(s'_1,s'_2)$): This is the symmetric case of case (2);

\end{enumerate}

\end{itemize}

\item If $((s_1,s_2), s_3) \rTo{c}_? ((s'_1,s'_2), s'_3)$, then there exist $(s'_1,(s'_2, s'_3))$ such that
\[
(s_1,(s_2, s_3)) \rTo{c}_? (s'_1,(s'_2, s'_3)) \quad \text{and} \quad (((s'_1,s'_2), s'_3),(s'_1,(s'_2, s'_3))) \in \mcal{R}
\]
\begin{itemize}
 \item It follows similarly by case analysis on Def.~\ref{def:comp}.

\end{itemize}

%
%
%

\end{enumerate}

\end{itemize}

\end{description}

\end{proof}

Hence, parallel composition is well defined for arbitrary numbers of TSs. 
\vspace{-7pt}
\section{Our Distribution  Approach}\label{sec:over}
\vspace{-7pt}
We first show that communication-closed TSs admit trivial decomposition:

\begin{lemma}[Trivial Decomposition]\label{lem:triv} Given a deterministic and communication-closed TS
${T} = \langle S
,s^0,\och,O,\listen, L,\Delta \rangle$ and a partition of its interface $\och=\bigcup_{k\in K} \och_k$  and
$O=\bigcup_{k\in K} O_k$ for $K=\{1,\dots,n\}$ such that $\forall j,k\in K.(j\neq k)$ implies $(\och_k\cap \och_j)=\emptyset$ and $(O_k\cap O_j)=\emptyset$. We decompose $T$ into a
set of  TSs with
	interfaces $\{\langle \och_k,O_k\rangle\}_{k\in K}$ such that the following holds:
\begin{compactitem}
\item The composition $T_1\| \dots \| T_n$ is a communication-closed
TS.
\item For every $T_k$, we have that $T_k$ is a TS and $T_k$ is isomorphic to $T$.
%
%
\item $T$ is strongly bisimilar to the composition $T_1\| \dots \| T_n$.
\end{compactitem}
\end{lemma}
\begin{proof}  We construct each $T_k=\langle S_k,s^0_k,\och_k,O_k, \listen_k,L_k,
\Delta_k\rangle$ as follows:
\begin{compactitem}
%
\item
For each $T_k$ we set
  $S_k = {S}$, \,
  $s_k^0 = { s^0}$, \,   $\Delta_k=\Delta$, \, and  $\listen^k(s) =
  \listen(s)$
%
%
\item
For each $s\in S$ with $L(s)=(\och',O')$ where $\och'\subseteq \och$ and
$O'\subseteq O$ , we project the label of $s$ in $T$ on the
corresponding $s$ in $T_k$ as follows: $L_k(s)= ((\och'\cap \och_k),(O'\cap
O_k))$.\qed
%
\end{compactitem}
%
\end{proof}
%


Lemma~\ref{lem:triv} produces fully synchronized transition systems (TSs) with isomorphic structures. That is, each resulting TS has the same states and transitions as the original TS $T$, differing only in their state labeling and interfaces. The decomposition assumes that for all distinct $j,k \in K$, the initiating channel sets are disjoint: \(\forall j,k\in K.(j\neq k). (\och_j \cap \och_k = \emptyset)\). This means that different TSs are responsible for initiating on different channels.

Concretely, each initiate transition \((s, c, s') \in \Delta\) of $T$ is assigned to a unique agent $T_k$ such that \(c \in \och_k\) and \((s, c, s') \in \Delta_k\). 
Every other agent $T_j$ must provide a corresponding react transition \((s, c, s') \in \Delta_j\), since \(c \notin \och_j\). Given this fully synchronous behavior, it is unsurprising that the composition \(T_1 \| \dots \| T_n\) is strongly bisimilar to the original system $T$.

Lemma~\ref{lem:triv} therefore establishes an upper bound on the number of communications each agent $T_k$ must react to in the composition, while preserving semantic equivalence with $T$. Thus, if each agent $T_k$ is fully informed of the global state, the decomposition guarantees correctness of the distribution.

To illustrate, consider the states of a single agent, say $T_1$, in relation to the composite system \(\mathcal{C} = T_2 \| \dots \| T_n\), where $\mathcal{E}$ denotes the set of states of $\mathcal{C}$. There exists a bijective mapping \({\bf f} : S_1 \to \mathcal{E}\) from the states of $T_1$ to companion states of $\mathcal{C}$, such that \({\bf f}(s^0_1) = \epsilon^0\) (i.e., the bijection holds for initial states), and for every \(s \in S_1\), the pair \((s, \epsilon)\) is a reachable state in the full composition iff \(\epsilon = {\bf f}(s)\).

We want to reduce this communication overhead (implied by Lemma~\ref{lem:triv}) while preserving the correctness of the decomposition. We aim to construct partially informed TSs that interact only \emph{by demand}. This is achieved by exploiting the reconfiguration in our TS (Def.~\ref{def:shadow}) and introducing a novel notion of \emph{Reconfigurable Bisimulation}. The idea is to use the TS’s state-dependent listening function $\listen$ to only listen when further input from the composition is required to proceed; otherwise, the TS disconnects from the corresponding channels.


Note that the decomposition given by Lemma~\ref{lem:triv} is strongly bisimilar to the original TS, and strong bisimilarity is preserved under parallel composition (Lemma~\ref{lem:bsimclose}). Therefore, we may apply Lemma~\ref{lem:triv} iteratively, treating each agent interface in relation to the remainder of the system. 

That is, we start from a communication-closed transition system (TS) $T$, and a partition $\{\langle \och_k, O_k \rangle\}_{k \in K}$ of the global channels $\och$ and outputs $O$. 

\begin{compactenum}
  \item We select an interface $\langle \och_k, O_k \rangle$ and trivially decompose with respect to the remainder $\langle \bigcup_{j \neq k} \och_j, \bigcup_{j \neq k} O_j \rangle$ to obtain the agent $T_k$ and the rest of the system as a single TS, which we denote by $\mathcal{C}$. We refer to $\mathcal{C}$ as the \emph{parameter}.
  
  \item We minimise the transition relation $\Delta_k$ of agent $T_k$ with respect to $\mathcal{C}$, yielding a minimised agent denoted by $[T_k]^{\mathcal{C}}$.
  
  \item We repeat steps (1)--(2) for each $k \in K$.
  \item The composition of all minimised TSs is bisimilar to the original system $T$.\vspace{0.3cm}
  
\end{compactenum}



The key novelty of reconfigurable bisimulation lies in its ability to reduce unnecessary \emph{react} transitions. Moreover, the reduction in state space is a direct consequence of eliminating such redundant input transitions. 
This stands in contrast to other bisimulation notions~\cite{CastellaniH89,MilnerS92,Sangiori93,GlabbeekW96}, which focus on reducing the number of states or eliminating \emph{internal} (hidden) $\tau$-transitions—transitions that are unobservable and thus irrelevant to our  distribution problem.

\vspace{-9pt}

\section{Parametrized Reconfigurable Bisimulation}\label{sec:bisim}
\vspace{-6pt}

We consider a system ${T} = \langle S
,s^0,\och_T,O_T,\listen, L,\Delta \rangle$  and
 a parameter  $\mathcal{C} = \langle \bcal{E}
,\epsilon^0,\och_{\bcal{C}},\break O_{\bcal{C}},\listen,
L_{\bcal{C}},\Delta
 \rangle$, constructed according to Lemma~\ref{lem:triv}. That is, $T$ and $\mathcal{C}$ only differ in their interfaces (i.e., initiate channels $Y$ and outputs $O$), and state labelling $L$. Moreover,  their composition $T\|\mathcal{C}$ is a deterministic and communication closed TS, and  there is a bijective mapping ${\bf f}$ from $S$ to $\mathcal{E}$ as explained above. 

Intuitively, we consider a system $T$  operating (synchronously) in an environment modeled by $\mathcal{C}$. Thus, bisimulation is itself parameterized and must capture the behavior of $T$ under every environmental state $\epsilon$. The goal is to identify those states of $T$ that can be made indistinguishable by removing unnecessary react transitions, and without affecting the correctness of the joint interaction.



%
%
%
%

We want to check when two states \(s_1\) and \(s_2\) of \(T\) are behaviorally equivalent within a given environment, namely under a specific parameter state \(\epsilon \in \mathcal{E}\).

\begin{definition}[Reconfigurable Bisimulation]\label{def:bisim}
Consider a TS ${T}$  to be minimised with respect to a parameter
$\mathcal{C}$, as defined above, and let ${\bf f}$ be the bijection from $S$ to $\mathcal{E}$.
An $\mathcal{E}$-indexed family of relations
$
\mathcal{R}=\{\mathcal{R}_{\epsilon}\}_{\epsilon\in\mathcal{E}},
\ 
\mathcal{R}_{\epsilon}\subseteq S\times S,
$
is a reconfigurable bisimulation if, for every $\epsilon\in\mathcal{E}$, $(s_1,s_2)\in\mathcal{R}_{\epsilon}$ implies that:


\begin{compactenum}
 \item $L(s_1) = L(s_2)$
\item For all $c\in (\och_T\cup \och_{\bcal{C}})$ we have that:
\begin{compactenum}
\item[ 2.1] 
$\mathcal{C}$ does not engage on $c$ if 
for every $\epsilon'$ we have
$(\epsilon,c,\epsilon')\notin\Delta$ 
then

$\ s_1\rTo{c}_! s'_1$  \quad implies\quad $\exists s'_2.$  $\
s_2\rTo{c}_! s'_2$ and $(s'_1,s'_2)\in\mathcal{R}_{\epsilon}$;
\item[ 2.2] 
$\mathcal{C}$ engages on $c$ if for some $\epsilon'$ we have
$(\epsilon,c,\epsilon')\in\Delta$ then
\begin{compactitem}
\item[ a.]  $\ s_1\rTo{c}_! s'_1$  \quad implies\quad $\exists s'_2.$  $\ s_2\rTo{c}_! s'_2$ and $(s'_1,s'_2)\in\mathcal{R}_{\epsilon'}$
\item[ b.] $\ s_1\rTo{c}_? s'_1$\quad  implies either
\begin{compactenum}
	\item[\rom{1}]
	$c\in\listen(s_2), \ \ \exists s'_2.\ \ s_2\
	\rTo{c}_?\  s'_2\  \ \mbox{and}\
	(s'_1,s'_2)\in\mathcal{R}_{\epsilon'}$, or
	\item[\rom{2}]
	$c\notin\listen(s_2)$ and
$
\left (
\begin{array}{cc}
	(s'_1,s_2)\notin\mathcal{R}_{\epsilon}\quad
	\mbox{implies}\quad  (s'_1,{\bf f}^{-1}(\epsilon')) \in
	\mathcal{R}_{\epsilon'}  &
%
\end{array}
\right )
$\\[1ex]
\end{compactenum}
\end{compactitem}
\end{compactenum}
\end{compactenum}

$\mathcal{R}$ is a reconfigurable bisimulation if $\mathcal{R}$ and $\mathcal{R}^{-1}$ are reconfigurable bisimulations. Two states $s_1$ and $s_2$ are reconfigurable bisimilar
with respect to a parameter state $\epsilon\in\mathcal{E}$, written
$s_1\sim_{\epsilon} s_2$,  \emph{\bf iff} there exists a
reconfigurable bisimulation $\mathcal{R}$ such that
$(s_1,s_2)\in\mathcal{R}_{\epsilon}$, i.e., $\sim_{\epsilon}=\bigcup{}{}{\set{\mathcal{R}_{\epsilon}~|~  \mathcal{R}\ \mbox{is a}\ \mbox{reconfigurable bisimulation}  }}$.
\end{definition}

\begin{figure}[t!]
\centering
\begin{tabular}{c}

$
\begin{array}{cc}\qquad
\begin{array}{c}
 \includegraphics[scale=.38]{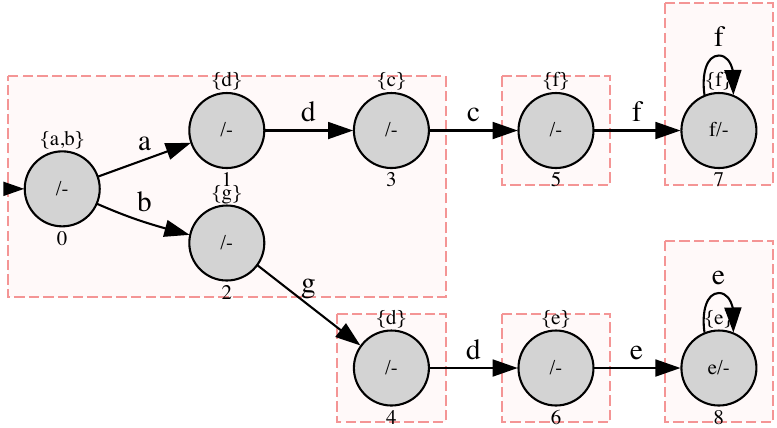}  \\

\mbox{(a) $T$  with  interface $\conf{\set{\msf{e,f}},\emptyset}$}
\end{array}
\end{array}$
$
\begin{array}{c}
 \includegraphics[scale=.38]{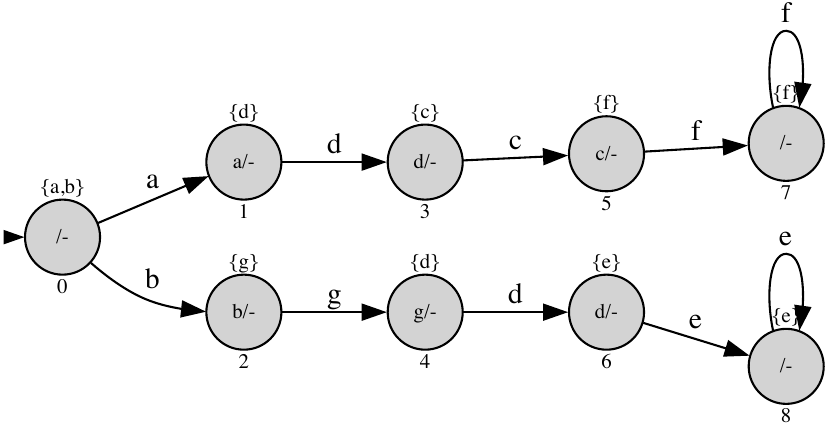}\\
\mbox{(b) $\mathcal{C}$  with  interface $\conf{\set{\msf{a,b,c,d,g}},\emptyset}$}
\end{array}$
\end{tabular}
\vspace{-2mm}
\caption{Minimizing a TS $T$ with respect to a parameter $\mathcal{C}$, illustrated with red markings  \label{fig:bsimex}
}
\vspace{-2mm}
\end{figure}

We use Fig.~\ref{fig:bsimex} to provide an intuitive explanation of Def.~\ref{def:bisim}. There, the parameter $\mathcal{C}$ in panel~(b) is used to analyze the system $T$ in panel~(a), assuming the bijection ${\bf f}$ is the identity function.

The relation $\mathcal{R}_\epsilon$  captures which pairs of system states are behaviorally equivalent, given a parameter state $\epsilon$. Namely, whenever $(s_1, s_2) \in \mathcal{R}_\epsilon$ then: (1) the two states must have the same observable label, i.e., $L(s_1) = L(s_2)$; and (2) for every channel $c \in \och_T \cup \och_{\mathcal{C}}$, we distinguish two main cases.

\vspace{-10pt}

\paragraph{2.1 Parameter non-engagement.} Whenever the \emph{System initiates and the parameter cannot react} from $\epsilon$, then every  \emph{$c$-labeled initiate transition} of $s_1$ must be matched by a corresponding  \emph{$c$-labeled initiate transition} of $s_2$, and the successors $s'_1$ and $s'_2$  remain equivalent in $\mathcal{R}_\epsilon$. That is, the parameter in $\epsilon $ remains fully uninformed about such transition; and thus the reached states $s'_1$ and $s'_2$ must remain indistinguishable in $\mathcal{R}_\epsilon$.
\begin{itemize}
\item[]
Consider system states $5$ and $6$ in Fig.~\ref{fig:bsimex}(a) w.r.t. parameter state $0$ in Fig.~\ref{fig:bsimex}(b). States $5$ and $6$ can initiate interactions, but on different channels. They are therefore distinguished, since they cannot match each other's transitions, even though parameter state $0$ cannot engage on either channel.
\end{itemize}
\vspace{-7pt}

\paragraph{2.2 Parameter engagement.} Whenever the parameter reacts or initiates on $c$ from $\epsilon$ and reaches $\epsilon'$, we distinguish two cases depending on who is the initiator:

{ a.} \emph{System initiates:} If $s_1$ performs a $c$-labeled initiate transition, then $s_2$ must also perform a matching initiate transition, and their successors must be related in $\mathcal{R}_{\epsilon'}$. Namely, the parameter is informed and reacts, and the successors $s'_1$ and $s'_2$ must be equivalent for the successor parameter state $\epsilon'$.

\begin{itemize}
\item[] 
For instance, consider system states $5$ and $7$ w.r.t. parameter state $5$. States $5$ and $7$ can initiate interactions on the same channel, with each interaction leading to same successor state. Thus, parameter state $5$ reacts to both transitions in the same manner, and the successor parameter state $7$ cannot distinguish system state $7$ from itself. Nevertheless, states $5$ and $7$ are distinguished because their state labels do not match.
\end{itemize}

{ b.} \emph{Parameter initiates:} If $s_1$ performs a $c$-labeled react transition, two cases arise. \rom{1} If $s_2$ is listening to $c$, it must also perform a matching react transition, and the successors must be related in $\mathcal{R}_{\epsilon'}$. Namely, the parameter in state $\epsilon$ is able to influence $s_2$, and thus such reaction is necessary;

\begin{itemize}
\item[] 
For instance, system states $1$ and $4$ are distinguished w.r.t. parameter state $1$ despite having matching labels and $\mathsf{d}$-transitions, since these transitions lead to states $3$ and $6$, which cannot be equated under any successor parameter state $\epsilon'$: state $6$ can initiate on $\mathsf{e}$, whereas state $3$ cannot match it.
\end{itemize}

\rom{2} If $s_2$ is not listening to $c$ then the parameter cannot directly influence $s_2$ by initiating on $c$. Thus, we must examine the necessity of the reaction in this situation. 
Recall that react transitions are blocking—that is, they can only be executed when enabled by the parameter. 
We have two cases:

\begin{compactitem}
\item \emph{The premise $(s'_1,s_2)\notin\mathcal{R}_{\epsilon}$ of the outer implication fails to hold—i.e., $s'_1$ and $s_2$ remains equivalent in $\mathcal{R}_\epsilon$}. Namely, the parameter' state $\epsilon$ deems the reaction $s_1\rTo{c}_? s'_1$ as unnecessary since $s_1$ and $s'_1$ are still equivalent from the perspective of $\epsilon$. Intuitively, a reconfiguration by removing the reaction $s_1\rTo{c}_? s'_1$  and merging $s_1$ and $s'_1$ into a single state containing $\{s_1,s'_1\}$ would result in a state that is equivalent to $s_2$ from the perspective of $\epsilon$.

\begin{itemize}
\item[] For instance, consider system states $0$ and $1$ w.r.t. parameter state $0$. By synchrony, when the parameter initiates on $\mathsf{a}$ ($0 \rTo{\mathsf{a}}_! 1$), the system $T$ must be in state $0$ to react, and thus $\mathsf{a} \notin \listen(1)$. Similarly, if the parameter initiates on $\mathsf{d}$ via $1 \rTo{\mathsf{d}}_! 3$, $T$ has already left state $0$, and $\mathsf{d} \notin \listen(0)$. In both cases, the parameter cannot distinguish states $0$ and $1$.
\end{itemize}

\item  \emph{The premise $(s'_1,s_2)\not\in\mathcal{R}_{\epsilon}$ holds.}  
In this case the suggested reconfiguration by $\mathcal{R}_\epsilon$ instead merges $s_1$, $s_2$ into a new single state $\set{s_1,s_2}$, while keeping the reaction $s_1\rTo{c}_? s'_1$.
Notice that since the parameter is the initiator on $c$ then $(\epsilon,c,\epsilon')\in\Delta$  must imply ${\bf f}^{-1}(\epsilon)\rTo{c}_? {\bf f}^{-1}(\epsilon')$ by the synchrony of $T$ and $\bcal{C}$ (Lemma~\ref{lem:triv}). That is, there is at least one state  ${\bf f}^{-1}(\epsilon)$ that can react to $c$. The latter is not necessarily $s_1$. 
Accordingly, $(s'_1,{\bf f}^{-1}(\epsilon')) \in \mathcal{R}_{\epsilon'}$ must hold to ensure that such states reach with $c$-transitions to equivalent states. 

\begin{itemize}
\item[] For instance, consider system states $2$ and $3$ under parameter state $3$. When the parameter initiates on $\mathsf{c}$ via $3 \rTo{\mathsf{c}}_! 5$, $T$ reacts from $3$ by synchrony ($3 \rTo{\mathsf{c}}_? 5$), yet $\mathsf{c} \notin \listen(2)$. Since ${\bf f}$ is the identity function, it follows that ${\bf f}^{-1}(3) \rTo{\mathsf{c}}_? {\bf f}^{-1}(5)$ and $(5, {\bf f}^{-1}(5)) \in \mathcal{R}_5$. Similarly, for the case when the parameter initiates on $\mathsf{g}$ via $2 \rTo{\mathsf{g}}_! 4$.\end{itemize}
\end{compactitem}

We use the standard lattice structure of bisimulations to derive the properties of 
reconfigurable-bisimulation.
Given two $\mathcal{E}$-indexed families of binary
relations $\mathcal{R}$ and $\mathcal{R}'$ for some set $\mathcal{E}$, we
define operations on them componentwise:
	\begin{compactitem}
		\item $\mathcal{R}\subseteq \mathcal{R}'$ iff
		$\mathcal{R}_{\epsilon}\subseteq \mathcal{R}'_{\epsilon}$ for all
		$\epsilon\in\mathcal{E}$.

		\item $\mathcal{R}\cup \mathcal{R}'$ is the $\mathcal{E}$-indexed
		family with $(\mathcal{R}\cup
		\mathcal{R}')_{\epsilon}=\mathcal{R}_{\epsilon}\cup
		\mathcal{R}'_{\epsilon}$

	\end{compactitem}

\begin{theoremrep} For all $\epsilon\in\mathcal{E}$, $\sim_{\epsilon}$ is an equivalence relation and $\{\sim_\epsilon\}_{\epsilon\in\mathcal{E}}$
is the largest reconfigurable  bisimulation.
\end{theoremrep}
\begin{proof}
\begin{compactitem}
\item[]
\item $\sim_{\epsilon}$ is an equivalence relation: It is sufficient to prove that $\sim_{\epsilon}$  is reflexive, symmetric, and transitive.
\item[]
\item $\{\sim_\epsilon\}_{\epsilon\in\mathcal{E}}$ is the largest reconfigurable  bisimulation. By definition of $\sim_{\epsilon}$, we have that:
\[\sim_{\epsilon}=\bigcup{}{}{\set{\mathcal{R}_{\epsilon}~|~  \mathcal{R}\ \mbox{is a}\ \mbox{reconfigurable bisimulation}  }}\]

That is, $\sim_{\epsilon}$ is the supremum of the set\\
$A_\epsilon=\set{\mathcal{R}_{\epsilon}~|~ \mathcal{R}\ \mbox{is a}\
\mbox{reconfigurable bisimulation}  }$. That is, every
$\mathcal{R}_{\epsilon}\in A_\epsilon$ is included in
$\sim_{\epsilon}$. It remains to show that the $\mathcal{E}$-indexed
family  $\set{\bigcup{}{}{A_\epsilon}}_{\epsilon\in\mathcal{E}}$ is a reconfigurable bisimulation.

\item[] By symmetry of $\sim_{\epsilon}$, we only need to prove that if
$(s_1,s_2)\in (\bigcup{}{}{A_\epsilon})$ for arbitrary $\epsilon$ and
$s_1$ satisfies the conditions of Def.~\ref{def:bisim} and evolves to a
new state $s'_1$ then there is a state $s'_2$ such that also $s_2$
satisfies the conditions and evolves to $s'_2$ and $(s'_1,s'_2)\in
(\bigcup{}{}{A_{\epsilon'}})$. Let us assume that the statement holds.
 Since $(s_1,s_2)\in (\bigcup{}{}{A}_{\epsilon})$ then exists a
 reconfigurable  bisimulation $\mathcal{R}$ such that
 $(s_1,s_2)\in \mathcal{R}_{\epsilon}$. But $\mathcal{R}$ is a
 reconfigurable  bisimulation, and thus if $s_1$ satisfies
 the conditions of Def.~\ref{def:bisim} and evolves to a new state
 $s'_1$ then there is a state $s'_2$ such that also $s_2$ satisfies the
 conditions and evolves to $s'_2$ and $(s'_1,s'_2)\in
 \mathcal{R}_{\epsilon'}$. That is, $(s'_1,s'_2)\in
 (\bigcup{}{}{A}_{\epsilon'})$, and thus the whole
 $\mathcal{E}$-indexed family
 $\{\sim_\epsilon\}_{\epsilon\in\mathcal{E}}$ is
 reconfigurable  bisimulation as required.
\end{compactitem}
\end{proof}

We show that there is a fixed point characterization of reconfigurable  bisimulation, by defining a monotonic function $\mcal{F}$ as follows.
Let $\mathcal{R}$ be an $\mathcal{E}$-indexed family of binary relations $\mathcal{R}_{\epsilon}\subseteq S\times S$ where $\epsilon\in\mathcal{E}$, we let $\mcal{F}(\mathcal{R})$ be the $\mathcal{E}$-indexed family of binary relations over $S$ such that, for all $s_1,s_2\in S$, we have that $(s_1,s_2)\in(\mcal{F}(\mathcal{R}))_{\epsilon}$ \emph{\bf iff} $(s_1,s_2)$ satisfies the conditions of Def.~\ref{def:bisim}.

\begin{lemmarep}\label{lem:inclusion}  $\mathcal{R}$ is a reconfigurable bisimulation iff  for all $\epsilon\in\mathcal{E},\ \mathcal{R}_{\epsilon}\subseteq \mcal{F}(\mathcal{R})_{\epsilon}$
\end{lemmarep}

\begin{proof}
\begin{compactitem}
\item[]
\item[] $\Rightarrow$: Consider that $(s_1,s_2)\in \mathcal{R}_{\epsilon}$ for arbitrary $\epsilon\in\mathcal{E}$. Since $\mathcal{R}$ is a reconfigurable bisimulation, we have that $(s_1,s_2)$  satisfies all items of Def.~\ref{def:bisim} and evolves to a pair $(s'_1,s'_2)$ such that $(s'_1,s'_2)\in \mathcal{R}_{\epsilon'}$. By  definition of $\mcal{F}(\mathcal{R})_{\epsilon}$, this implies that $(s_1,s_2)\in \mcal{F}(\mathcal{R})_{\epsilon}$. That is, $\mathcal{R}_{\epsilon}\subseteq \mcal{F}(\mathcal{R})_{\epsilon}$ for all $\epsilon\in\mathcal{E}$.
\item[]
\item[] $\Leftarrow$: Consider that $\mathcal{R}_{\epsilon}\subseteq \mcal{F}(\mathcal{R})_{\epsilon}$ for arbitrary $\epsilon\in\mathcal{E}$. This would imply that all conditions that hold for $\mcal{F}(\mathcal{R})_{\epsilon}$ also hold for $\mathcal{R}_{\epsilon}$ for all $\epsilon\in\mathcal{E}$, and thus $\mathcal{R}$ is a reconfigurable bisimulation.
\end{compactitem}
\end{proof}

\begin{lemmarep} \label{lem:mono} Function $\mcal{F}$ is monotonic on the complete lattice of $\mathcal{E}$-indexed family of binary relations ordered by componentwise inclusion, and we have that:
\begin{compactitem}
\item $\mcal{F}$ has a maximal fixed point, equivalent to $\bigcup{}{}{\set{\mathcal{R}~|~ \mathcal{R}\subseteq \mcal{F}(\mathcal{R})}}$
\item $\bigcup{}{}{\set{\mathcal{R}~|~ \mathcal{R}\subseteq \mcal{F}(\mathcal{R})}}$ = $\set{\sim_{\epsilon}}_{\epsilon\in\mathcal{E}}$
\end{compactitem}
\end{lemmarep}

\begin{proof}
\begin{compactitem}
\item[]
\item[] $\mcal{F}$ is a monotone: We need to prove that if $\mathcal{R}_{\epsilon}\subseteq \mathcal{R}'_{\epsilon}$ then $\mcal{F}(\mathcal{R})_{\epsilon}\subseteq \mcal{F}(\mathcal{R}')_{\epsilon}$ for all $\epsilon\in\mathcal{E}$. The proof derives immediately from the definition of $\mcal{F}(\mathcal{R})_{\epsilon}$.

\item[] Assume
\[
\mathcal{R}\subseteq \mathcal{R}'.
\]
We prove that
\[
\mcal{F}(\mathcal{R})_{\epsilon}\subseteq \mcal{F}(\mathcal{R}')_{\epsilon}.
\]

By Definition, we have that $\mathcal{R}\subseteq \mathcal{R}'$ iff
		$\mathcal{R}_{\epsilon}\subseteq \mathcal{R}'_{\epsilon}$ for all
		$\epsilon\in\mathcal{E}$. (Line 408)

Fix an arbitrary environment state $\epsilon\in \mathcal{E}$ and suppose
\[
(s_{1},s_{2})\in \mcal{F}(\mathcal{R})_{\epsilon}.
\]
We show that
\[
(s_{1},s_{2})\in \mcal{F}(\mathcal{R}')_{\epsilon}.
\]

Since $(s_{1},s_{2})\in \mcal{F}(\mathcal{R})_{\epsilon}$, all conditions of
Definition~5.1 hold with respect to $R$.

The equality
\[
L(s_{1})=L(s_{2})
\]
is independent of the indexed family and therefore also holds for $\mathcal{R}'$.

Now fix an arbitrary channel $c$.

Suppose first that the parameter does not engage on $c$ from
$\epsilon$. Then every initiate transition
\[
s_{1}\xrightarrow{c!}s_{1}'
\]
has a matching transition
\[
s_{2}\xrightarrow{c!}s_{2}'
\]
such that
\[
(s_{1}',s_{2}')\in \mathcal{R}_{\epsilon}.
\]
Since
\[
\mathcal{R}_{\epsilon}\subseteq \mathcal{R}'_{\epsilon},
\]
we obtain
\[
(s_{1}',s_{2}')\in \mathcal{R}'_{\epsilon},
\]
so the same witness satisfies the corresponding condition for $\mathcal{R}'$.

Suppose next that the parameter engages on $c$, namely
\[
(\epsilon,c,\epsilon')\in\Delta.
\]

If $T$ initiates on $c$, then every transition
\[
s_{1}\xrightarrow{c!}s_{1}'
\]
has a matching transition
\[
s_{2}\xrightarrow{c!}s_{2}'
\]
such that
\[
(s_{1}',s_{2}')\in \mathcal{R}_{\epsilon'}.
\]
Since
\[
\mathcal{R}_{\epsilon'}\subseteq \mathcal{R}'_{\epsilon'},
\]
it follows that
\[
(s_{1}',s_{2}')\in \mathcal{R}'_{\epsilon'},
\]
and the same witness satisfies the corresponding condition for $\mathcal{R}'$.

Assume now that the parameter initiates on $c$ and
\[
s_{1}\xrightarrow{c?}s_{1}'.
\]

If
\[
c\in ls(s_{2}),
\]
then there exists a matching transition
\[
s_{2}\xrightarrow{c?}s_{2}'
\]
such that
\[
(s_{1}',s_{2}')\in \mathcal{R}_{\epsilon'}.
\]
Again,
\[
\mathcal{R}_{\epsilon'}\subseteq \mathcal{R}'_{\epsilon'}
\]
implies
\[
(s_{1}',s_{2}')\in \mathcal{R}'_{\epsilon'},
\]
so this clause is preserved.

Finally, suppose
\[
c\notin ls(s_{2}).
\]
Then Definition~5.1 requires
\[
(s_{1}',s_{2})\notin \mathcal{R}_{\epsilon}
\Longrightarrow
(s_{1}',f^{-1}(\epsilon'))\in \mathcal{R}_{\epsilon'}.
\]
Using the logical equivalence
\[
P\Longrightarrow Q
\equiv
\neg P\vee Q,
\]
this implication is equivalent to
\[
(s_{1}',s_{2})\in \mathcal{R}_{\epsilon}
\;\vee\;
(s_{1}',f^{-1}(\epsilon'))\in \mathcal{R}_{\epsilon'}.
\]
Hence every occurrence of the indexed family is positive. Since
\[
\mathcal{R}_{\epsilon}\subseteq \mathcal{R}'_{\epsilon}
\qquad\text{and}\qquad
\mathcal{R}_{\epsilon'}\subseteq \mathcal{R}'_{\epsilon'},
\]
whichever disjunct is satisfied for $R$ is also satisfied for $R'$. Therefore,
\[
(s_{1}',s_{2})\notin \mathcal{R}'_{\epsilon}
\Longrightarrow
(s_{1}',f^{-1}(\epsilon'))\in \mathcal{R}'_{\epsilon'},
\]
and the final clause is preserved as well.

Thus every condition required for
\[
(s_{1},s_{2})\in \mcal{F}(\mathcal{R})_{\epsilon}
\]
also holds for $\mathcal{R}'$. Hence
\[
(s_{1},s_{2})\in \mcal{F}(\mathcal{R}')_{\epsilon}.
\]

Since $\epsilon$ and $(s_{1},s_{2})$ were arbitrary,
\[
\mcal{F}(\mathcal{R})_{\epsilon}\subseteq \mcal{F}(\mathcal{R}')_{\epsilon}
\qquad
\text{for every }\epsilon\in E,
\]
that is,
\[
\mcal{F}(\mathcal{R})\subseteq F(\mathcal{R}').
\]
Therefore, $\mcal{F}$ is monotone.
\item[]
\item $\mcal{F}$ has a maximal fixed point: This is a direct consequence of the monotonicity of $\mcal{F}$,  Lemma~\ref{lem:inclusion}, and Tarski's fixed point theorem.

\item[]
\item $\bigcup{}{}{\set{\mathcal{R}~|~ \mathcal{R}\subseteq \mcal{F}(\mathcal{R})}}$ = $\{\sim_\epsilon\}_{\epsilon\in\mathcal{E}}$:
\item[]
\item[]  By defintion, we have that: \[\sim_{\epsilon}\ =\bigcup{}{}{\set{\mathcal{R}_{\epsilon}~|~  \mathcal{R}\ \mbox{is an}\ \mbox{reconfigurable bisimulation}  }}\]

By Lemma~\ref{lem:inclusion}, we can rewrite $\sim_{\epsilon}$ as follows:

\[\sim_{\epsilon} =\bigcup{}{}{\set{\mathcal{R}_{\epsilon} ~|~ \mathcal{R} \subseteq \mcal{F}(\mathcal{R}) }}\]

But the largest fixed point is
$\bigcup{}{}{\set{\mathcal{R}~|~ \mathcal{R}\subseteq \mcal{F}(\mathcal{R})}}$, and thus this is equivalent to $\{\sim_\epsilon\}_{\epsilon\in\mathcal{E}}$.
\end{compactitem}
\end{proof}

Using Lemma~\ref{lem:mono}, we can  (albeit in a na\"ive manner) solve reconfigurable bisimulation equivalence by computing the largest fixed point of the monotonic function $\mcal{F}$:
initialise
$\set{\mathcal{R}^0_{\epsilon}}$ to
$\mathcal{R}^0_\epsilon=S \times S$ for every $\epsilon$ and compute
$\set{\mathcal{R}^{i+1}_\epsilon}=\mcal{F}(\set{\mathcal{R}^{i}_\epsilon})$
until stability, i.e., for some $n$, every $\epsilon\in\mathcal{E}$, ${\mathcal{R}^{n}_{\epsilon}}={\mathcal{R}^{n+1}_{\epsilon}}$.  The algorithm returns the
indexed family $\set{\mathcal{R}_{\epsilon}~|~
\epsilon\in\mathcal{E}}$, which determines the equivalence from the point of view of each parameter state $\epsilon$. In practice, representing each $\mathcal{R}_\epsilon$ as $S \times S$ is inefficient, as it results in $O(|S|^2)$ pairs per parameter state $\epsilon$. 


A more efficient way is to use a set of  partitions of
$T$-states from the point of view of every state $\epsilon$ of the
parameter, i.e., $\set{S/{\sim}_\epsilon}_{\epsilon\in\mathcal{E}}
=\set{\set{[s]^{{\sim_\epsilon}} ~|~ s\in S}_{\epsilon}~|~
\epsilon\in\mathcal{E}}$, where $[s]^{{\sim}_\epsilon}$ is the
equivalence class of $s$ based on $\sim_\epsilon$. The latter is an equivalent representation but costs $O(|S|)$ per parameter state $\epsilon$, and $O(|S|^2)$ across all $\mathcal{R}_\epsilon$.

\vspace{-7pt}

\paragraph{\bf Extracting the Minimized Transition System:}

Let $\{\mathcal{R}_\epsilon\}_{\epsilon\in \mathcal{E}}$  be the maximal reconfigurable bisimulation of $T$ with respect to  $\mathcal{C}$, inducing for each $\epsilon$ a partition $\{ [s]^{{\sim_\epsilon}} ~|~ s \in S \}$.
Our goal is to derive a single partition of $S$ that consistently minimizes $T$ with respect to all relevant parameter states.

Recall that $T$ and $\mathcal{C}$ arise from the trivial decomposition and that each state $s\in S$ has a unique parameter state ${\bf f}(s)\in\mathcal{E}$ enabling same transitions with dual roles. Moreover, the composition is deterministic and synchronous, so whenever $T$ reaches $s$, $\mathcal{C}$ is uniquely at ${\bf f}(s)$. Hence, only the partition induced by ${\bf f}(s)$ is relevant for $s$, and $s$ must belong to some block in this partition, i.e., $[s]^{\sim_{{\bf f}(s)}}$.

However, $[s]^{\sim_{{\bf f}(s)}}$ may contain another state $s'$ whose own $[s']^{\sim_{{\bf f}(s')}}$  disagrees about $s$ being equivalent to $s'$.
Such states can never be equivalent to $s$, as equivalence must be symmetric. 
To ensure global consistency, we retain only \emph{mutual} equivalences. For each $s\in S$, define
\[
C_s=\{s'\mid s'\in[s]^{\sim_{{\bf f}(s)}} \text{ and } s\in[s']^{\sim_{{\bf f}(s')}}\}.
\]

We then compute a maximal common partition, by a reduction to a clique partitioning problem in an agreement graph~\cite{GJ79,Bodlaender94,Gramm01} as follows:
\begin{compactenum}
\item Construct the  graph $G=(S,E)$, where $\{s,s'\}\in E$ iff $s'\in C_s$ and $s\in C_{s'}$.
\item Partition $S$ into cliques of $G$, yielding a partition $\bcal{P}^{\rulename{nd}}$ that captures maximal mutual agreement. However, this step may incur inconsistencies in $\bcal{P}^{\rulename{nd}}$. That is, whenever an equivalence class $C_s$ is refined into distinct blocks $B_1,B_2\in \bcal{P}^{\rulename{nd}}$ due to disagreement, such blocks may act as splitters for a third block $B_3\in \bcal{P}^{\rulename{nd}}$. Thus, we do the following step.
\item Apply Paige--Tarjan refinement~\cite{PaigeTarjan1987} to $\bcal{P}^{\rulename{nd}}$, splitting blocks only w.r.t. transitions targeting different blocks, to obtain a consistent partition $\bcal{P}$.
\end{compactenum}

\noindent
Finally, the refined partition $\bcal{P}$ is used to construct the minimized quotient transition system of $T$. The clique partitioning problem is NP-hard~\cite{GJ79}; we therefore use a standard polynomial-time greedy algorithm in our tool-support.

\begin{definition}[Compression]\label{def:quo}
Given a TS ${T} = \langle S
,s^0,\och_T, O_T,\listen_T, L_T,\Delta_T \rangle$  and
 a parameter TS $\mathcal{C} = \langle \mathcal{E}
,\epsilon^0,\och_{\mathcal{C}},O_{\mathcal{C}},\listen_{\mathcal{C}},
L_{\mathcal{C}},\Delta_{\mathcal{C}}
 \rangle$, construct a  compressed TS
$[T]^{{\mathcal{C}}}=\langle \mcal{B},B^0,\och_T,O_T$,
$ \listen,L,\Delta_{[T]^{{\mathcal{C}}}}\rangle$ based on the obtained  partition $\bcal{P}$:
\begin{compactitem}
\item $\mcal{B}=\set{B~|~ B\in \bcal{P}}$ and $B^{0}\in\bcal{P}$, s.t., $s^0\in B^{0}$

\item $\och_T$ and $O_T$ are preserved.
\item $\Delta_{[T]^{{\mathcal{C}}}}=
 \{  (
\begin{array}{c}
	B,c,B'
\end{array} 
 )  |
\begin{array}{l}
B\neq B',\ \exists s\in B,\ s'\in B',\  \mbox{s.t.,}\
 (s,c,s')\in\Delta_T
\end{array}
 \}
$

\qquad\quad\ \ $\bcup{}{}{} \{ (
\begin{array}{c}
	B,c,B
\end{array} 
 )
 |
\begin{array}{l}
 c\in \och_T,\
 \forall s\in B, \exists s'\in B,\  \mbox{s.t.,}\ (s,c,s')\in\Delta_T
\end{array}
	\}
$

\item
  $\listen(B) = \set{c~|~(B,
	c,
	B')\in\Delta_{[T]^{{\mathcal{C}}}}}$
\item $L(B)=L_T(s)$ for an arbitrary $s\in B$
\end{compactitem}
\end{definition}

The compressed TS
$[T]^{{\mathcal{C}}}$ preserves state labelling and the interface of $T$, but removes unnecessary react transitions by modifying both the listening function and the transition relation.
The function $\listen(B)$ is constructed based on the transition
relation $\Delta_{[T]^{{\mathcal{C}}}}$.
The number of states of $[T]^{{\mathcal{C}}}$  is bounded
by $|S|$.

 The following lemma shows that, unlike react transitions, initiate transitions cannot be eliminated: a block \(B\) implements an initiate transition only if all its states do, and therefore all such transitions—including those internal to \(B\)—must be implemented. In contrast, a react transition is implemented by \(B\) if at least one state enables it and the transition leads to a different block \(B'\); internal react transitions (within \(B\)) are thus omitted. The lemma also establishes determinism of the induced transition system \( [T]^{\mathcal{C}} \), requiring that all states in a block enabling the same react transition reach the same successor block.

\begin{lemmarep}\label{lem:prop}
Given a compressed TS $[T]^{\mathcal{C}}$ then for all $B\in\mcal{B}$, we have that:
\begin{compactenum}
\item $B \rTo{c}_! B'$ implies $\forall s\in B.\ s\rTo{c}_! s'$ and $s'\in B'$ 
\item $B \rTo{c}_? B'$ implies $\exists s\in B.\ s\rTo{c}_? s'$ s.t., $s'\in B'$ and $B\neq B'$
\item $B \rTo{c}_? B'$ implies $\forall s_1,s_2\in B.\ s_1\rTo{c}_? s'_1$ and $s_2\rTo{c}_? s'_2$ implies $s'_1,s'_2\in B'$
\end{compactenum}
\end{lemmarep}
\begin{proof}
The proof follows from the definition of $\rTo{c}_!$, the definition of $\Delta_{[T]^{\mathcal{C}}}$, and the construction of the partition $\bcal{P}$. We prove each statement separately as follows:

\begin{compactenum}
\item 
By the definition of the initiate transition, we have that $B \rTo{c}_! B'$ implies $(B, c, B') \in \Delta_{[T]^{\bcal{C}}}$ and $c \in \och_{[T]^{\bcal{C}}}$. That is, $[T]^{\bcal{C}}$ must have $c$ in its interface channels to be able to initiate on $c$. Moreover, $B \in \bcal{P}$ by the construction of $\Delta_{[T]^{\bcal{C}}}$, and $B \neq \emptyset$ by the construction of $\bcal{P}$ (otherwise $\bcal{P}$ cannot be a partition). Now, we distinguish two cases:

\begin{compactenum}
\item 
\emph{Self-loops} $(B = B')$: This case follows immediately from $B \neq \emptyset$, the definition of $\rTo{c}_!$, and the construction of $\Delta_{[T]^{\bcal{C}}}$ (Def.~\ref{def:quo}). Note that the Paige-Tarjan refinement step during the construction of $\bcal{P}$ ensures that all states in a block enabling the same transition exit to the same block. 

\item 
\emph{Exit transitions} $(B \neq B')$: 

By the construction of $\Delta_{[T]^{\bcal{C}}}$ (Def.~\ref{def:quo}), there exist $s \in B$, $s' \in B'$ such that $(s, c, s') \in \Delta_T$. Moreover, since $\och_T = \och_{[T]^{\bcal{C}}}$, we conclude $c \in \och_T$. By the definition of $\rTo{c}_!$, this implies $s \rTo{c}_! s'$, i.e., there is at least one state $s \in B$ that can initiate on $c$.

\vspace{1ex}

It remains to show that this also holds for every other state $s'' \in B$. By the construction of $\bcal{P}$, we have $B \neq \emptyset$ and for all $s, s'' \in B$, $s\in C_{s''}\mbox{ and } s''\in C_s$ holds, equivalently, 
\[
(s, s'') \in \mathcal{R}_{\mathbf{f}(s)} \quad \text{and} \quad (s, s'') \in \mathcal{R}_{\mathbf{f}(s'')}.
\]

Since $s \rTo{c}_! s'$, by reconfigurable bisimulation (Def.~\ref{def:bisim}, item (3a)) and $(s, s'') \in \mathcal{R}_{\mathbf{f}(s)}$, there exists $s'''$ such that
\[
s'' \rTo{c}_! s''', \quad \text{and} \quad (s', s''') \in \mathcal{R}_{\mathbf{f}(s')}.
\]
Similarly, from $(s, s'') \in \mathcal{R}_{\mathbf{f}(s'')}$, we also get $(s', s''') \in \mathcal{R}_{\mathbf{f}(s''')}$.

\vspace{1ex}

Now, we must show that $s''' \in B'$. Suppose for contradiction that $s' \in B'$, $s''' \in B''$, and $B' \neq B''$. From above, we have:
\[
(s', s''') \in \mathcal{R}_{\mathbf{f}(s')} \quad \text{and} \quad (s', s''') \in \mathcal{R}_{\mathbf{f}(s''')},
\]
which implies:
\[
s''' \in C_{s'} \quad \text{and} \quad s' \in C_{s'''}.
\]
By symmetry, the same holds for $s''$ and $s'''$. 

If $B' \neq B''$, then there must exist some state $\hat{s} \in C_{s'}$ such that $\hat{s} \notin C_{s'''}$. The clique partitioning step in the construction of $\bcal{P}$ would then either:
\begin{itemize}
  \item Problematic case: merge $\hat{s}$ with $s'$ and separate $s'''$ into a different block, 
  \item Symmetric case: merge $\hat{s}$ with $s$ and separate $s'''$ into a different block, or
  \item Safe case: merge $s'$ with $s'''$ directly.
\end{itemize}

Focus on the first case: if $\hat{s}$ is merged with $s'$ but not with $s'''$, then this must trigger the  Paige-Tarjan refinement step as the predecessors of $s'$ and $s'''$ (i.e., $s$ and $s''$) exist in the same block $B$, enabling two $c$-labeled transitions, one would now feed $B'$ and one would feed $B''$. Thus,  $B'$ and $B''$ are both splitter for ${B}$. Consequently, ${B}$ must have been split and $s$ and $s''$ could not co-exist in $B$, or $\hat{s}\in C_{s'''}$ which contradicts the assumption that $\hat{s} \notin C_{s'''}$.

Hence, such a split is forbidden, and $s''' \in B'$ as required.
\end{compactenum}

\item 
Follows immediately from the definition of $\rTo{c}_?$ and the construction of $\Delta_{[T]^{\bcal{C}}}$ (Def.~\ref{def:quo}).

\item 
Assume, towards contradiction, that $s_1 \rTo{c}_? s_1'$ and $s_2 \rTo{c}_? s_2'$ with $s_1' \in B' , s_2'\in B''$ respectively, and $B' \neq B''$. Since $s_1, s_2 \in B$, we must have  $s_1 \in C_{s_2}$ and $s_2 \in C_{s_1}$, or equivalently,
\[
(s_1, s_2) \in \mathcal{R}_{\mathbf{f}(s_1)} \quad \text{and} \quad (s_1, s_2) \in \mathcal{R}_{\mathbf{f}(s_2)}.
\]

By the definition of reconfigurable bisimulation (Def.~\ref{def:bisim}), since both $s_1$ and $s_2$ react to the same action $c$, their successors $s_1'$ and $s_2'$ must be related. That is,
\[
(s_1', s_2') \in \mathcal{R}_{\mathbf{f}(s_1')} \quad \text{and} \quad (s_1', s_2') \in \mathcal{R}_{\mathbf{f}(s_2')},
\]
or equivalently, \[s_1' \in C_{s_2'}\mbox{ and }s_2' \in C_{s_1'}\]

If $B' \neq B''$, then there must exist $s \in C_{s_1'}$ but $s \notin C_{s_2'}$ (or the symmetric case).  The clique partitioning step in the construction of $\bcal{P}$ would then merge $s$ with $s_1'$ and separate $s_2'$ into a different block (or merge $s_1'$ with $s_2'$). Consider again the problematic case where $s$ is merged with $s_1'$ but not with $s_2'$.

In this case, this must trigger the  Paige-Tarjan refinement step as the predecessors of $s'_1$ and $s'_2$ (i.e., $s_1$ and $s_2$) exist in the same block $B$, enabling two $c$-labeled transitions, one would now feed $B'$ and one would feed $B''$. Thus,  $B'$ and $B''$ are both splitter for ${B}$. Consequently, ${B}$ must have been split and $s_1$ and $s_2$ could not co-exist in $B$, or ${s}\in C_{s'_2}$ which contradicts the assumption that ${s} \notin C_{s'_2}$.
\end{compactenum}
\end{proof}

We prove our main result by reducing each agent with respect to the remainder of the composition. The resulting composed system is then bisimilar to the original. We define the context in which bisimilarity is established. We focus on the set of communications initiated internally by individual agents in the composition. Communication initiated externally is not relevant, as the distribution problem assumes a fixed number of agents and no external interactions. 
\vspace{-5pt}
\paragraph{\bf Initiate Transitions.}
For every pair of transition systems $T_i$ and $T_j$ with disjoint initiate channel sets $\och_i \cap \och_j = \emptyset$, we define the set of internally initiated transitions as:\quad 
$
\Init{T_i}{T_j} = \left\{ (s, c, s') ~\middle|~ c \in \Yij,\ (s, c, s') \in \Deltaij \right\}
$

\begin{theoremrep} \label{corollary:main}
Let $T$ be a deterministic and communication-closed transition system, trivially decomposed via Lemma~\ref{lem:triv} into the set $\set{T_1,\dots,T_n}$ for $n\geq 2$. Then: $
(T_1 \| \dots\|T_n)\ \sim ([T_1]^{[T_2 \| \dots\|T_n]} \| \dots\| [T_n]^{[T_1 \| \dots\|T_{n-1}]}) ,
$
where $\sim$ is interpreted under the corresponding initiated transitions $\InitN{T_1}{T_n}$.

\end{theoremrep}
\begin{toappendix}
To prove Theorem~\ref{corollary:main}, we need to introduce the following auxiliary theorem and lemma.

\begin{theoremrep}[Compression preserves $\sim$]\label{thm:main}
Let $T$ be a deterministic and communication-closed transition system, trivially decomposed via Lemma~\ref{lem:triv} into $T_1$ and $T_2$, with $\och_1 \cap \och_2 = \emptyset$. Then:
\[
T_1 \| T_2\sim [T_1]^{[T_2]} \| T_2 \sim T_1 \| [T_2]^{[T_1]} ,
\]
where each $\sim$ is interpreted under the corresponding initiated transitions $\Init{T_1}{T_2}$.
\end{theoremrep}


\begin{proof}
It is sufficient to prove that the initial state of $(T_1 \| T_2)$ is strongly bisimilar to the initial state of $(T_1 \| [T_2]^{[T_1]})$, where $\sim$ is interpreted under the corresponding initiated transitions $\Init{T_1}{T_2}$.  The dual case $([T_1]^{T_2} \| T_2)$ follows by commutativity of parallel composition $\|$.

 We will use the symbol $s_1$ to range over the states of $T_1$, $s_2$ for the states of $T_2$,  and $B$ for the states of $[T_2]^{[T_1]}$. That is, we need to prove that
\[
(s^0_1, s^0_2) \sim (s^0_1, B^0),
\]
and we only focus on initiation from the set $\Init{T_1}{T_2}$. This limits the verification of strong bisimilarity to the first two items of Def.~\ref{def:sbisim}.

Recall that the original composition $(T_1 \| T_2)$ is communication-closed; that is, all transitions are self-initiated by construction (cf. Lemma~\ref{lem:triv}). The objective of the proof is to show that compression does not alter the behaviour with respect to a specific parameter. This is fundamentally different from showing that compression preserves behaviour with respect to \emph{any} parameter — the latter is clearly false and unnecessarily restrictive.


We write
\[
s_2 \sim_{s_1} B
\]
as shorthand for $[s_2]^{\sim_{s_1}} \supseteq B$. That is, if we denote the reconfigurable bisimulation by the family of relations $\{ \mathcal{R}_{s \in T_1} \}$, then the equivalence class of $s_2$ under $\mathcal{R}_{s_1}$ contains the set $B$. Equivalently, for every $s \in B$, we have $s_2 \sim_{s_1} s$.

By the construction of the compression operator (Def.~\ref{def:quo}), all compressed states $B$ belongs to the partition $\bcal{P}$ constructed before, and satisfy the property: 
\[
B \neq \emptyset \quad \text{and} \quad \forall s, s' \in B,\  (s'\in C_s \mbox{ and } s\in C_s') \text{ holds}.
\]
Moreover,
\[
(s'\in C_s \mbox{ and } s\in C_s') \mbox{ implies } (s, s') \in \mathcal{R}_{\mathbf{f}(s)} \text{ and } (s, s') \in \mathcal{R}_{\mathbf{f}(s')}.
\]

By trivial decomposition Lemma~\ref{lem:triv}, we have that $T_1$ and $T_2$ are isomorphic and fully synchronous transition systems with complementary roles. That is, every state $s_2$ of $T_2$ has a unique companion state $s_1$ in the composition $T_1\| T_2$ such that $s_1={\bf f}(s_2)$, and both $s_2$ and ${\bf f}(s_2)$ enable the same set of transitions, but with dual interaction roles (i.e., if $s_2$ reacts to an action $c$, then ${\bf f}(s_2)$ initiates on $c$, and vice versa).

 Given $s_1 = \mathbf{f}(s_2)$ and $s_2 \in B$, there exists a reconfigurable bisimulation $\mathcal{R}_{s_1}$ that relates all states in $B$. Thus, $B$ is a sub-block in the partition induced by $\mathcal{R}_{s_1}$ (i.e., the partition before computing maximal agreement, and thus $[s_2]^{\sim_{s_1}} \supseteq B$), and so $s_2 \sim_{s_1} B$.  Note that $[s_2]^{\sim_{s_1}} \supseteq B$ because $B$ might have been refined.

 The rationale is that  for all states $s \in B$, the associated reconfigurable bisimulations $\mathcal{R}_{\mathbf{f}(s)}$ must agree with every other $\mathcal{R}_{\mathbf{f}(s')}$ such that $s' \in B$; otherwise, they cannot coexist in $B$ due to the consistency property above.

In this proof, we will rely on this property, since the composition $T_1 \| T_2$ is constructed with a built-in bijective mapping $\mathbf{f}$ between the states of $T_1$ and $T_2$.

\vspace{1ex}
Now, everything is in place to complete the proof. We construct the following relation:
\[
\mathcal{R} = \{\, ((s_1, s_2), (s_1, B)) \mid \ s_2 \in B \mbox{ and } s_2 \sim_{s_1} B \}.
\]

This relation $\mathcal{R}$ is defined based on the reconfigurable bisimulation induced by:

\begin{itemize}
  \item[] The quotient structure of $T_2$ with $T_1$ as parameter, resulting in $[T_2]^{[T_1]}$.
\end{itemize}

We additionally require that $(s_1, s_2)$ be a reachable state in the composition $T_1 \| T_2$, as we aim to faithfully mimic  the behaviour of $T_1 \| T_2$.

Hence, for the reconfigurable bisimulation on $T_2$ with $T_1$ as parameter, yielding $[T_2]^{[T_1]}$, we have $s_1 = \mathbf{f}(s_2)$ (again by Lemma~\ref{lem:triv}).

We now prove:
\begin{enumerate}
\item $((s^0_1, s^0_2), (s^0_1, B^0)) \in \mathcal{R}$,
\item $\mathcal{R}$ is a strong bisimulation when $\sim$ is interpreted under the corresponding initiated transitions $\Init{T_1}{T_2}$.
\end{enumerate}

\paragraph{(1) Base case:}
By construction of $[T_2]^{[T_1]}$, we have that  $s^0_2 \in B^0$, and thus $s^0_2 \sim_{s^0_1} B^0$. Since $s^0_1 = \mathbf{f}(s^0_2)$, we get that $(s^0_1, s^0_2)$ is a reachable state in the composition $T_1\|T_2$. Hence, $((s^0_1, s^0_2), (s^0_1, B^0)) \in \mathcal{R}$.

\paragraph{(2) Inductive step:}
Assume $((s_1, s_2), (s_1, B)) \in \mathcal{R}$ with $(s_1, s_2)$ reachable in $T_1 \| T_2$. We show label preservation and mutual simulation of transitions in both directions:

\begin{enumerate}
\item $L(s_1,s_2)=L(s_1,B)$: This follows by composition (Def.~\ref{def:comp}) and the construction  $B$ (Def.~\ref{def:quo}). The latter preserves labelling of states. 

\item Forward direction: $(s_1, s_2) \rTo{c}_! (s'_1, s'_2)$ implies $(s_1, B) \rTo{c}_! (s'_1, B')$ and $((s'_1, s'_2), (s'_1, B')) \in \mathcal{R}$. 

\item[] By definition $(s_1,s_2)\rTo{c}_!(s'_1,s'_2)$ {\bf iff} $c\in (\och_{T_1}\cup \och_{T_2}),\ c\in\listen((s_1,s_2))$ and $((s_1,s_2),c,(s'_1,s'_2))\in\Delta_{T_1\| T_2}$. We need to show that also $(s_1, B) \rTo{c}_! (s'_1, B')$, i.e.,  $c\in\listen(s_1,B)$, $c\in (\och_{T_1}\cup \och_{[T_2]^{[T_1]}})$ and $((s_1,B),c,(s'_1,B'))\in \Delta_{T_1\| [T_2]^{[T_1]}}$, and $((s'_1, s'_2), (s'_1, B')) \in \mathcal{R}$.

\item[] We conduct a case analysis on
$((s_1,s_2),c,(s'_1,s'_2))\in\Delta_{T_1\| T_2}$. By composition
(Def.~\ref{def:comp}), we have the following cases:
\begin{itemize}
  \item 
  If $c \in \och_{T_1}$ and $(s_1, c, s'_1) \in \Delta_{T_1}$: That is, $s_1 \rTo{c}_! s'_1$.

  By assumption, $(s_1, s_2)$ is a reachable state in the composition $T_1 \| T_2$, i.e., $s_1 = \mathbf{f}(s_2)$ for the parameter $T_1$. By trivial decomposition (Lemma~\ref{lem:triv}), this implies that $s_2 \rTo{c}_? s'_2$, i.e., $s_2$ must react. By the determinism of the original TSs, we also have $s'_1 = \mathbf{f}(s'_2)$.

  Since $s_2 \in B$, we get $s_2 \sim_{s_1} B$. Now, by reconfigurable bisimulation $s_2 \sim_{s_1} B$, we distinguish two cases:

  \begin{itemize}
    \item 
    $c \in \listen(B)$ and $(B, c, B') \in \Delta_{[T_2]^{[T_1]}}$.

    \item[] 
    By reconfigurable bisimulation (Def.~\ref{def:bisim}), we have that $B \rTo{c}_? B'$ and $s'_2 \sim_{s'_1} B'$. It remains to show that $s'_2$ is still in $B'$ and is not filtered out by the Clique partitioning refinement step.

    Recall from Lemma~\ref{lem:prop} that $B \rTo{c}_? B'$ implies that there exists $s \in B$ such that $s \rTo{c}_? s'$ with $s' \in B'$ and $B \neq B'$. 

    Assuming $s \ne s_2$, we must show that $s'_2$ also ends up in $B'$. It is sufficient to argue that $B'$ cannot be split in this case, and both $s'$ and $s'_2$ coexist in $B'$. This follows from item~3 in Lemma~\ref{lem:prop}, hence $s'_2 \in B'$ as required.

    \item[] 
    By parallel composition (Def.~\ref{def:comp}), we have that $c \in \listen(s_1, B)$, $c \in (\och_{T_1} \cup \och_{[T_2]^{[T_1]}})$, and
    \[
    ((s_1, B), c, (s'_1, B')) \in \Delta_{T_1 \| [T_2]^{[T_1]}}.
    \]

    \item[] 
    That is, $(s_1, B) \rTo{c}_! (s'_1, B')$ and $((s'_1, s'_2), (s'_1, B')) \in \mathcal{R}$, as required.

    \item 
    $c \notin \listen(B)$ and $B = B'$.

    \item[] 
    That is, $B$ did not implement the transition $s_2 \rTo{c}_? s'_2$. By assumption $s_2 \sim_{s_1} B$, and thus by reconfigurable bisimulation (Def.~\ref{def:bisim}), this is only possible when $s'_2 \sim_{s_1} s_2$, as otherwise $B$ would have implemented the react transition. 

    Therefore, the premise of the top implication in Def.~\ref{def:bisim}~2.2(b)(ii) is false and the implication holds vacuously. This means $s'_2$ is still in $B$.

    However, we also need to show that $s'_2 \sim_{s'_1} s_2$ holds. This follows from the fact that $s'_2 \in B$ and $s'_1 = \mathbf{f}(s'_2)$. 

    By construction of $B$, all states $s \in B$ must have reconfigurable bisimulations $\mathcal{R}_{\mathbf{f}(s)}$ that agree with every other $\mathcal{R}_{\mathbf{f}(s')}$ such that $s' \in B$, or else they cannot coexist in $B$. 

    Since $s_2, s'_2 \in B$ and $s'_1 = \mathbf{f}(s'_2)$, it must be the case that $s'_2 \sim_{s'_1} s_2$ holds.

    \item[] 
    By parallel composition (Def.~\ref{def:comp}), we have that $c \in \listen(s_1, B)$, $c \in (\och_{T_1} \cup \och_{[T_2]^{[T_1]}})$, and
    \[
    ((s_1, B), c, (s'_1, B)) \in \Delta_{T_1 \| [T_2]^{[T_1]}}.
    \]
    That is, $B$ discarded the communication and did not change.

    \item[] 
    Therefore, $(s_1, B) \rTo{c}_! (s'_1, B)$ and $((s'_1, s'_2), (s'_1, B)) \in \mathcal{R}$, as required.
  \end{itemize}
 \item 
  If $c \in \och_{T_2}$ and $(s_2, c, s'_2) \in \Delta_{T_2}$: That is, $s_2 \rTo{c}_! s'_2$.

  By assumption, $(s_1, s_2)$ is a reachable state in the composition $T_1 \| T_2$, i.e., $s_1 = \mathbf{f}(s_2)$ for the parameter $T_1$. By trivial decomposition (Lemma~\ref{lem:triv}), this implies that $s_1 \rTo{c}_? s'_1$, i.e., $s_1$ must react. By the determinism of the original TSs, we also have $s'_1 = \mathbf{f}(s'_2)$.

  Since $s_2 \in B$, we get $s_2 \sim_{s_1} B$. 
  
  Now, by reconfigurable bisimulation $s_2 \sim_{s_1} B$, we have that:
    \item[] 
     $B \rTo{c}_? B'$ and $s'_2 \sim_{s'_1} B'$. It remains to show that $s'_2$ is still in $B'$ and is not filtered out by the Clique partitioning refinement step. This follows from item~1 in Lemma~\ref{lem:prop}, hence $s'_2 \in B'$ as required.

    \item[] 
    By parallel composition (Def.~\ref{def:comp}), we have that $c \in \listen(s_1, B)$, $c \in (\och_{T_1} \cup \och_{[T_2]^{[T_1]}})$, and
    \[
    ((s_1, B), c, (s'_1, B')) \in \Delta_{T_1 \| [T_2]^{[T_1]}}.
    \]

    \item[] 
    That is, $(s_1, B) \rTo{c}_! (s'_1, B')$ and $((s'_1, s'_2), (s'_1, B')) \in \mathcal{R}$, as required.

\end{itemize}

\item  Backward direction: $(s_1, B) \rTo{c}_! (s'_1, B')$  implies $(s_1, s_2) \rTo{c}_! (s'_1, s'_2)$ and $((s'_1, s'_2), (s'_1, B')) \in \mathcal{R}$. 

\item[] By definition $(s_1, B) \rTo{c}_! (s'_1, B')$ {\bf iff} $c\in (\och_{T_1}\cup \och_{[T_2]^{[T_1]}})$, $c\in\listen(s_1,B)$ and $((s_1,B),c,(s'_1,B'))\in \Delta_{T_1\| [T_2]^{[T_1]}}$. We need to show that also $c\in (\och_{T_1}\cup \och_{T_2}),\ c\in\listen((s_1,s_2))$ and $((s_1,s_2),c,(s'_1,s'_2))\in\Delta_{T_1\| T_2}$, and and $((s'_1, s'_2), (s'_1, B')) \in \mathcal{R}$

\item[] We conduct a case analysis on
$((s_1,B),c,(s'_1,B'))\in \Delta_{T_1\| [T_2]^{[T_1]}}$. By composition
(Def.~\ref{def:comp}), we have the following cases:
\begin{itemize}
  \item 
  If $c \in \och_{T_1}$ and $(s_1, c, s'_1) \in \Delta_{T_1}$: That is, $s_1 \rTo{c}_! s'_1$.

  We distinguish two cases:

  \begin{itemize}
    \item 
    $c \in \listen(B)$ and $(B, c, B') \in \Delta_{[T_2]^{[T_1]}}$.

    \item[] 
    Then it must be the case that $B \rTo{c}_? B'$.  By assumption, $(s_1, s_2)$ is a reachable state in the composition $T_1 \| T_2$, i.e., $s_1 = \mathbf{f}(s_2)$ for the parameter $T_1$. By trivial decomposition (Lemma~\ref{lem:triv}), this implies that $s_2 \rTo{c}_? s'_2$, i.e., $s_2$ must react. By the determinism of the original TSs, we also have $s'_1 = \mathbf{f}(s'_2)$.

  Since $s_2 \in B$, we get $s_2 \sim_{s_1} B$, and by reconfigurable bisimulation $s'_2 \sim_{s'_1} B'$. It remains to show that $s'_2$ is still in $B'$ and is not filtered out by the Clique partitioning refinement step.
 This follows from item~3 in Lemma~\ref{lem:prop}, hence $s'_2 \in B'$ as required.

    \item[] 
    By parallel composition (Def.~\ref{def:comp}), we have that $c \in \listen(s_1, s_2)$, $c \in (\och_{T_1} \cup \och_{T_2})$, and
    \[
    ((s_1, s_2), c, (s'_1, s'_2)) \in \Delta_{T_1 \| T_2}.
    \]

    \item[] 
    That is, $(s_1, s_2) \rTo{c}_! (s'_1, s'_2)$ and $((s'_1, s'_2), (s'_1, B')) \in \mathcal{R}$, as required.

    \item 
    $c \notin \listen(B)$ and $B = B'$.

    \item[] 
    That is, $B$ did not implement the transition $s_2 \rTo{c}_? s'_2$, but we have already established that $s_2 \rTo{c}_? s'_2$ must happen. By assumption $s_2 \sim_{s_1} B$, and thus by reconfigurable bisimulation (Def.~\ref{def:bisim}), this is only possible when $s'_2 \sim_{s_1} s_2$, as otherwise $B$ would have implemented the react transition. In other words, the reaction  $s_2 \rTo{c}_? s'_2$ did not exit $B$ and by construction of $\Delta_{[T_2]^{[T1]}}$ self-loop reactions are not implemented. 

However, we also need to show that $s'_2 \sim_{s'_1} s_2$ holds. This follows from the fact that $s'_2 \in B$ and $s'_1 = \mathbf{f}(s'_2)$. 

    By construction of $B$, all states $s \in B$ must have reconfigurable bisimulations $\mathcal{R}_{\mathbf{f}(s)}$ that agree with every other $\mathcal{R}_{\mathbf{f}(s')}$ such that $s' \in B$, or else they cannot coexist in $B$. 

    Since $s_2, s'_2 \in B$ and $s'_1 = \mathbf{f}(s'_2)$, it must be the case that $s'_2 \sim_{s'_1} s_2$ holds.

    \item[] 
    By parallel composition (Def.~\ref{def:comp}), we have that $c \in \listen(s_1, s_2)$, $c \in (\och_{T_1} \cup \och_{T_2})$, and
    \[
    ((s_1, s_2), c, (s'_1, s'_2)) \in \Delta_{T_1 \| T_2}.
    \]

    \item[] 
    Therefore, $(s_1, s_2) \rTo{c}_! (s'_1, s'_2)$ and $((s'_1, s'_2), (s'_1, B)) \in \mathcal{R}$, as required.
  \end{itemize}
 \item 
  If $c \in \och_{[T_2]^{[T_1]}}$ and $(B, c, B) \in \Delta_{[T_2]^{[T_1]}}$: That is, $B \rTo{c}_! B'$.

  Since $s_2 \in B$, we get $s_2 \sim_{s_1} B$, and by reconfigurable bisimulation, we must have that  $s_2 \rTo{c}_! s'_2$ and $B'\sim_{s'_1} s'_2$. However, the latter is only true if $s'_2$ ended up in $B'$ and was not filtered out by the Clique partitioning refinement step. This follows from item~1 in Lemma~\ref{lem:prop}, hence $s'_2 \in B'$ as required.
  
  By assumption, $(s_1, s_2)$ is a reachable state in the composition $T_1 \| T_2$, i.e., $s_1 = \mathbf{f}(s_2)$ for the parameter $T_1$. By trivial decomposition (Lemma~\ref{lem:triv}), this implies that $s_1 \rTo{c}_? s'_1$, i.e., $s_1$ must react. By the determinism of the original TSs, we also have $s'_1 = \mathbf{f}(s'_2)$.
  
    \item[] 
    By parallel composition (Def.~\ref{def:comp}), we have that $c \in \listen(s_1, s_2)$, $c \in (\och_{T_1} \cup \och_{T_2})$, and
    \[
    ((s_1, s_2), c, (s'_1, s'_2)) \in \Delta_{T_1 \| T_2}.
    \]

    \item[] 
    That is, $(s_1, s_2) \rTo{c}_! (s'_1, s'_2)$ and $((s'_1, s'_2), (s'_1, B')) \in \mathcal{R}$, as required.

\end{itemize}

\end{enumerate}

\paragraph{Conclusion:}
The relation $\mathcal{R}$ satisfies the  conditions of strong bisimulation  when $\sim$ is interpreted under the corresponding initiated transitions $\Init{T_1}{T_2}$. Therefore, the initial state $(s^0_1, s^0_2)$ of $T_1 \| T_2$ is strongly bisimilar to the initial state $(s^0_1, B^0)$ of $T_1 \| [T_2]^{[T_1]}$.
\end{proof}

\begin{lemmarep}
Let $T_1, T_2, T_3, T_4$ be deterministic transition systems such that for all defined compositions $T_i \| T_j$, we have $\och_i \cap \och_j = \emptyset$.

Assume:
\[
(T_1 \| T_2) \sim (T_1 \| T_3) \quad \text{and} \quad (T_1 \| T_2) \sim (T_2 \| T_4).
\]

Then it follows that:
\[
(T_1 \| T_2) \sim (T_3 \| T_4),
\]
where each $\sim$ is interpreted under the corresponding initiated transitions $\Init{T_i}{T_j}$.
\label{lem:gen}
\end{lemmarep}

\begin{proof}

To prove that $({T}_1 \| {T}_2)\ \sim\ ({T}_3 \| {T}_4)$, it is sufficient to prove that their initial states are equivalent by the definition of $\sim$. That is, we need to show:
\[
(s^0_1,s^0_2)\sim (s^0_3,s^0_4)
\]
under all transitions in $\msf{Init}[{T}_1 \| {T}_2]$ and respectively $\msf{Init}[{T}_3 \| {T}_4]$.

We construct a relation $\mcal{R}$ as follows:
\[
\mcal{R} = \left\{((s,r),(p,q))~\middle|~ (s,r)\sim(s,p) \text{ and } (s,r)\sim (r,q)\right\}
\]
\medskip

Now, we must show:

\begin{enumerate}
    \item $((s^0_1,s^0_2),(s^0_3,s^0_4)) \in \mcal{R}$, and
    \item $\mcal{R}$ is a (strong, closed) bisimulation under the initiated transitions in $\msf{Init}[T_1 \| T_2]$ and $\msf{Init}[T_3 \| T_4]$.
\end{enumerate}

\paragraph{(1)} The initial state pair is in $\mcal{R}$ by construction, since:
\[
(s^0_1,s^0_2)\sim(s^0_1,s^0_3) \text{ and } (s^0_1,s^0_2)\sim(s^0_2,s^0_4)
\]
are guaranteed by the hypothesis.

\paragraph{(2)} To show that $\mcal{R}$ is a bisimulation, let $((s_1,s_2),(s_3,s_4)) \in \mcal{R}$ and verify the bisimulation conditions.

\begin{itemize}

\item \textbf{Label agreement:} Show that $L((s_1,s_2)) = L((s_3,s_4))$.\\
By the assumption of $\mcal{R}$, we have:
\[
(s_1,s_2)\sim(s_1,s_3) \quad \text{and} \quad (s_1,s_2)\sim(s_2,s_4)
\]
From $(s_1,s_2)\sim(s_1,s_3)$ and Def.~\ref{def:comp}, we deduce $L(s_2)=L(s_3)$.\\
From $(s_1,s_2)\sim(s_2,s_4)$ and Def.~\ref{def:comp}, we deduce $L(s_1)=L(s_4)$.\\
Thus:
\[
L((s_1,s_2)) = L((s_3,s_4))
\]

\item \textbf{Transition preservation:} Show that whenever $((s_1,s_2),(s_3,s_4)) \in \mcal{R}$, the following two directions hold:

\begin{enumerate}
\item If $(s_1,s_2) \rTo{c}_! (s'_1,s'_2)$, then there exist $(s'_3, s'_4)$ such that
\[
(s_3,s_4) \rTo{c}_! (s'_3,s'_4) \quad \text{and} \quad ((s'_1,s'_2),(s'_3,s'_4)) \in \mcal{R}
\]

\begin{itemize}

\item Case 1: $s_1 \rTo{c}_! s'_1$ and $s_2$ reacts or discards\\
From the assumption $(s_1,s_2)\sim(s_1,s_3)$, we deduce:
\[
(s_1,s_3)\rTo{c}_! (s'_1,s'_3)
\]
Similarly, $(s_1,s_2)\sim(s_2,s_4)$ gives:
\[
(s_2,s_4)\rTo{c}_! (s'_2,s'_4)
\]
Since $\och_1 \cap \och_2 = \emptyset$ and $s_1$ initiates, $s_2$ cannot initiate. Thus, $s_4$ must be the initiator in $(s_2,s_4)$, and $s_3$ must react or discard in $(s_1,s_3)$. In both cases, by Def.~\ref{def:comp}, we get:
\[
(s_3,s_4)\rTo{c}_! (s'_3,s'_4)
\]
Finally, from the transitions, we have:
\[
(s'_1,s'_2)\sim(s'_1,s'_3) \quad \text{and} \quad (s'_1,s'_2)\sim(s'_2,s'_4)
\]
so that $((s'_1,s'_2),(s'_3,s'_4)) \in \mcal{R}$.

\item Case 2: $s_1$ reacts or discards and $s_2 \rTo{c}_! s'_2$\\
Symmetric to Case 1. $(s_2,s_4)\rTo{c}_! (s'_2,s'_4)$ and $(s_1,s_3)\rTo{c}_! (s'_1,s'_3)$, giving:
\[
(s_3,s_4)\rTo{c}_! (s'_3,s'_4)
\]
and again:
\[
(s'_1,s'_2)\sim(s'_1,s'_3),\quad (s'_1,s'_2)\sim(s'_2,s'_4)
\Rightarrow ((s'_1,s'_2),(s'_3,s'_4)) \in \mcal{R}
\]

\end{itemize}

\item If $(s_3,s_4) \rTo{c}_! (s'_3,s'_4)$, then there exist $(s'_1, s'_2)$ such that
\[
(s_1,s_2) \rTo{c}_! (s'_1,s'_2) \quad \text{and} \quad ((s'_1,s'_2),(s'_3,s'_4)) \in \mcal{R}
\]

\begin{itemize}

\item Case 1: $s_3 \rTo{c}_! s'_3$ and $s_4$ reacts or discards\\
By $(s_1,s_2)\sim(s_1,s_3)$, we have:
\[
(s_1,s_3)\rTo{c}_! (s'_1,s'_3)
\Rightarrow (s_1,s_2)\rTo{c}_! (s'_1,s'_2)
\]
with $(s'_1,s'_2)\sim(s'_1,s'_3)$.\\
Also, $(s_1,s_2)\sim(s_2,s_4) \Rightarrow (s_2,s_4)\rTo{c}_! (s'_2,s'_4)$ with $(s'_1,s'_2)\sim(s'_2,s'_4)$.\\
So $((s'_1,s'_2),(s'_3,s'_4)) \in \mcal{R}$.

\item Case 2: $s_3$ reacts or discards and $s_4 \rTo{c}_! s'_4$\\
By $(s_1,s_2)\sim(s_2,s_4)$, we have:
\[
(s_2,s_4)\rTo{c}_! (s'_2,s'_4)
\Rightarrow (s_1,s_2)\rTo{c}_! (s'_1,s'_2)
\]
with $(s'_1,s'_2)\sim(s'_2,s'_4)$.\\
Also, $(s_1,s_2)\sim(s_1,s_3) \Rightarrow (s_1,s_3)\rTo{c}_! (s'_1,s'_3)$ with $(s'_1,s'_2)\sim(s'_1,s'_3)$.\\
So again, $((s'_1,s'_2),(s'_3,s'_4)) \in \mcal{R}$.

\end{itemize}

\end{enumerate}

\item \textbf{Other cases:} Transitions of the form $(s_1,s_2)\rTo{c}_?$ or $(s_3,s_4)\rTo{c}_?$ are excluded, as we only consider communication \emph{initiated} from the composition. That is, we restrict to $\msf{Init}[T_1 \| T_2]$ and $\msf{Init}[T_3 \| T_4]$.

\end{itemize}

\end{proof}

Now, using Theorem~\ref{thm:main} and Lemma~\ref{lem:gen}, we show how to generalise this result to $n$-agent decomposition.
\end{toappendix}

\begin{proof}
The proof proceeds by induction on the length of the composition $n$.
\begin{itemize}
\item \textbf{Base case $n=2$:} in this case, we need to prove the following:
  \[
  (T_1 \|T_2)\ \sim ([T_1]^{[T_2] } \| [T_2]^{[T_1]})
  \]

  By Theorem~\ref{thm:main}, we have that $(T_1 \|T_2)\ \sim\ (T_1 \|[T_2]^{[T_1]})\ $ and $\ (T_1 \|T_2)\ \sim\ ([T_1]^{[T_2]} \|T_2)\ $. By commutativity of $\|$, we have that $(T_1 \|T_2)\ \sim\ (T_1 \|[T_2]^{[T_1]})\ $ and $\ (T_1 \|T_2)\ \sim\ (T_2 \|[T_1]^{[T_2]})\ $

  By Lemma~\ref{lem:gen}, we conclude that $(T_1 \|T_2)\ \sim\ ([T_1]^{[T_2]} \|[T_2]^{[T_1]})\ $ as required.

  \item \textbf{Induction Hypothesis (IH):} we assume that the lemma holds for all $k$ such that $2\leq k\leq n$.

  \item \textbf{Induction Step (IS):} we show that it holds for $k+1$.

By IH, we have the following:

\[
\underbrace{{T_1\ \| \dots \|\ T_k}\ \|\ {T_{k+1}}}_{T}\ \sim ({[T_1]^{[T_2 \| \dots\|T_{k+1}]}\ \| \dots\|\ [T]^{[T_1 \| \dots\| T_{k-1}\| T_{k+1}]}_k}\|\ {T_{k+1}})
\]
By associativity and commutativity of $\|$, we have

\[
{\underbrace{T_{k+1}}_{S}\ \|\ (\underbrace{T_1\ \| \dots\ \|\ T_{k}}_{R}})\ \sim  \underbrace{T_{k+1}}_{S}\| \underbrace{({[T_1]^{[T_2 \| \dots\|T_{k+1}]}\ \| \dots\|\ [T]^{[T_1 \| \dots\| T_{k-1}\ \|T_{k+1}]}_k})}_{P}\qquad\qquad~(1)
\]

By commutativity of $\|$, we have

\[
\underbrace{{T_1\ \| \dots \|\ T_k}\ \|\ {T_{k+1}}}_{T} \sim \ {\underbrace{T_{k+1}}_{S}\ \|\ (\underbrace{T_1\ \| \dots\ \|\ T_{k}}_{R}})
\]

By Theorem~\ref{thm:main}, we have

\[
 {\underbrace{T_{k+1}}_{S}\ \|\ (\underbrace{T_1\ \| \dots\ \|\ T_{k}}_{R}})\ \sim\ {\underbrace{[T]^{[T_1 \| \dots\| T_{k}]}_{k+1}}_{Q}\ \|\ (\underbrace{T_1\ \| \dots\ \|\ T_{k}}_{R}})
\]

By commutativity of $\|$, we have

\[
 {\underbrace{T_{k+1}}_{S}\ \|\ (\underbrace{T_1\ \| \dots\ \|\ T_{k}}_{R}})\ \sim\ {(\underbrace{T_1\ \| \dots\ \|\ T_{k}}_{R})\ \|\ \underbrace{[T]^{[T_1 \| \dots\| T_{k}]}_{k+1}}_{Q}}\qquad\qquad~(2)
\]

Apply Lemma~\ref{lem:gen} on $(1)+(2)$, we have

\[
 {\underbrace{T_{k+1}}_{S}\ \|\ (\underbrace{T_1\ \| \dots\ \|\ T_{k}}_{R}})\qquad \sim
\]

\[
 {\underbrace{({[T_1]^{[T_2 \| \dots\|T_{k+1}]}\ \| \dots\|\ [T]^{[T_1 \| \dots\| T_{k-1}\| T_{k+1}]}_k})}_{P}\ \|\ \underbrace{[T]^{[T_1 \| \dots\| T_{k}]}_{k+1}}_{Q}}\qquad
\]

By commutativity of $\|$, we have

\[
 (\underbrace{{T_1\ \| \dots \|\ T_k}\ \|\ {T_{k+1}}}_{T})\sim ([T_1]^{[T_2 \| \dots\|T_{k+1}]}\ \| \dots\|\ [T]^{[T_1 \| \dots\| T_{k-1}\| T_{k+1}]}_k\|\ [T]^{[T_1 \| \dots\| T_{k}]}_{k+1})\qquad
\]

as required.
\end{itemize}
\end{proof}

From Theorem~\ref{corollary:main}, our strategy is sound:
We perform the trivial decomposition of a TS based on separate interfaces of each agent.
We then minimise each agent separately based on reconfigurable bisimulation w.r.t. the rest of the system.
This removes some react transitions and accordingly restricts the listening function of the states it changes.
Then, we take the agents that were minimised separately and compose them together.
The result is bisimilar to the original.


\vspace{-10pt}

\section{Teamwork Synthesis \& Reconfigurable Coordination}\label{sec:app}
\vspace{-10pt}
%
%
%

Distributed synthesis~\cite{PnueliR90,DBLP:journals/lmcs/Gimbert22} is undecidable due to partial information imposed by a fixed communication structure~\cite{FinkbeinerS05}.
To address this, we recast the problem as reconfigurable multi-agent coordination, allowing agents to connect to information sources on demand.
We introduce \emph{teamwork synthesis}, which departs from classical distributed synthesis by removing the communication structure from the problem input and instead computes the necessary set of interactions.
\vspace{-7pt}

 \begin{definition}[Teamwork synthesis]\label{def:syntm}
	Given a sequential system $M$ with interface $\conf{Y,{O}}$, a partition of the interface to ${\{\langle Y_k,O_k\rangle\}_{k\in K}}$  for $K=\{1,\dots,n\}$ such that $Y=\bigcup_{k\in K}Y_k$ and $O=\bigcup_{k\in K}O_k$, the \emph{teamwork
	synthesis} problem is to find a set of 
    agents implemented by 
    asynchronous TSs $\{T_k\}_{k\in K}$
	such that $T=T_1\|\dots\| T_n$ satisfies
	$\mathcal{L}_M\cong \mathcal{L}_T$, where $\cong$ is language isomorphism.
\end{definition}

\vspace{-7pt}

Def.~\ref{def:syntm} expects a sequential system $M$ with interface $\conf{Y,{O}}$ as the input. Thus, if the original specifications are given as a formula, say $\varphi$ over $\conf{Y,{O}}$, then we first solve a single-agent synthesis of $\varphi$ over $\conf{Y,{O}}$ as a preprocessing step. The solution of the latter is the sequential system $M$. If $\varphi$ is not realizable for a single-agent, then it is clearly impossible to distribute.

We use a \emph{specialised} Mealy machine as the sequential system input.
\vspace{-7pt}

\begin{definition}\label{def:mealy} A Mealy machine $M=\langle Q,\ Y,\ {O},\ \iota,\delta\rangle$ is defined below: 
	\begin{compactitem}
		\item $Q$ is the set of states of $M$.

		\item $Y$ is the input alphabet and ${O}$ is a set of output variables.

		\item $\delta \subseteq Q\times({Y}\times \Exp{O})\times Q$ is the
		\emph{deterministic} transition (and output) of $M$.

	\item $\iota=(c_0,O_0),q_0$ is the initial labeled transition, where
	$q_0\in Q$ is the initial state, $c_0\in {Y}$ is the initial input, and
	$O_0\in \Exp{O}$ is the output assignment.
	\end{compactitem}
\end{definition}

A run $r$ of $M$ is an infinite sequence
$(c_0,O_0)q_0(c_1,O_1)q_1(c_2,O_2)q_2\dots$ such that for
all
$k\geq 0: (q_k,c_{k+1},O_{k+1},q_{k+1})\in \delta$, $q_0$ is the
initial state, and $c_0$ and $O_0$ are the initial input and
output assignment.
A word of $M$ is the projection of a run $r$ to transition labels.
That is, the word $w$ generated by $r$ is as follows
$w=(c_0,O_0)(c_1,O_1)(c_2,O_2)(c_3,O_3)\dots$.
The language of $M$, denoted $\mathcal{L}_M$, is the set of infinite
sequences in $({Y}\times \Exp{O})^{\omega}$ generated by $M$.

Given a Mealy machine $M$,
we say that a TS $T$ implements $M$ {\bf iff} $\mathcal{L}_T \cong
\mathcal{L}_M$.

\noindent
Unlike the Mealy machine with interface $\conf{Y,O}$, whose language is in $({Y}\times \Exp{O})^{\omega}$, the language of a TS (Def.~\ref{def:shadow}) over the same interface $\conf{Y,O}$ is in $(\Exp{Y}\times \Exp{O})^{\omega}$. Thus, we need to define a language isomorphism mapping languages of the form $({Y}\times \Exp{O})^{\omega}$ to  the form $(\Exp{Y}\times \Exp{O})^{\omega}$. Formally, given two alphabets $\Sigma$ and $\Pi$, we say that $\Sigma$ is \emph{isomorphic} to $\Pi$ if there exists an isomorphism $I: \Sigma \rightarrow \Pi$. We extend $I$ to sequences over $\Sigma$ and to sets of sequences.
We say that language $L\subseteq \Sigma^*$ is isomorphic to $L'\subseteq \Pi^*$ if there exists an isomorphism between $\Sigma$ and $\Pi$ such that $I(L)=L'$.
In such a case we write $L\cong L'$.

\vspace{-7pt}

\begin{lemma}[TS implementation]\label{lem:mtots} For Mealy machine $M=\langle Q,\ Y,\ {O},\ \iota,\delta\rangle$, we construct a communication closed TS ${T} = \langle S
,s^0,Y, O,\listen, L,\Delta \rangle$ satisfying:
\begin{compactenum}
\item For all $s\in S\backslash\set{s^0}$, we have that $L^c(s)$ is a singleton.
\item $|S|=|\delta|+1$ and $s^0\notin \Delta(s,c)$ for all $c\in Y$ and $s\in S$
\item $\mathcal{L}_T\cong\mathcal{L}_M$
\end{compactenum}

\end{lemma}
\begin{proof}
We construct  $T$ as follows:
\begin{compactitem}
\item $S=\delta\cup \{s^0\}$, where $s^0=\iota$. 


\item The interface is $\conf{Y, O}$. That is, the input alphabet is the set of channel names for $T$ and $O$ is a matching set of variables to represent the output.

\item $L((q,(c,O),q'))=(\set{c},O)$ and $L(s^0)=(\set{c_0},O_0)$


\item $\Delta = \left \{ \left . \left (
\begin{array}{c}
	((c_0,O_0),q_0),
	c,
	(q_0,(c,O),q')
\end{array} 
\right )
\right |
\begin{array}{l}
 (q_0,(c,O),q')\in\delta \\[2ex]
\end{array}
\right	\}
 \quad \bcup{}{}{}
$
\\

\qquad 
 $\left \{ \left . \left (
\begin{array}{c}
	(q,(c,O),q'),\
c',
	(q',(c',O'),q'')
\end{array} 
\right )
\right |
\begin{array}{l}
 (q,(c,O),q')\in\delta, \\
 (q',(c',O'),q'')\in\delta
\end{array}
\right	\}
$

\item $\listen(s)=\set{c~|~(s,c,s')\in \Delta}$.
\end{compactitem}

By construction, $T$ is communication-closed, i.e., all transitions in $\Delta$ are self-initiated transitions (owned) because all channels are included in $T$'s interface. That is, $T$ can run independently. Clearly $\mathcal{L}_T\cong\mathcal{L}_M$ with the isomorphism $I:Y\times 2^O \rightarrow Y_{\{\}}\times 2^O$, where $Y_{\{\}}=\set{\set{c} ~|~ c\in Y}$ and $I(c,O)=(\set{c},O)$.\qed
\end{proof}

This construction transforms the mealy machine transitions into states in the TS implementation as follows: given every two \emph{consecutive transitions} in $\delta$, e.g., $(q,(c,O),q')$ and $(q',(c',O'),q'')$, we transform them into two  states in the TS $T$, such that the label of a TS state is isomorphic to the label of the corresponding mealy transition, e.g., $L((q,(c,O),q'))=(\set{c},O)$. Moreover, we create a corresponding transition in $\Delta$ of the TS $T$ to relate such states, and label it with the reached state's channel label, e.g., $((q,(c,O),q'),{\bf c'},(q',({\bf c'},O'),q''))\in\Delta$. By doing so, we have injected artefact communications that explain the transitioning  among states in the TS $T$ as environment signals.

\begin{figure}[h!]
\centering
\begin{tabular}{c}
$
\begin{array}{c}
 \includegraphics[scale=.5]{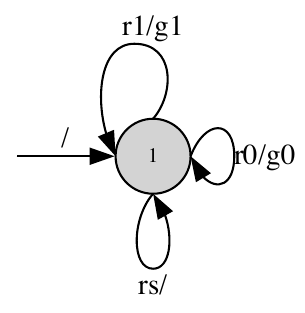}\\
\mbox{(a) $M$  with  interface $\conf{\set{\msf{rs,r0,r1}},\set{\msf{g0,g1}}}$}
\end{array}$
$
\begin{array}{cc}\quad
\begin{array}{c}
 \includegraphics[scale=.4]{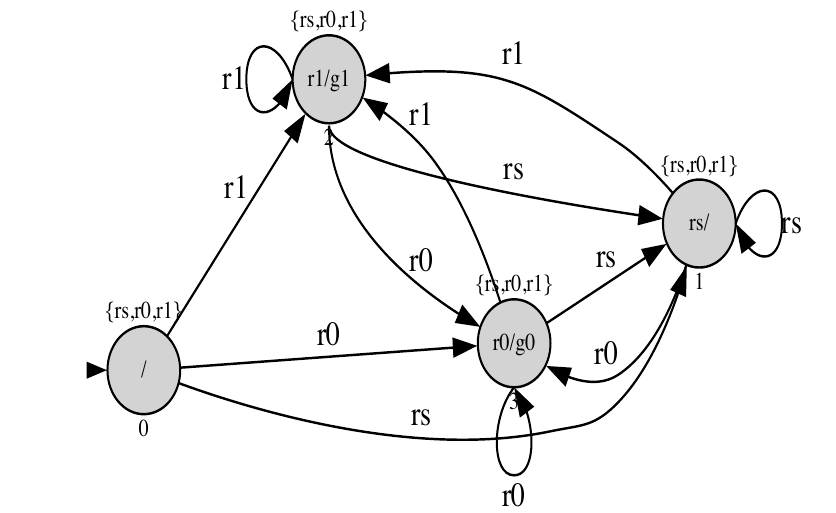}  \\

\mbox{(b) $T$  with  interface $\conf{\set{\msf{rs,r0,r1}},\set{\msf{g0,g1}}}$}
\end{array}
\end{array}$
\end{tabular}
\vspace{-2mm}
\caption{Mealy machine $M$ and its TS implementation $T$\label{fig:spec}
}
\vspace{-2.5mm}
\end{figure}

Since the language of the TS $T$ is projected onto state labels, a transition is interpreted as follows: from a state labeled $(\set{c},O)$, the system initiates a signal $c'$ and transitions to a state labeled $(\set{c'},O')$, recording the signal $c'$ and the induced output assignment $O'$. Thus, the sequence $(\set{c},O),(\set{c'},O')$ in the language of $T$ is isomorphic to $(c,O),(c',O')$ in the Mealy language.
 Figure~\ref{fig:spec} shows Arbiter Reset 2 benchmark (Sect.~\ref{sec:eval}) and its TS implementation. 

We can cast the teamwork synthesis as a multi-agent coordination problem as follows: we partition the input signals $Y$ and output variables $O$ among agents according to the supplied partition. 
Each agent initiates on its owned signals and reacts to those owned by others, possibly modifying its own outputs. Thus, every agent acts as part of the environment for the others.
More precisely:
 we distribute $T$ constructed by Lemma~\ref{lem:mtots} on a partition of agent interfaces ${\{\langle Y_k,O_k\rangle\}_{k\in K}}$ trivially using
Lemma~\ref{lem:triv}; 
and we reduce the unnecessary interactions per agent from the previous step
using
reconfigurable bisimulation. The result is a set of asynchronous systems whose joint language is isomorphic to the original. 

\vspace{-10pt}

\section{Experimental Evaluation}\label{sec:eval}
\vspace{-10pt}
We implemented our approach in the prototype synthesis engine \texttt{SynTM}~\cite{abdalrahman202621807991}, which supports both TS-based distribution and \ltl{}-based teamwork synthesis. \texttt{SynTM} accepts either a TS or a general \ltl{} formula~$\varphi$ with interface-distribution, and it minimally rewrites~$\varphi$ to fit our specialized Mealy-machine model (Def.~\ref{def:mealy}). Using \texttt{Strix}~\cite{strix}, the tool first synthesizes a single-agent implementation; if $\varphi$ is realizable, the resulting Mealy machine is converted to a TS (Def.~\ref{lem:mtots}) and then distributed and minimized. \texttt{SynTM} further supports strong bisimulation checking, connectivity metrics, and semantic simulation.

We evaluated \texttt{SynTM} on benchmarks selected for their clean, distributable interfaces, including the $\mathsf{Path}_n$ family~\cite{GenestGMW10} for highly connected systems, Arbiter and AMBA specifications~\cite{strix,seml,bosy}, custom multi-agent scenarios inspired by mutual exclusion, product-line literature~\cite{rs,DIppolitoBPU13} and contributed case studies (e.g., Sect.~\ref{sec:case}), and regular-expression-based cases for analyzing qualitative properties. A key limitation is the lack of suitable distributed benchmarks, since existing suites target single-agent synthesis and are hard to adapt to multi-agent settings. For example, most AMBA benchmarks hardcode general LTL formulas using auxiliary variables that are not part of the system’s actual interface.

\begin{table}[t!]
\caption{%
Evaluation results.
}
\vspace{-1pt}
\label{tab:all}
\begin{minipage}[t]{.50\linewidth}
\centering
\scalebox{0.65}{
\begin{tabular}[t]{l|c|ccc|r|r|r}
\hline
Name \& number & $|T|$ & \multicolumn{3}{c|}{{rBsim}} & sBsim & Zielonka & $\listen$ \\
\cline{3-5}
& & $|min|$ & $|max|$ & $|avg|$ & $|avg|$ & & \\
\hline
$\mathsf{Path}_{3}$~\cite{GenestGMW10} & 4 & 2 & 2 & 2 & 4 & \(\mathrm{4^{3^4} . |T|^{3^2}}\) & 0.250 \\
$\mathsf{Path}_{4}$ & 7 & 3 & 5 & 4 & 5.5 & \(\mathrm{4^{4^4} . |T|^{4^2}}\) & 0.379 \\
$\mathsf{Path}_{5}$ & 11 & 6 & 6 & 6 & 11 & \(\mathrm{4^{5^4} . |T|^{5^2}}\) & 0.375 \\
$\mathsf{Path}_{6}$ & 16 & 6 & 11 & 8.5 & 16 & \(\mathrm{4^{6^4} . |T|^{6^2}}\) & 0.336 \\
$\mathsf{Path}_{7}$ & 22 & 11 & 11 & 11 & 22 & \(\mathrm{4^{7^4} . |T|^{7^2}}\) & 0.295 \\
$\mathsf{Path}_{8}$ & 29 & 11 & 16 & 13.5 & 29 & \(\mathrm{4^{8^4} . |T|^{8^2}}\) & 0.255 \\
$\mathsf{Path}_{9}$ & 37 & 16 & 16 & 16 & 37 & \(\mathrm{4^{9^4} . |T|^{9^2}}\) & 0.223 \\
$\mathsf{Path}_{10}$ & 46 & 16 & 22 & 19 & 46 & \(\mathrm{4^{10^4} . |T|^{10^2}}\) & 0.199 \\
$\mathsf{Path}_{11}$ & 56 & 22 & 22 & 22 & 56 & \(\mathrm{4^{11^4} . |T|^{11^2}}\) & 0.179 \\
$\mathsf{Path}_{12}$ & 67 & 22 & 29 & 25.5 & 67 & \(\mathrm{4^{12^4} . |T|^{12^2}}\) & 0.162 \\
\hline
Arbiter Simple 3~\cite{strix} & 4 & 2 & 2 & 2 & 2 & n/a & 0.250 \\
Arbiter Simple 4 & 5 & 2 & 2 & 2 & 2 & n/a & 0.200 \\
Arbiter Reset 2 & 4 & 2 & 3 & 2.5 & 2.5 & n/a & 0.375 \\
Arbiter Reset 3 & 5 & 2 & 3 & 2.33 & 2.33 & n/a & 0.267 \\
Arbiter Reset 4 & 6 & 2 & 3 & 2.25 & 2.25 & n/a & 0.208 \\
Arbiter Full 2 & 9 & 6 & 9 & 7.5 & 7.5 & n/a & 0.722 \\
Arbiter Full 3 & 27 & 17 & 23 & 19.67 & 19.67 & n/a & 0.691 \\
Arbiter Priority 2 & 5 & 3 & 5 & 4 & 4 & n/a & 0.600 \\
Arbiter Priority 3 & 28 & 9 & 23 & 16 & 16 & n/a & 0.452 \\
\hline
\end{tabular}}
\end{minipage}%
\begin{minipage}[t]{.57\linewidth}
\centering
\scalebox{0.65}{
\begin{tabular}[t]{l|c|ccc|r|r}
\hline
Name \& number & $|T|$ & \multicolumn{3}{c|}{{rBsim}} & {sBsim} & $\listen$ \\
\cline{3-5}
& & $|min|$ & $|max|$ & $|avg|$ & $|avg|$ & \\
\hline
Product Line 3 & 9 & 3 & 5 & 4.33 & 7.67 & 0.210 \\
AMBA Decomposed 2 & 4 & 3 & 3 & 3 & 3.33 & 0.417 \\
AMBA Decomposed 4 & 5 & 3 & 3 & 3 & 3.25 & 0.300 \\
\hline
\rulename{Alg}: Dekker 5 & 147 & 3 & 9 & 5.4 & 114.2 & 0.002 \\
\rulename{Alg}: Peterson 5 (\cite{rs}) & 78 & 3 & 8 & 5 & 49.6 & 0.004 \\
Time-Sharing v1~(Sect.~\ref{sec:case}) & 23 & 3 & 6 & 4.75 & 17 & 0.027 \\
Time-Sharing v2 & 30 & 3 & 7 & 5 & 27 & 0.019 \\
Product Line 3 (\cite{DIppolitoBPU13}) & 13 & 3 & 7 & 4.67 & 11.33 & 0.222 \\
Product Line 4 & 13 & 3 & 3 & 3 & 10.75 & 0.167 \\
\hline
$(a^{2}b^{2}c(af^{*}+be^{*}+cg^{*}))$ & 12 & 7 & 10 & 8.5 & 12 & 0.097 \\
$(a^{2}b^{2}a(af^{*}+be^{*}+cg^{*}))$ & 12 & 7 & 10 & 8.5 & 12 & 0.153 \\
$(a^{6}b(af^{*}+be^{*}+cg^{*}))$ & 14 & 7 & 12 & 9.5 & 14 & 0.083 \\
$(ad^{3}cf^{*}+bd^{3}ge^{*})$ & 13 & 4 & 13 & 8.5 & 12 & 0.054 \\
$(aca^{2}(af^{*}+be^{*}+cg^{*}))$ & 11 & 7 & 9 & 8 & 11 & 0.136 \\
$(acab(af^{*}+be^{*}+cg^{*}))$ & 11 & 7 & 9 & 8 & 11 & 0.106 \\
$(aca^{5}b(af^{*}+be^{*}+cg^{*}))$ & 15 & 7 & 13 & 10 & 15 & 0.078 \\
$(adcb(af^{*}+be^{*}+cg^{*}))$ & 11 & 7 & 9 & 8 & 11 & 0.114 \\
$(ab(af^{*}+be^{*}+cg^{*}))$ & 9 & 7 & 7 & 7 & 9 & 0.130 \\
$(a^{2}b(af^{*}+be^{*}+cg^{*}))$ & 10 & 7 & 8 & 7.5 & 10 & 0.117 \\
$(da^{5}b(af^{*}+be^{*}+cg^{*}))$ & 14 & 7 & 12 & 9.5 & 14 & 0.071 \\
\hline
\end{tabular}}
\end{minipage}
\vspace{-.5cm}
\end{table}

\vspace{-7pt}

\paragraph{\bf Results and Analysis:}\label{sec:results}

Our results are summarized in Table~\ref{tab:all}, grouped by the four benchmark categories. We start from the size of the sequential system $T$ that is either the input or the result of single-agent synthesis. We trivially decompose $T$ into a number of agents (the postfix of the benchmark's name). We then report the size of the minimised agents relative to $|T|$ under both reconfigurable bisimulation (rBsim) and strong bisimulation (sBsim). The former shows the min/max/avg size of the minimized agents while the latter only shows the average size. For the $\mathsf{Path}_n$ benchmarks, we additionally include the known bound on the size of distribution according to Zielonka’s construction if applicable due to lack of tool support. Finally, we include the weighted connectivity metric~$\listen$, which reflects how frequently agents must react within the distributed system and provides an aggregated measure of inter-agent communication.

Measuring connectivity is subtle, as it depends on the number of react transitions per state, the agent’s share of the global state space, and how often the global composition visits its states. Our metric approximates these factors:
\bigskip

$\qquad\qquad\qquad
\listen = \frac{1}{|Ag| \cdot |\overline{S}|} \sum_{T_k \in Ag}
\frac{\sum_{s \in S_k} |\listen(s) \setminus Y_k|}{|C| - |Y_k|}
$

\medskip

The inner fraction counts an agent’s react transitions in each state, $|\listen(s)\setminus Y_k|$, normalized by the maximum possible, $|C|-|Y_k|$. We average this over all agents $|Ag|$ and scale by the reduced system size $|\overline{S}|$ according to strong bisimulation, which removes graph symmetries irrelevant to connectivity.

\vspace{-10pt}

\paragraph{The $\msf{Path}_n$ benchmark.}
Across $10$ instances, \texttt{SynTM} produced compact distributed agents, typically between $\tfrac{1}{3}|T|$ and $\tfrac{1}{2}|T|$, in sharp contrast to Zielonka’s construction, which grows exponentially to \(\mathrm{4^{12^4} \cdot |T|^{12^2}}\) when $n=12$. Connectivity reduction stabilized after initial fluctuations, demonstrating the robustness of our approach even in highly connected systems. As expected, strong bisimulation offered little reduction due to limited symmetry.

\vspace{-10pt}

\paragraph{Benchmarks from synthesis tools.}
We selected 12 instances of Arbiter and AMBA benchmarks of varying sizes and configurations from the \texttt{Strix} synthesis engine~\cite{strix}. These benchmarks are beyond the capacity of  Zielonka's distribution, as their sequential implementations are not $I$-diamond closed~\cite{GenestGMW10}. Our results indicate that, in terms of state-space reduction, our approach performs comparably to strong bisimulation. However, agents were not fully connected, illustrating that pruning react transitions does not necessarily lead to proportional state-space reduction. One contributing factor is the relatively small size of the sequential implementation of these instances, which imposes a lower bound on individual agent size, i.e., restricted by the number of distinct state labels.

Connectivity was higher than expected because metrics based on averages can overemphasize rarely visited yet highly reactive states in small systems.
\vspace{-10pt}

\paragraph{Custom-designed scenarios.}
Our contributed scenarios, including a case study in Sect.~\ref{sec:case}, showed substantial reductions and extremely low connectivity (down to $0.027$). For Peterson and Dekker algorithms, we produced agents smaller than handcrafted ones.  Even when manually developed, composed and supplied to the engine, the tool eliminated unnecessary  transitions, demonstrating that connectivity should not be left to the designer. Moreover, our tool was far outperforming strong bisimulation while reducing synchronizations. 
\vspace{-10pt}

\paragraph{Regular-expression-based benchmarks.} 
We evaluated the distribution of regular expressions to investigate qualitative aspects of our approach, focusing on how global counting and nondeterminism are handled. Benchmarks were analyzed with respect to communication connectivity and its correlation with the produced agent size. Here, we report on only $11$ instances.

Expressions were defined over the channel set $\{a, b, c, d, e, f, g\}$ and distributed across two interfaces: one initiating on $\{e, f, g\}$, the other on the rest. Each followed a pattern of an ordered sequence over $\{a, b, c, d\}$ followed by a nondeterministic choice enabling an arbitrary subset of these channels.

The first two instances, a reacting agent observes a sequence of two $a$'s, two $b$'s, then either $c$ or $a$, followed by arbitrary $a$, $b$, and $c$. In the first one, $c$ uniquely marks the nondeterminisim, allowing the agent to ignore earlier actions and drop connectivity to $0.097$. In the second, the agent must count three $a$'s before the choice, preventing early disconnection and increasing connectivity to $0.153$.

In both cases, the other agent safely disconnects after reaching terminal deadlock states, no longer participating in communication—thanks to the non-blocking semantics of the transition system.

A small variation in the second instance—allowing only one $b$ after a sequence of $a$'s—makes $b$ the disambiguating channel, again reducing connectivity (to $0.083$) since counting is no longer needed. The fourth benchmark confirms that resolving nondeterminism early helps lower connectivity.

The remaining benchmarks explore how global counting impacts connectivity. Notably, even without major size reductions, low connectivity is often achievable.

\vspace{-10pt}

\section{Conclusions, Related works, and Future works}\label{sec:conc}
\vspace{-10pt}
We introduced an alternative approach to the problem of distributing a sequential system into a set of asynchronous systems recognising the same language, a problem initially introduced by Zielonka~\cite{Zielonka87}. The novelty of our approach is that it removes the rigid communication structure of the original formulation as input to the problem, and instead dynamically introduces a reconfigurable communication structure. Our distribution reduces synchronizations in the produced systems and their size is, in worst case, the size of the sequential system. 
To enable our distribution, we introduced a novel parametric bisimulation that is able to reduce both state-space and unnecessary transitions. We showed how to compute our bisimulation and proved the correctness of the approach. We showed how to use our distribution to enable teamwork synthesis from global specifications, a reformulation of distributed synthesis. 
Distributed synthesis is known to be undecidable, which is bypassed by using reconfigurable communication. Experimental evaluation shows, practically, a huge reduction in size w.r.t. the sequential system and that the needed asynchrony can be attained. 
 
%
\medskip

\noindent
{\bf Related works:} 
Distribution techniques span three related, but distinct lines: synthesis~\cite{Zielonka87} (constructing implementations from ``declarative" specifications), choreographies~\cite{DBLP:conf/esop/CarboneHY07} (projecting ``communication" protocols), and multiparty session types~\cite{DBLP:conf/popl/HondaYC08} (verifying conformance via typing). Our work is in synthesis: the specification defines ``what" behavior is required, not ``how" it is coordinated. 

In choreography, a global protocol is “the specification”, and already specifies send/receive structure and the task is to decompose it. In our setting, communication is not part of the specification, but rather is dynamically computed to guarantee distributed realizability. This distinction makes these approaches incomparable. Indeed, we do not know a priori which react-actions are irrelevant to each agent; instead, we use reconfigurable bisimulation to semantically analyze which react actions can be safely removed without introducing nondeterminism.

We use the TS transition system for our distribution, which builds on existing reconfigurable semantics approaches from CTS~\cite{AbdAlrahmanP21,rcp,AlrahmanMP22}. However, our TS is more asynchronous in that it relaxes the semantics of CTS and allows nonblocking message initiation (or send). Thus, unlike multicast formalisms (such as  CTS~\cite{AbdAlrahmanP21,rcp}, Zielonka automata~\cite{Zielonka87,GenestGMW10}, Hoare's CSP calculus~\cite{Hoare21a}, or synchronous automata~\cite{ramadge89}), an agent cannot block a communication on a shared multicast channel just because it is not ready to participate in, e.g., rendezvous.  

Our notion of bisimulation is novel with respect to existing literatures~\cite{CastellaniH89,MilnerS92,Sangiori93,GlabbeekW96,NicolaV95}. 
It is the only bisimulation that is able to reduce synchronizations while preserving language equivalence. Note that stutter bisimulation~\cite{NicolaV95} only preserves language equivalence of stutter traces, and that is why it does not preserve the next operator of \ltl. Our bisimulation sophisticatedly decides when a react/receive transition can be abstracted safely. Parametric bisimulation already exists in the literature~\cite{Larsen87}, but it does not provide a mechanism to reduce synchronizations. Moreover, the asynchronous reconfigurable semantics that our notion deals with makes the problem more challenging. This is because the parameter cannot alone decide which transitions to enable as in~\cite{Larsen87}. Notably, our bisimulation is equipped with a non-trivial compression technique to reduce synchronizations. The latter is challenging because the solution of a parametric bisimulation is a set of (possibly conflicting) partitions, one for each parameter state. Accordingly, we established a conflict-resolution by a reduction to the clique-partitioning problem~\cite{GJ79,Bodlaender94,Gramm01} .
\medskip

\noindent
{\bf Future work.}
We want to generalise the Mealy machine in Def.~\ref{def:mealy} to represent general $\omega$-regular languages. Also, generalise the trivial decomposition (Lemma \ref{lem:triv}) to allow different agents to initiate on the same channel. We already have plans to relax the translation from Mealy machines to TS (Def.~\ref{lem:mtots}) to accept input as general sets rather than singletons. In fact, this is only apparent in the latter construction, but not the rest of the distribution. 
However, Lemma \ref{lem:triv} still does not support initiation on a shared channel.


Lastly, we want to see possible applications of our distribution for techniques around distributed knowledge dissemination for multi-agent systems, e.g., Knowledge reasoning 
\cite{FHMV95,GochetG06} and strategic reasoning~\cite{CHP10}, and also in distributed supervisory control~\cite{Thistle05,ramadge89}.
\clearpage

\begin{credits}
\subsubsection{\ackname} Yehia Abd Alrahman is funded by VR grants  SynTM (2020-03401) and S$^2$MAS (2025-05071) and Nir Piterman is funded by ERC consolidator grant D-SynMA (772459) and VR grant
(2025-05419)

\subsubsection{Data Availability Statement} The artifact contains executable binaries, scripts, benchmark data, and result-processing tools required to reproduce the experimental results reported in our paper. It is publicly available on Zenodo~\cite{abdalrahman202621807991}.

\subsubsection{\discintname}
The authors have no competing interests to declare.
\end{credits}

%
%
%
\bibliographystyle{splncs04}
\bibliography{biblio}


\begin{toappendix}
\section{Additional Details about Overview Example}
\label{app:more overview}
We revisit the example of the four-element system presented in  Section~\ref{sec:new overview}.
We give the missing details regarding the quotient systems for arm 1, arm 2, and the machine. 
Recall that the system (along with details for arm 3) is depicted in Figure~\ref{fig:fourarm}.
    
We show the equivalence classes created for arm 1 and the resulting quotient in Figure~\ref{fig:threearmintray}.
We can see that, here, states $1a$ and $2f$ have a different label and cannot be united. 
If we were to remove the state label, they could be unified. 
We show the equivalence classes created for arm 2 and the resulting quotient in Figure~\ref{fig:fourarmfwd}. 
As before, the label of $2f$ and $2d$ are different and they cannot be united. 
Finally, we show the equivalence classes created for the machine and the resulting quotient in Figure~\ref{fig:fourarmproc}.
We note that the labels of $4p$ and $5a$ are different. 

\begin{figure}[bt]
\begin{center}

\begin{tikzpicture}[
    ->, >=stealth',
    auto,
    semithick,
    node distance=2cm,
    state/.style={circle, draw, minimum size=18pt, inner sep=1.5pt},
    num/.style={font=\scriptsize}
]

\node[state] (q0dash) at (5,0) {0\textcolor{gray}{-}};
\node[state] (q0d) at (5,1) {0\textcolor{gray}{d}};
\node[state] (q1a) at (6.5,0) {1\textcolor{blue}{a}};
\node[state] (q1d) at (6.5,1) {1\textcolor{gray}{d}};
\node[state] (q2f) at (8,0) {2\textcolor{gray}{f}};
\node[state] (q2d) at (8,1) {2\textcolor{gray}{d}};
\node[state] (q3a) at (10,0) {3\textcolor{blue}{a}};
\node[state] (q3d) at (10,1) {3\textcolor{gray}{d}};
\node[state] (q4) at (8,2.2) {4\textcolor{gray}{p}};
\node[state] (q6) at (9,2.2) {6\textcolor{gray}{f}};
\node[state] (q7) at (10,2.2) {7\textcolor{blue}{a}};
\node[state] (q5p) at (11,2.2) {5\textcolor{gray}{p}};
\node[state] (q5a) at (11,3) {5\textcolor{blue}{a}};

\node[draw=green,dotted,fit=(q0dash)] {};
\node[draw=green,dotted,fit=(q2f) ] {};
\node[draw=green,dotted,fit=(q6) ] {};
\node[draw=green,dotted,fit=(q2d) ] {};
\node[draw=green,dotted,fit=(q4) ] {};
\node[draw=green,dotted,fit=(q0d) ] {};
\node[draw=blue,dotted,fit=(q1a)] {};
\node[draw=blue,dotted,fit=(q3a)] {};
\node[draw=blue,dotted,fit=(q5a)] {};
\node[draw=blue,dotted,fit=(q7)] {};
\node[draw=red,dotted,fit=(q3d)] {};
\node[draw=red,dotted,fit=(q5p)] {};
\node[draw=red,dotted,fit=(q1d)] {};

\node[coordinate] (init) at (4.3,0) {};
\draw[->] (init) -- (q0dash);


\draw (q0dash)  edge node {a} (q1a);
\draw (q0d)  edge node [pos=0.3,below] {a} (q1a);
\draw (q1a)  edge node {f} (q2f);
\draw (q1d)  edge node [below,pos=0.3] {f} (q2f);
\draw (q2f)  edge node {a} (q3a);
\draw (q2d)  edge node {a} (q3a);
\draw (q2f)  edge[bend left=30] node [left] {p} (q4);
\draw (q2d)  edge[bend left=30] node [left] {p} (q4);
\draw (q3a)  edge[bend right=30] node [right] {p} (q5p);
\draw (q3d)  edge[bend right=30] node [right] {p} (q5p);
\draw (q4)  edge[bend left=60] node [pos=0.3,below] {a} (q5a);
\draw (q4)  edge[bend right=30] node [pos=0.2,above] {d} (q0d);
\draw (q5p)  edge[bend right=40] node [pos=0.5,above] {f} (q6);
\draw (q5a)  edge[bend right=40] node [pos=0.8,above] {f} (q6);
\draw (q5p)  edge[out=45,in=-225,looseness=2] node [above,pos=0.7] {d} (q1d);
\draw (q5a)  edge[bend right=110] node [pos=0.7,above] {d} (q1d);
\draw (q6)  edge node [below] {a} (q7);
\draw (q6)  edge node [left] {d} (q2d);
\draw (q7)  edge node {d} (q3d);

\node[state] (state1) at (12,0) {0\textcolor{gray}{-}};
\node[num]   [left=0pt of state1,yshift=10pt,xshift=5pt]  {$\{a\}$};
\node[state] (state2) at (13,1) {1\textcolor{blue}{a}};
\node[num]   [above=0pt of state2,xshift=-3pt,yshift=-3pt]  {$\{p,d,f\}$};
\node[state] (state3) at (14,2) {2\textcolor{gray}{f}};
\node[num]   [above=0pt of state3,yshift=-2pt,xshift=0pt]  {$\{f\}$};
\node[coordinate] (init) at (11.3,0) {};
\draw[->] (init) -- (state1);

\draw (state1) edge[bend left=30] node {a} (state2);
\draw (state2) edge node [pos=0.7] {d,p} (state3);
\draw (state2) edge[bend left=30] node [pos=0.2] {f} (state1);
\draw (state3) edge[bend left=50] node  {f} (state1);

\end{tikzpicture}
\end{center}
\vspace*{-7mm}
\caption{\label{fig:threearmintray}Controller with states labeled according to arm 1's view (i.e., $-$, $f$, $p$, or $d$ as one label and $a$ as the other label) and partitioned to arm 1's equivalence relations: 
the $0-$, $0d$, $2d$, $2f$, $4p$ and $6f$ as one equivalence class (green); 
$1a$, $3a$, $5a$, and $7a$ as one equivalence class (blue);
and 
$1d$, $3d$ and $5p$ as the last equivalence class (red) (left)
and the quotient system for arm 1 with states labeled by connected channels (right).}
\end{figure}

\begin{figure}[bt]
\begin{center}

\begin{tikzpicture}[
    ->, >=stealth',
    auto,
    semithick,
    node distance=2cm,
    state/.style={circle, draw, minimum size=18pt, inner sep=1.5pt},
    num/.style={font=\scriptsize}
]

\node[state] (q0dash) at (5,0) {0\textcolor{gray}{-}};
\node[state] (q0d) at (5,1) {0\textcolor{gray}{d}};
\node[state] (q1a) at (6.5,0) {1\textcolor{gray}{a}};
\node[state] (q1d) at (6.5,1) {1\textcolor{gray}{d}};
\node[state] (q2f) at (8,0) {2\textcolor{blue}{f}};
\node[state] (q2d) at (8,1) {2\textcolor{gray}{d}};
\node[state] (q3a) at (10,0) {3\textcolor{gray}{a}};
\node[state] (q3d) at (10,1) {3\textcolor{gray}{d}};
\node[state] (q4) at (8,2.2) {4\textcolor{gray}{p}};
\node[state] (q6) at (9,2.2) {6\textcolor{blue}{f}};
\node[state] (q7) at (10,2.2) {7\textcolor{gray}{a}};
\node[state] (q5p) at (11,2.2) {5\textcolor{gray}{p}};
\node[state] (q5a) at (11,3) {5\textcolor{gray}{a}};

\node[draw=green,dotted,fit=(q0dash) (q0d)] {};
\node[draw=green,dotted,fit=(q4)] {};
\node[draw=blue,dotted,fit=(q1d) (q1a)] {};
\node[draw=blue,dotted,fit=(q5a) (q5p)] {};
\node[draw=black,dotted,fit=(q6)] {};
\node[draw=black,dotted,fit=(q2f)] {};
\node[draw=red,dotted,fit=(q2d)] {};
\node[draw=red,dotted,fit=(q7) (q3a) (q3d)] {};

\node[coordinate] (init) at (4.3,0) {};
\draw[->] (init) -- (q0dash);


\draw (q0dash)  edge node {a} (q1a);
\draw (q0d)  edge node [pos=0.3,below] {a} (q1a);
\draw (q1a)  edge node {f} (q2f);
\draw (q1d)  edge node [below,pos=0.3] {f} (q2f);
\draw (q2f)  edge node {a} (q3a);
\draw (q2d)  edge node {a} (q3a);
\draw (q2f)  edge[bend left=30] node [left] {p} (q4);
\draw (q2d)  edge[bend left=30] node [left] {p} (q4);
\draw (q3a)  edge[bend right=30] node [right] {p} (q5p);
\draw (q3d)  edge[bend right=30] node [right] {p} (q5p);
\draw (q4)  edge[bend left=60] node [pos=0.3,below] {a} (q5a);
\draw (q4)  edge[bend right=30] node [pos=0.2,above] {d} (q0d);
\draw (q5p)  edge[bend right=40] node [pos=0.5,above] {f} (q6);
\draw (q5a)  edge[bend right=40] node [pos=0.8,above] {f} (q6);
\draw (q5p)  edge[out=45,in=-225,looseness=2] node [above,pos=0.7] {d} (q1d);
\draw (q5a)  edge[bend right=110] node [pos=0.7,above] {d} (q1d);
\draw (q6)  edge node [below] {a} (q7);
\draw (q6)  edge node [left] {d} (q2d);
\draw (q7)  edge node {d} (q3d);

\node[state] (state1) at (12,0) {0\textcolor{gray}{-}};
\node[num]   [left=0pt of state1,yshift=10pt,xshift=5pt]  {$\{a\}$};
\node[state] (state2) at (13,0) {1\textcolor{gray}{a}};
\node[num]   [below=0pt of state2,xshift=-3pt,yshift=3pt]  {$\{f\}$};
\node[state] (state3) at (14,0) {2\textcolor{blue}{f}};
\node[num]   [below=0pt of state3,yshift=2pt,xshift=0pt]  {$\{a,p,d\}$};
\node[state] (state4) at (14,1) {2\textcolor{gray}{d}};
\node[num]   [above=0pt of state4,yshift=-3pt,xshift=-6pt]  {$\{a,p\}$};
\node[state] (state5) at (15,0) {3\textcolor{gray}{a}};
\node[num]   [above=0pt of state5,yshift=-3pt,xshift=6pt]  {$\{p\}$};
\node[coordinate] (init) at (11.3,0) {};
\draw[->] (init) -- (state1);

\draw (state1) edge node {a} (state2);
\draw (state2) edge node {f} (state3);
\draw (state3) edge[bend right=50] node [above] {p} (state1);
\draw (state3) edge node [above] {a} (state5);
\draw (state3) edge node [right] {d} (state4);
\draw (state4) edge[bend right=50] node [left] {p} (state1);
\draw (state4) edge node {a} (state5);
\draw (state5) edge[bend left=70] node {p} (state2);

\end{tikzpicture}
\end{center}
\vspace*{-7mm}
\caption{\label{fig:fourarmfwd}
Controller with states labeled according to arm 2's view (i.e., $-$, $a$, $p$, or $d$ as one label and $f$ as the other label) and partitioned to its equivalence relations: 
the two red boxes are one class each and there are three additional classes: $4p$ one class (green), $1d$, $1a$, $5a$, and $5p$ one class (blue), and $2f$ and $6f$ on class (black) (left)
and the quotient system for arm 2 with states labeled by connected channels (right).}
\vspace*{-7mm}
\end{figure}

\begin{figure}[bt]
\begin{center}

\begin{tikzpicture}[
    ->, >=stealth',
    auto,
    semithick,
    node distance=2cm,
    state/.style={circle, draw, minimum size=18pt, inner sep=1.5pt},
    num/.style={font=\scriptsize}
]

\node[state] (q0dash) at (5,0) {0\textcolor{gray}{-}};
\node[state] (q0d) at (5,1) {0\textcolor{gray}{d}};
\node[state] (q1a) at (6.5,0) {1\textcolor{gray}{a}};
\node[state] (q1d) at (6.5,1) {1\textcolor{gray}{d}};
\node[state] (q2f) at (8,0) {2\textcolor{gray}{f}};
\node[state] (q2d) at (8,1) {2\textcolor{gray}{d}};
\node[state] (q3a) at (10,0) {3\textcolor{gray}{a}};
\node[state] (q3d) at (10,1) {3\textcolor{gray}{d}};
\node[state] (q4) at (8,2.2) {4\textcolor{blue}{p}};
\node[state] (q6) at (9,2.2) {6\textcolor{gray}{f}};
\node[state] (q7) at (10,2.2) {7\textcolor{gray}{a}};
\node[state] (q5p) at (11,2.2) {5\textcolor{blue}{p}};
\node[state] (q5a) at (11,3) {5\textcolor{gray}{a}};

\node[draw=red,dotted,fit=(q0dash) (q0d) (q1d) (q1a)] {};
\node[draw=red,dotted,fit=(q5a)] {};
\node[draw=red,dotted,fit=(q6) (q7)] {};
\node[draw=blue,dotted,fit=(q5p)] {};
\node[draw=blue,dotted,fit=(q4)] {};
\node[draw=red,dotted,fit=(q3d) (q2f) (q2d) (q3a)] {};

\node[coordinate] (init) at (4.3,0) {};
\draw[->] (init) -- (q0dash);


\draw (q0dash)  edge node {a} (q1a);
\draw (q0d)  edge node [pos=0.3,below] {a} (q1a);
\draw (q1a)  edge node {f} (q2f);
\draw (q1d)  edge node [below,pos=0.3] {f} (q2f);
\draw (q2f)  edge node {a} (q3a);
\draw (q2d)  edge node {a} (q3a);
\draw (q2f)  edge[bend left=30] node [left] {p} (q4);
\draw (q2d)  edge[bend left=30] node [left] {p} (q4);
\draw (q3a)  edge[bend right=30] node [right] {p} (q5p);
\draw (q3d)  edge[bend right=30] node [right] {p} (q5p);
\draw (q4)  edge[bend left=60] node [pos=0.3,below] {a} (q5a);
\draw (q4)  edge[bend right=30] node [pos=0.2,above] {d} (q0d);
\draw (q5p)  edge[bend right=40] node [pos=0.5,above] {f} (q6);
\draw (q5a)  edge[bend right=40] node [pos=0.8,above] {f} (q6);
\draw (q5p)  edge[out=45,in=-225,looseness=2] node [above,pos=0.7] {d} (q1d);
\draw (q5a)  edge[bend right=110] node [pos=0.7,above] {d} (q1d);
\draw (q6)  edge node [below] {a} (q7);
\draw (q6)  edge node [left] {d} (q2d);
\draw (q7)  edge node {d} (q3d);

\node[state] (state1) at (12,0) {0\textcolor{gray}{-}};
\node[num]   [left=0pt of state1,yshift=10pt,xshift=5pt]  {$\{f\}$};
\node[state] (state2) at (13,0) {2\textcolor{gray}{f}};
\node[num]   [right=0pt of state2,xshift=-3pt,yshift=-3pt]  {$\{p\}$};
\node[state] (state3) at (13,1) {4\textcolor{blue}{p}};
\node[num]   [above=0pt of state3,yshift=-3pt,xshift=-8pt]  {$\{a,d,f\}$};
\node[state] (state4) at (14,1) {6\textcolor{gray}{f}};
\node[num]   [right=0pt of state4,yshift=-3pt,xshift=-3pt]  {$\{d\}$};
\node[state] (state5) at (15,2) {5\textcolor{gray}{a}};
\node[num]   [above=0pt of state5,yshift=-3pt,xshift=6pt]  {$\{d\}$};
\node[coordinate] (init) at (11.3,0) {};
\draw[->] (init) -- (state1);

\draw (state1) edge node [below] {f} (state2);
\draw (state2) edge node {p} (state3);
\draw (state3) edge[bend left=30] node {f} (state4);
\draw (state3) edge[bend right=30] node [above] {d} (state1);
\draw (state3) edge[bend left=30] node [above] {a} (state5);
\draw (state4) edge node [right] {d} (state2);
\draw (state5) edge[bend left=90] node {d} (state1);
\draw (state5) edge node {f} (state4);

\end{tikzpicture}
\end{center}
\vspace*{-7mm}
\caption{\label{fig:fourarmproc}
Controller with states labeled according to the machine's view (i.e., $-$, $a$, $f$, or $d$ as one label and $p$ as the other label) and partitioned to its equivalence relations: 
every red box is one class and there is an additional blue class including $4p$ and $5p$ (left)
and the quotient system for the machine with states labeled by connected channels (right).
}
\vspace*{-7mm}
\end{figure}

We also consider Zielonka's distribution in this case.
As a first difference, we have to understand the problem well enough to assign connectivity for all channels.
In this simple system, it is clear that arm 1 needs information about $a$ and $f$, arm 2 needs information about $a$, $f$, and $p$, the machine needs information about $f$, $p$, and $d$, and arm 3 needs information about $p$ and $d$. 
The communication configuration does looks as follows:
$$\fbox{\mbox{arm 1}} \stackrel{a,f}{\longleftrightarrow}
\fbox{\mbox{arm 2}}  \stackrel{f,p}{\longleftrightarrow}
\fbox{\mbox{machine}}  \stackrel{p,d}{\longleftrightarrow}
\fbox{\mbox{arm 3}}
$$
Obviously, there are channels that are connected to more than two agents.
Thus, this communication configuration does \emph{not} correspond to a tree \cite{KrishnaM13}.
However, it is \emph{tree-like}, as the connectivity of all non-binary channels is continuous along the tree (e.g., $p$ spans arm 2, the machine, and arm 3), and hence the quadratic distribution construction applies to it  \cite{DBLP:conf/fossacs/LehautMP26} .
We include the systems resulting for the four components in Figure~\ref{fig:fourarmzielonka} and their composition in Figure~\ref{fig:fourarmzielonkacomposition}.
Notice that transitions cannot be defined by looking at a single component and one must consider all the components connected to one channel in order to define a transition.
Thus, the full definition of transitions is available only in Figure~\ref{fig:fourarmzielonkacomposition}.
Notice that arm 2 by itself is larger than the original transition system.
Furthermore, the product of the four components is 2.5 times larger than the original.

\begin{figure}[bt]
\begin{center}
\scalebox{0.5}{
\begin{tikzpicture}[
    ->, >=stealth',
    auto,
    semithick,
    node distance=1cm,
    state/.style={circle, draw, minimum size=18pt, inner sep=1.5pt},
    num/.style={font=\scriptsize}
]

\node[state] (q0) at (2,10) {(0,0)};
\node[state] (q1) at (2,8) {(1,1)};
\node[state] (q2) at (2,6) {(2,2)};
\node[state] (q3) at (2,4) {(3,3)};
\node[state] (q6) at (2,2) {(6,6)};
\node[state] (q7) at (2,0) {(7,7)};
\node[state] (q5) at (0,4) {(5,5)};

\node[coordinate] (init) at (1.3,10) {};
\draw[->] (init) -- (q0);
\coordinate[label={\Large Arm 1}] (name) at (0.5,8);


\draw (q0)  edge node {a} (q1);
\draw (q1)  edge node [left] {f} (q2);
\draw (q2)  edge node [left] {a} (q3);
\draw (q2)  edge[bend right=30] node [right] {a} (q1);
\draw (q2)  edge[bend right=30] node [left] {a} (q5);
\draw (q3)  edge node {f} (q6);
\draw (q3)  edge[bend right=30] node [right] {f} (q2);
\draw (q6)  edge node {a} (q7);
\draw (q6)  edge node {a} (q5);
\draw (q6)  edge[bend right=50,looseness=1.3] node [right]{a} (q1);
\draw (q7)  edge[bend left=90,looseness=1.5] node {f} (q2);

\node[state] (q00) at (7,12) {(0,0)};
\node[state] (q01) at (7,10) {(0,1)};
\node[state] (q22) at (7,8) {(2,2)};
\node[state] (q23) at (7,6) {(2,3)};
\node[state] (q55) at (7,4) {(5,5)};
\node[state] (q66) at (7,2) {(6,6)};
\node[state] (q67) at (7,0) {(6,7)};
\node[state] (q44) at (9,6) {(4,4)};
\node[state] (q45) at (9,4) {(4,5)};

\node[coordinate] (init) at (6.3,12) {};
\draw[->] (init) -- (q00);
\coordinate[label={\Large Arm 2}] (name) at (5.5,10);


\draw (q00)  edge node {a} (q01);
\draw (q01)  edge node [left] {f} (q22);
\draw (q22)  edge node [left] {a} (q23);
\draw (q22)  edge[bend right=20] node [right] {p}(q44);
\draw (q23)  edge node {p} (q55);
\draw (q55)  edge[bend left=30] node [left] {f} (q22);
\draw (q55)  edge node [left] {f} (q66);
\draw (q66)  edge node {a} (q67);
\draw (q66)  edge[bend left=10] node [pos=0.3,right]{p} (q44);
\draw (q67)  edge[bend left=30,looseness=1.3] node {p} (q55);
\draw (q44)  edge node {a} (q45);
\draw (q45)  edge[bend right=90,looseness=1.3] node [right]{f} (q22);

\node[state] (q0) at (12,12) {0};
\node[state] (q2) at (12,10) {2};
\node[state] (q4) at (12,8) {4};
\node[state] (q5) at (14,10) {5};
\node[state] (q6) at (14,8) {6};
\node[state] (q1) at (14,12) {1};

\node[coordinate] (init) at (11.3,12) {};
\draw[->] (init) -- (q0);
\coordinate[label={\Large Machine}] (name) at (15.5,10);


\draw (q0)  edge node {f} (q2);
\draw (q2)  edge node [left] {p} (q4);
\draw (q2)  edge node [above] {p} (q5);
\draw (q4)  edge[bend left=30] node {d} (q0);
\draw (q5)  edge node {f} (q6);
\draw (q5)  edge node {d} (q1);
\draw (q6)  edge[bend right=0] node [right]{d} (q2);
\draw (q1)  edge[bend right=0] node [right]{f} (q2);

\node[state] (q00) at (12,4) {(0,0)};
\node[state] (q44) at (12,2) {(4,4)};
\node[state] (q22) at (14,2) {(2,2)};
\node[state] (q55) at (14,4) {(5,5)};
\node[state] (q11) at (16,4) {(1,1)};

\node[coordinate] (init) at (11.3,4) {};
\draw[->] (init) -- (q00);
\coordinate[label={\Large Arm 3}] (name) at (13,5);


\draw (q00)  edge node {p} (q44);
\draw (q00)  edge node [above] {p} (q55);
\draw (q44)  edge[bend left=30] node [left] {d} (q00);
\draw (q55)  edge[bend left=0] node {d} (q22);
\draw (q55)  edge node {d} (q11);
\draw (q22)  edge node {p} (q44);
\draw (q11)  edge[bend right=30] node [above]{p} (q55);

\end{tikzpicture}}
\end{center}
\vspace*{-5mm}
\caption{\label{fig:fourarmzielonka} Zielonka distribution for the arms and the machine.
The choice between seemingly nondeterministic transitions from the same state depend on the state of other components participating in the communication and this can be deduced from the full composition below. For example, the $a$ transition from $(6,6)$ to $(5,5)$ in arm 1 is taken when arm 2 is in state $(4,4)$ and the $a$ transition to $(1,1)$ is taken when arm 2 is in state $(0,0)$.}
\vspace*{-7mm}
\end{figure}

\begin{figure}[bt]
\scalebox{0.7}{
\begin{tikzpicture}[
    ->, >=stealth',
    auto,
    semithick,
    node distance=2cm,
    state/.style={rectangle, draw, minimum size=18pt, inner sep=1.5pt},
    num/.style={font=\scriptsize}
]

\node[state] (q0-0000000) at (12,0) {(0,0)(0,0)0(0,0)};
\node[state] (q0-2200000) at (8,0) {(2,2)(0,0)0(0,0)};
\node[state] (q0-6600000) at (10,1) {(6,6)(0,0)0(0,0)};
\node[state] (q1-1101000) at (12,3) {(1,1)(0,1)0(0,0)};
\node[state] (q1-3355111) at (8,3) {(3,3)(5,5)1(1,1)};
\node[state] (q1-5545000) at (12,4) {(5,5)(4,5)0(5,5)};
\node[state] (q1-7755111) at (8,4) {(7,7)(5,5)1(1,1)};
\node[state] (q2-2222200) at (12,6) {(2,2)(2,2)2(0,0)};
\node[state] (q2-6666222) at (8,7) {(6,6)(6,6)2(2,2)};
\node[state] (q2-2222211) at (8,6) {(2,2)(2,2)2(1,1)};

\node[state] (q3-7767222) at (8,11) {(7,7)(6,7)2(2,2)};
\node[state] (q3-3323200) at (12,11) {(3,3)(2,3)2(0,0)};
\node[state] (q3-3323211) at (10,12) {(3,3)(2,3)2(1,1)};

\node[state] (q4-2244444) at (4,6) {(2,2)(4,4)4(4,4)};
\node[state] (q4-6644444) at (4,7) {(6,6)(4,4)4(4,4)};
\node[state] (q6-6666655) at (4,9) {(6,6)(6,6)6(5,5)};
\node[state] (q7-7767655) at (4,11) {(7,7)(6,7)6(5,5)};

\node[state] (q5-3355555) at (2,14) {(3,3)(5,5)5(5,5)};
\node[state] (q5-5545444) at (0,13) {(5,5)(4,5)4(4,4)};
\node[state] (q5-7755555) at (4,13) {(7,7)(5,5)5(5,5)};

\node[draw=gray,dotted,fit=(q0-0000000) (q0-2200000) (q0-6600000),label={0}] {};
\node[draw=gray,dotted,fit=(q1-1101000) (q1-3355111) (q1-5545000),label={1}] {};
\node[draw=gray,dotted,fit=(q2-2222200) (q2-6666222) (q2-2222211),label={2}] {};
\node[draw=gray,dotted,fit=(q3-7767222) (q3-3323200) (q3-3323211),label={3}] {};
\node[draw=gray,dotted,fit=(q4-2244444) (q4-6644444),label={4}] {};
\node[draw=gray,dotted,fit=(q6-6666655),label={6}] {};
\node[draw=gray,dotted,fit=(q7-7767655),label={7}] {};
\node[draw=gray,dotted,fit=(q5-3355555) (q5-5545444) (q5-7755555),label={5}] {};

\node[coordinate] (init) at (10.5,0) {};
\draw[->] (init) -- (q0-0000000);


\draw (q0-0000000) edge node {a} (q1-1101000);
\draw (q0-2200000) edge[bend left=30] node {a} (q1-1101000);
\draw (q0-6600000) edge node {a} (q1-1101000);

\draw (q1-1101000) edge[out=0,in=0] node {f} (q2-2222200);
\draw (q1-3355111) edge[out=180,in=180] node {f} (q2-2222211);
\draw (q1-5545000) edge[out=90,in=270] node {f} (q2-2222200);
\draw (q1-7755111) edge[out=90,in=270] node {f} (q2-2222211);

\draw (q2-2222200) edge[out=-135,in=-45,looseness=0.2] node [pos=0.3] {p} (q4-2244444);
\draw (q2-2222211) edge[out=170,in=-10] node [above] {p} (q4-2244444);
\draw (q2-6666222) edge[out=180,in=0] node {p} (q4-6644444);
\draw (q2-2222200) edge[out=90,in=-90,looseness=0.2] node [pos=0.3] {a} (q3-3323200);
\draw (q2-2222211) edge[out=0,in=-90] node [right] {a} (q3-3323211);
\draw (q2-6666222) edge[out=90,in=-90] node {a} (q3-7767222);

\draw (q3-3323200) edge[out=90,in=0] node {p} (q5-3355555);
\draw (q3-7767222) edge[out=90,in=0] node [right] {p} (q5-7755555);
\draw (q3-3323211) edge[bend right=20] node {p} (q5-3355555);

\node[coordinate] (midedge1) at (1.7,6) {};
\node[coordinate] (midedge2) at (1.7,0) {};
\draw (q4-2244444) -- (midedge1) -- (midedge2) -- node {d} (q0-2200000);
\node[coordinate] (midedge1) at (1.3,7) {};
\node[coordinate] (midedge2) at (1.3,1) {};
\draw[->] (q4-6644444) -- (midedge1) -- (midedge2) -- node  {d} (q0-6600000);
\draw (q4-2244444) edge[out=180,in=290] node {f} (q5-5545444);
\draw (q4-6644444) edge[out=180,in=290] node {f} (q5-5545444);

\draw (q5-3355555) edge[out=280,in=135] node {f} (q6-6666655);
\draw (q5-7755555) edge[out=-15,in=15,looseness=0.5] node {f} (q6-6666655);
\draw (q5-5545444) edge node [right,pos=0.1] {f} (q6-6666655);
\node[coordinate] (midedge1) at (2,3) {};
\draw (q5-3355555) -- (midedge1) -- node [below] {d} (q1-3355111);
\node[coordinate] (midedge1) at (1,11) {};
\node[coordinate] (midedge2) at (1,4) {};
\draw[->] (q5-7755555) [out=180] -- (midedge1) -- (midedge2) -- node [pos=0.7] {d} (q1-7755111);
\node[coordinate] (midedge1) at (0,3.5) {};
\node[coordinate] (midedge2) at (10,3.5) {};
\draw[->] (q5-5545444) [out=-120] -- (midedge1)  -- (midedge2) -- node {d}  (q1-5545000);

\draw (q6-6666655) edge node {a} (q7-7767655);
\draw (q6-6666655) edge node {d} (q2-6666222);

\draw (q7-7767655) edge[bend left=30] node {d} (q3-7767222);

\end{tikzpicture}
}
\vspace*{-3mm}
\caption{\label{fig:fourarmzielonkacomposition}The composition of the four asynchronous automata for the three arms and the machine. The label of a state corresponds to the state of arm 1, arm 2, the machine, and arm 3. For example, state $(3,3)(2,3)2(0,0)$ corresponds to arm 1 being in state $(3,3)$, arm 2 in state $(2,3)$, the machine in state $2$, and arm 3 in state $(0,0)$. 
Transitions here resolve the nondeterminism in the previous figure.
}
\vspace*{-4mm}
\end{figure}

We note that, in general, the distribution construction could lead to a situation where mismatches from the original protocol are discovered only after they occur and only once some part of the system collects enough information and is aware of the problem.
This can happen also when the language is a safety language and in tree architectures. 
However, in our case, there is enough connectivity to ensure that the components can jointly ensure safety.

\section{Detailed Comparison with Zielonka Distribution}
\label{sec:ts-zie}

We compare our approach, from a high-level perspective, to the optimal Zielonka-type construction~\cite{GenestGMW10}. 
We provide insights into how instances of Zielonka’s problem can be implemented within our framework, and highlight the key differences in terms of both benefits and trade-offs.


Our approach can be seen as a variant of the problem addressed by Zielonka’s distribution~\cite{Zielonka87}. Although both aim to distribute a centralized specification, the underlying theoretical challenges differ fundamentally. Zielonka’s approach focuses on achieving distribution while preserving a fixed communication structure. This is generally undecidable, and thus restrictions are imposed on the specification language to ensure compatibility with the required structure. However, it is also undecidable to determine whether a given specification language admits a sublanguage that satisfies these asynchrony constraints~\cite{DBLP:conf/concur/StefanescuEM03}. Efficient variants of Zielonka’s construction (cf.~\cite{DBLP:conf/concur/StefanescuEM03,KrishnaM13,GimbertMMW22}) introduce additional, often severe, restrictions on the shape of communication graphs and the type of events. Nonetheless, the computational complexity of these approaches remains high relative to the size of the sequential system.
Under Zielonka’s approach divergence from the target language is possible, however, this will be recognized once knowledge about the divergence is disseminated throughout the system or at the global synchronization that determines acceptance.
That is, given the fixed structure of communication, there could be cases where the prefix of computation seen by the system already does not belong to the target language but this will be recognized by the system later. 

Our approach is pragmatic and designed with practical applicability in mind. Unlike Zielonka’s setting, the degree of asynchrony is not provided as part of the input. Relying on the designer to specify this degree is often impractical, given the inherent complexity of distributed systems. Instead, we lift this restriction and dynamically inject or remove interactions while preserving correct distribution. The core challenge becomes minimizing these interactions without compromising system behavior. 
Moreover, our method imposes minimal restrictions on the specification language and produces highly asynchronous implementations without incurring significant overhead.
In particular, rather than delaying the recognition that violations have occured, our approach disseminates information to ensure that violations do not occur.
However, in the worst case, our approach could disseminate all information to all parts of the system. 
As our experimental results and the discussion below show, this is very unlikely to happen even in cases that are very hard for distribution. 

We note also a slight difference in the way communications occur.
In the asynchronous automata setting a communication can occur if and only if all parties to it are ready to participate in it.
In our setting, each channel has an owner who initiates communications on this channel.
All other agents are either disconnected from this channel or ready to participate in communications on this channel.
However, a communication on a channel cannot occur without the owner even if some other agents are connected to the channel and are ready to engage in communication on it. 

To illustrate how our approach compares to Zielonka's, we revisit the $\msf{Path}_n$ language  from~\cite{GenestGMW10}.
This language was used to derive an exponential lower bound for the optimal Zielonka-Type construction.
That is, for the $n$-th instance, each agent requires exponentially many states in order to distribute this language.

The setting involves $n$ agents, where each pair of agents is assigned a dedicated synchronization channel labeled by their identities---for example, agents $T_0$ and $T_1$ communicate via channel $01$. The language $\mathsf{Path}_n$ imposes that any two consecutive channels in a valid communication sequence must share at least one common identity. For instance, the sequence $01, 12, 20, \dots$ is valid, whereas $01, 23, 10, \dots$ is not. 

An instance of the $\mathsf{Path}_n$ language with $n = 4$ is implemented as a transition system (TS) $T$ shown in Fig.~\ref{fig:path4_ts}. Initially, all channels are possible (not shown in figure). 
Once an initial decision has been made, only the ``adjacent'' channels to the last channel used are available. 
Thus, after a communication on channel $01$, only channels involving either $0$ or $1$ are available.
Indeed, the outgoing transitions from $01$ lead to states $02$, $03$, $12$, and $13$. 

\begin{figure}[h!]
\begin{center}
    \includegraphics[scale=.5]{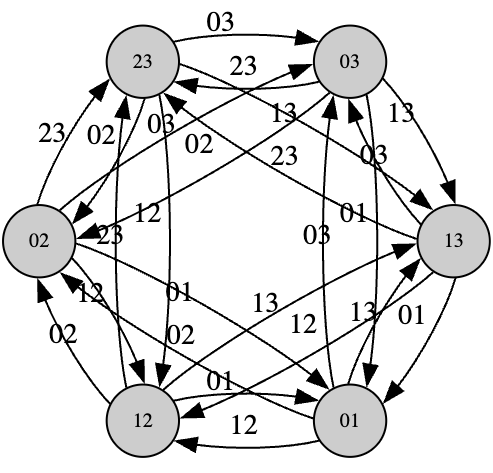}
\end{center}
\caption{\label{fig:path4_ts}Sequential $\msf{Path}_4$. States are labeled by the label of all incoming transitions into that state.}
\end{figure}
Observe that the states of $T$ are highly connected. In particular, state $01$ is connected to 4/5 of the other states. This connectivity becomes substantially more complex as $n$ increases. The difficulty of the example is not in its size but in its high degree of connectivity. Every pair of states is connected by a path of at most two steps (for every $n$).
There are 4 different paths of length 2 from $01$ to $23$ by replacing one of $0$ and $1$ in the first transition and then replacing the remaining one by the missing agent between $2$ and $3$.
For instance $01 \rightarrow 12 \rightarrow 23$. 

Zielonka's distribution requires that the sequential implementation is $I$-diamond closed, and clearly $T$ in Fig.~\ref{fig:path4_ts} satisfies such closure. Notably,  $T$ prohibits self-loops and regardless what is the current taken transition,  other choices will be available in the next state  to complete the path. 

A minimal implementation using the asynchronous automata framework is at least of size exponential in $n$, see~\cite{GenestGMW10}. 
However, the latter is only a lower bound, and the matching upper bound is \(\mathrm{4^{n^4} . |T|^{n^2}}\). 
Moreover, the asynchronous automata operate on the local rejecting principle, i.e, an asynchronous automaton allows all possible violations of the $\msf{Path}_4$ language for a finite time until the automaton recognizes that such an execution is not in $\msf{Path}_4$, and thus rejects it. 
Notice that this is unavoidable in this case. 
For example, after $01$ occurs, the next possible communications are $02$, $03$, $12$, and $13$. 
These can be partitioned to $02$ and $13$ or $03$ and $12$ and both pairs are independent.
It follows that in the asynchronous automata setting, it is not possible to allow both $02$ and $13$ without having the option for both occurring in some order, which is not permitted in the language.
This violation will be discovered later and the agents will reject the computation.
Furthermore, the framework does not provide a way to recover after rejection, which may need to synchronize all the affected local automata, and thus defeating the purpose. 
The latter is one of the main reasons why Zielonka's distribution is motivated to serve as an asynchronous run-time monitor rather than an actual system. 

In order to apply our approach to $\msf{Path}_4$ we have to decide on an owner for every channel.
We choose to partition the channels in the most balanced way possible (though other options are possible as well). 
Thus, we choose the following ownerships: agent $0$ owns channels $01$ and $02$, agent $1$ owns channels $12$ and $13$, agent $2$ owns channel $23$, and agent $3$ owns channel $03$. 
The distribution will choose what other channels each agent will be connected to (and when) in a way that will ensure that no violation of the $\msf{Path}_4$ occurs.

We then apply the four stages explained in the main paper:
\begin{itemize}
    \item
    Label each state by the labels of all incoming transitions (recall that the initial state not depicted in Figure~\ref{fig:path4_ts} is labeled by $-$).
    \item 
    Create four copies of the system, one per each agent, by trivial decomposition. 
    The copy of agent $i$ distinguishes between the labels corresponding to channels agent $i$ ``owns'' and all other labels are considered equal ($-$).
    \item 
    Compute reconfigurable bisimulation for each agent using the composition of the three other agents as the parameters.
    \item 
    Choose the equivalence classes such that all states in every equivalence classes agree on the equivalence and minimize each agent accordingly.
\end{itemize}
For the $\msf{Path}_4$ language and the ownership described above, our approach produces a matching distributed implementation in Fig.~\ref{fig:path4_dis}.
Recall that the composition of the four agents leads to a system that is bisimilar to the original.
In this composition, the owner of a channel initiates communication on a channel and effectively determines when communication can occur. Other participants either disconnect or allow the channel to operate. Thus, in our setting, the blocking of communication comes exactly from the owner of the channel and not from any participant who is not ready. 

\begin{figure}[h!]

\begin{minipage}{0.45\textwidth}
\begin{center}
  \includegraphics[width=\textwidth]{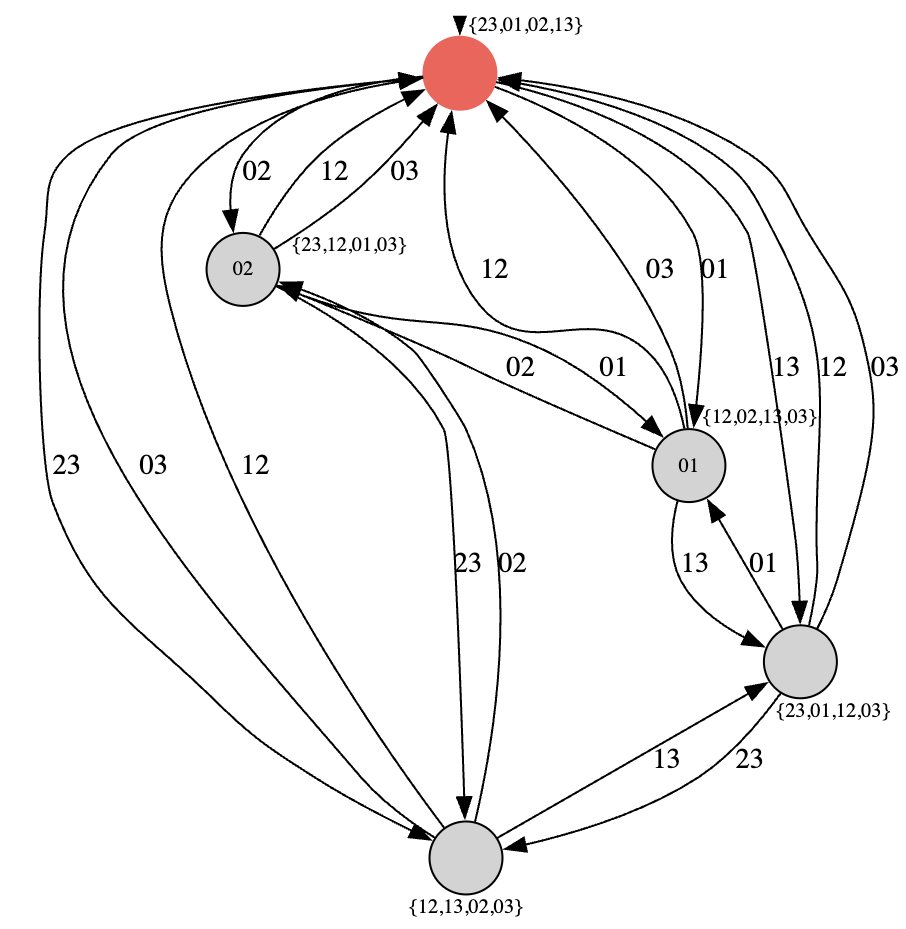} 

\mbox{(a) $T_0\mbox{, } \rulename{int}_0=\conf{\set{\msf{01,02}}}$}
\end{center}
\end{minipage}
\begin{minipage}{0.45\textwidth}
\begin{center}
 \includegraphics[width=\textwidth]{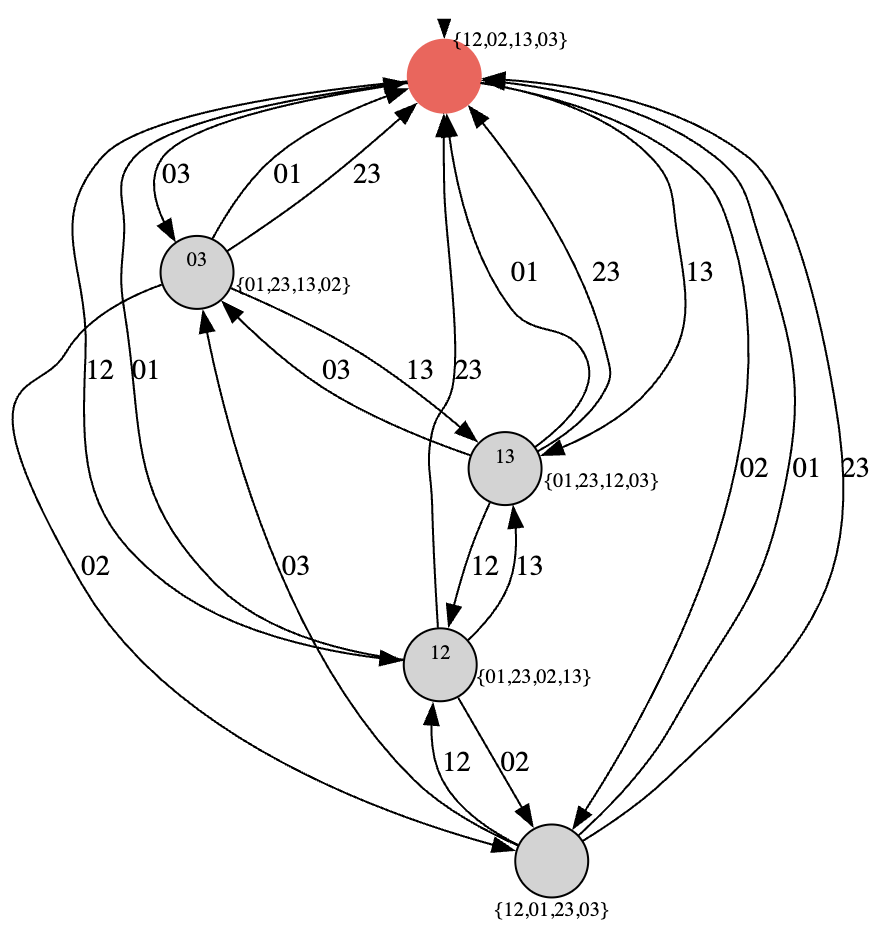} 

\mbox{(b) $T_1\mbox{, } \rulename{int}_1=\conf{\set{\msf{12,13}}}$}
\end{center}
\end{minipage}

\begin{minipage}{0.45\textwidth}
\begin{center}
 \includegraphics[width=\textwidth]{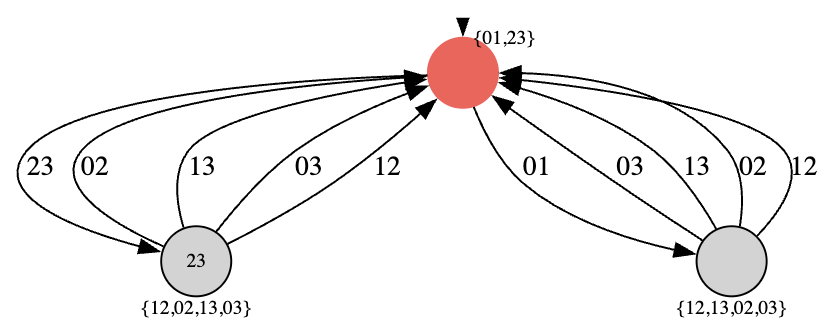} 

\mbox{(c) $T_2\mbox{, } \rulename{int}_2=\conf{\set{\msf{23}}}$}
%
\end{center}
\end{minipage}
\begin{minipage}{0.45\textwidth}
\begin{center}
 \includegraphics[width=\textwidth]{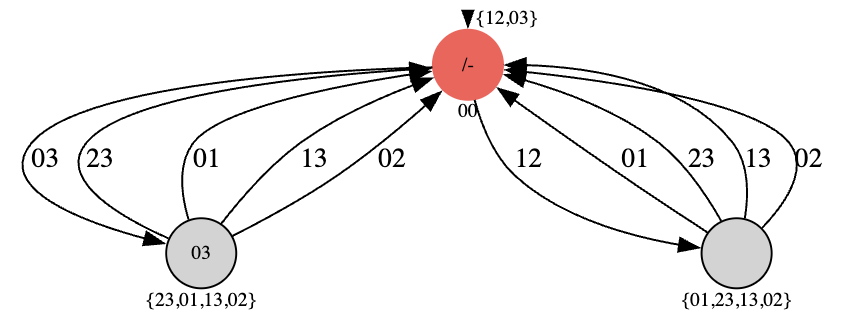} 

\mbox{(d) $T_3\mbox{, } \rulename{int}_3=\conf{\set{\msf{03}}}$}
\end{center}
\end{minipage}
\vspace{-2mm}
\caption{\label{fig:path4_dis}Distributed $\msf{Path}_4$.
Each state is labeled by the set of channels listened to in that state. Initial states are marked in red.
}
\end{figure}

Notably, the size of each local $T_i$ is smaller than the size of the sequential implementation $T$, with $T_2$ and $T_3$ with half the size.
Despite the high connectivity of this hard scenario, the maximal number of channels connected in any state is $4$ out of $6$ (however, between all states, each $T_i$ is connected to all channels).
Furthermore, the composition of Fig.~\ref{fig:path4_dis} is still asynchronous, and in many cases no more than two agents are communicating at any time instant. This is clear by looking at what channels each agent listens to in a specific state. Moreover, it supports our initial conjecture that it may not be wise to rely on the designer to decide the degree of asynchrony as communication-needs are dynamic. For instance, in the initial state of $T_2$ when it initiates $23$, only $T_0$ reacts and the rest stays unchanged. Similarly when $T_3$ initiates $03$,  only $T_1$ reacts, etc.

The tradeoff with respect to Zielonka's distribution is that each agent needs to listen to extra channels in some of its states to avoid breaking the language. 
This is clearly seen in $T_2$ and $T_3$.
For example, $T_2$ listens to $01$ and $23$ in its initial state expecting either to initiate on $23$ (and then wait for another communication involving $2$ or $3$ before doing this again) and the same after communication on $01$. 
Contrarily, $T_0$ and $T_1$ need to be aware of what happens to the two other agents in order to know when they can initiate. This makes their transition systems more complicated and keeps them more connected. 


\section{Case Study: Time-Sharing Service }
\label{sec:case}
We consider a more complicated case study referring directly to the full technical development.
We model a scenario including two clients, a server, and a service provider. 
The two clients ${C_1}$ and $C_2$ time share a service that is accessed through the server $S$. 
The server does not offer the service, but is responsible for forwarding the clients to a service provider $P$. 
The protocol proceeds as follows: $C_1$ and $C_2$ concurrently compete to reserve a connection with the server $S$ by issuing an initiate transition on $\msf{r1}$ (and correspondingly $\msf{r2}$). 
The server accepts one connection at a time, and thus the non-succeeding client must wait for a release transition $\msf{rl1}$ (or $\msf{rl2}$) from the other client.

Once the server $S$ accepts the connection, the succeeding client $C_1$ (or $C_2$) proceeds by initiating a service request transition $\msf{q1}$ (or $\msf{q2}$). 
The server forwards the request to a specific provider $P$.
The service provider $P$ initiates a connection transition $\msf{c}$ with the respective client. Once the client accepts the connection, $P$ supplies the service by issuing a serve transition $\msf{s}$. Now the respective  client $C_1$ (or $C_2$) is served and must release the session by issuing a release transition $\msf{rl1}$ (or $\msf{rl2}$). 
After such a release, it is the other client's turn to use the service. 

We show a sequential implementation of this scenario, ensuring that each client is infinitely often served. 
We will start directly from the sequential implementation as follows:

A sequential  TS $T$ with the interface $\conf{\set{\msf{r1,r2, q1,q2,f,c,s,rl1,rl2}},\emptyset}$ is shown in Fig.~\ref{fig:drlise}. Clearly, $T$ is a communication-closed TS and satisfies  the conditions of Def.~\ref{def:shadow}.
%

We want to automatically generate an equivalent distributed implementation on four agent interfaces as follows: server $S$ with $\conf{\set{\msf{f}}, \emptyset}$; service provider $P$ with  $\conf{\set{\msf{c,s}},\emptyset}$; client $C_1$ with  $\conf{\set{\msf{r1,q1,rl1}},\emptyset}$; client $C_2$ with  $\conf{\set{\msf{r2,q2,rl2}},\emptyset}$.
\begin{figure}[t!]
\includegraphics[scale=.5]{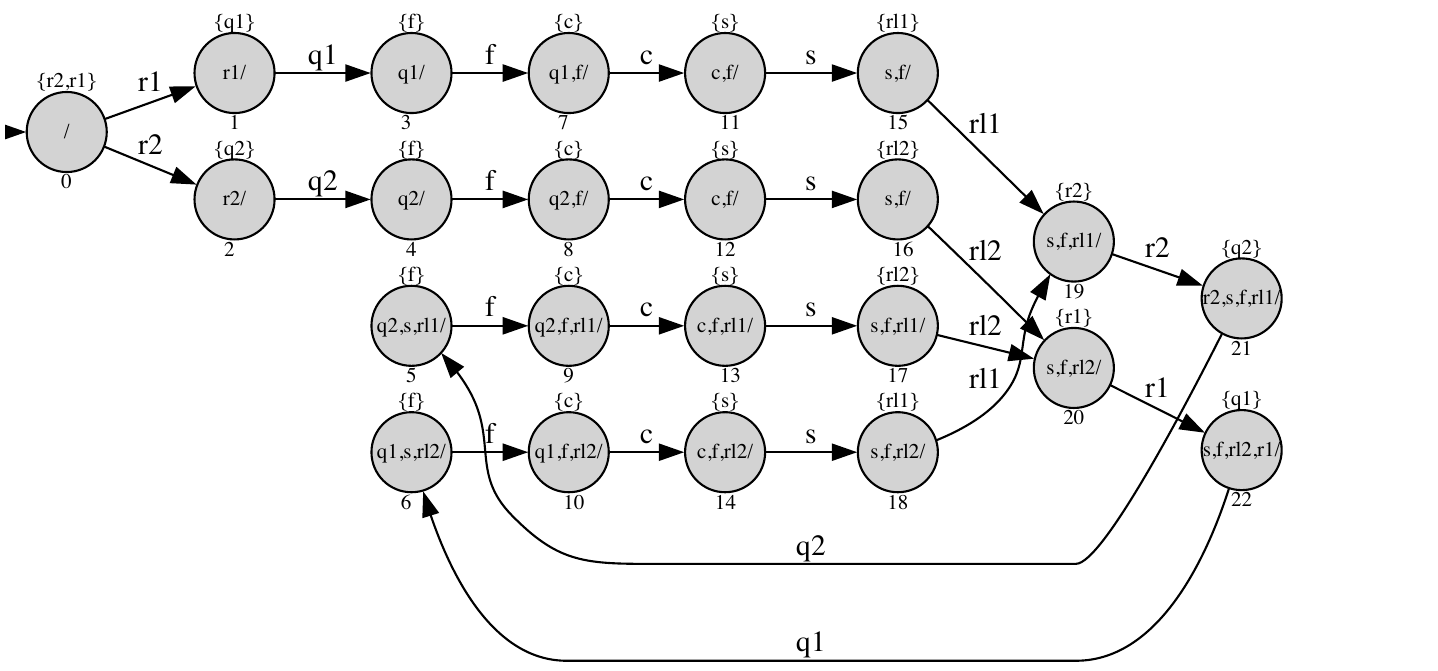}
\vspace*{-2mm}
\caption{Time-sharing service (sequential)}
\label{fig:drlise}
\vspace*{-2mm}
\end{figure}

\begin{figure}[t!]
\centering
\begin{tabular}{c}
$
\begin{array}{c}
 \includegraphics[scale=.5]{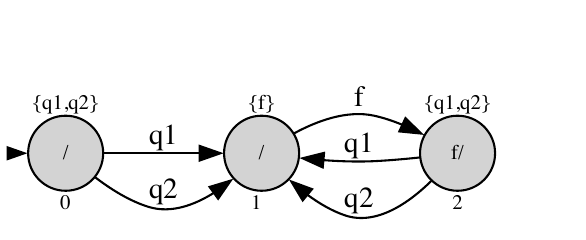}\\
\mbox{(a) $[S]^{\mathcal{C}_S}$ with interface $\conf{\set{\msf{f}}, \emptyset}$}\\
\end{array}$
\vspace*{-5mm}
$
\begin{array}{cc}
\begin{array}{c}
 \includegraphics[scale=.5]{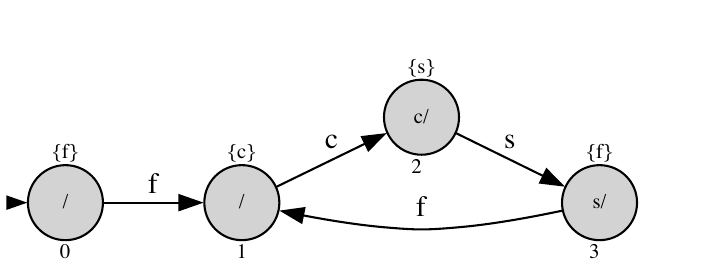}  \\
\mbox{(b) $[P]^{\mathcal{C}_P}$ with interface $\conf{\set{\msf{c,s}},\emptyset}$}
\end{array}
\end{array}$\\
\vspace*{-5mm}
$
\begin{array}{c}
 \includegraphics[scale=.5]{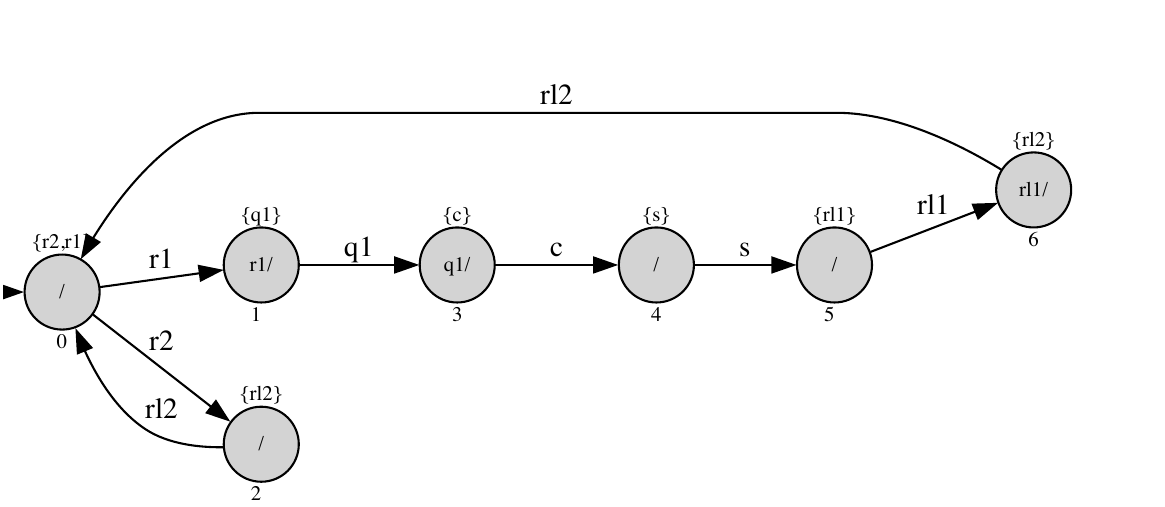}\\
\mbox{(c) $[C_1]^{\mathcal{C}_1}$ with  interface $\conf{\set{\msf{r1,q1,rl1}},\emptyset}$}
\end{array}$ \\
$
\begin{array}{cc}
\begin{array}{c}
 \includegraphics[scale=.5]{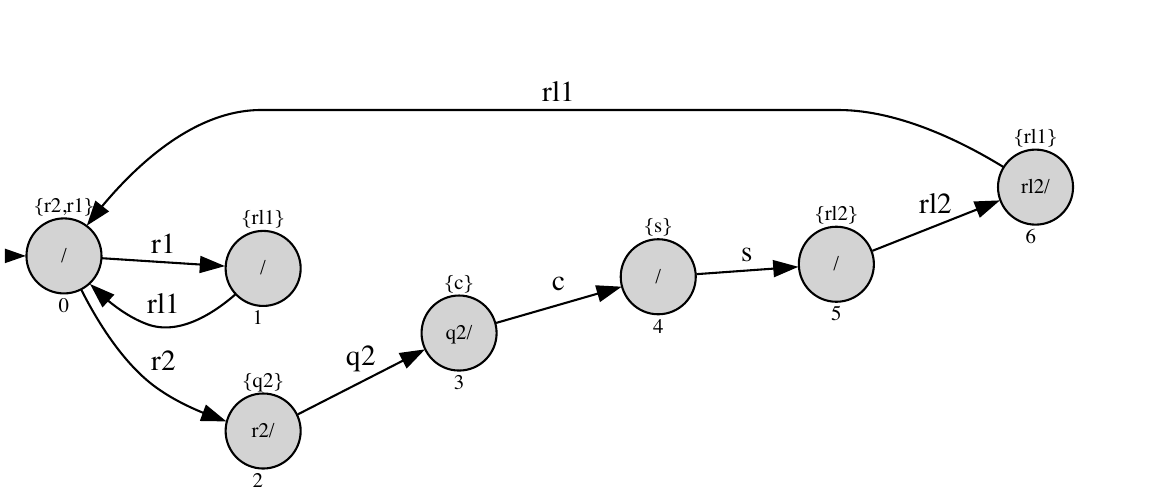}  \\
\mbox{(d) $[C_2]^{\mathcal{C}_2}$  with  interface $\conf{\set{\msf{r2,q2,rl2}},\emptyset}$}
\end{array}
\end{array}$
\end{tabular}
\vspace{-2mm}
\caption{Time-sharing service (distributed)\label{fig:sm}}
\vspace{-5mm}
\end{figure}


We can apply our distribution as follows: we consider the sequential TS $T$ in Fig.\ref{fig:drlise}; we pick one interface, say client $C_1$'s interface $\conf{\set{\msf{r1,q1,rl1}},\emptyset}$ and apply decomposition in Lemma~\ref{lem:triv} w.r.t.~the rest  $\conf{\set{\msf{q2,c,s,rl1,rl2}}, \emptyset}$ to get $C_1$ and the rest as a single TS $\mathcal{C}_1$ (the parameter of $C_1$). Both $C_1$ and $\mathcal{C}_1$ are isomorphic to the sequential  TS in Fig.\ref{fig:drlise}, and only differ in state labelling and interfaces by definition. That is, they only keep part of the label that is consistent with their interfaces. For instance, consider state $(17)$ in Fig.\ref{fig:drlise}, its label is projected on $C_1$ as $\msf{(\set{rl1},\emptyset)}$ and on $\mathcal{C}_1$ as $\msf{(\set{s,f},\emptyset)}$.  We can minimise/compress $\Delta_1$ of agent $C_1$ with respect to $\mathcal{C}_1$ using reconfigurable bisimulation to get $[C_1]^{\mathcal{C}_1}$. We repeat this process for all other interfaces; now the composition of all compressed TSs $[C_1]^{\mathcal{C}_1}\| [C_2]^{\mathcal{C}_2}\|[S]^{\mathcal{C}_S}\|[P]^{\mathcal{C}_P}$ is guaranteed to be equivalent to the original TS $T$.

 An automatically produced distributed implementation is reported in Fig.~\ref{fig:sm}. In terms of size, the size of the largest individual TS $C_1$ (equivalently $C_2$) is even less than $\frac{1}{3}$ of the size of the sequential implementation $T$. The smallest individual TS is almost $\frac{1}{8}$ the size of $T$. Moreover, using traditional bisimulation techniques would almost result in zero reduction of the state-space in the distribution problem we consider in this paper. This is because all transitions before and after the trivial composition have potentials to be composed with others. That is, they are either of type reaction (receive) or initiation (send), and we do not have any hidden ($\tau$) transitions, that existing notions can reduce. Reconfigurable bisimulation is unique in its ability to reduce unnecessary ``react'' (or receive)  transitions. Thus, all other notions can be as coarse as strong bisimulation in this context, which is known to be one of the finest equivalences.

 In terms of asynchrony, the distribution in Fig.~\ref{fig:sm} ensures that no more than two agents are communicating at any time instant, and all others are totally oblivious and not listening. This is clear by looking at what channels each agent listens to in a specific state. For instance, consider that all agents of Fig.~\ref{fig:sm} are in the initial state, the only agents that listen to channels in common and can initiate are $[C_1]^{\mathcal{C}_1}$ and $[C_2]^{\mathcal{C}_2}$. If $[C_1]^{\mathcal{C}_1}$ initiates on $\msf{r1}$ and moves to state ($1$) then only $[C_2]^{\mathcal{C}_2}$ may react and move to state ($1$). Later $[C_1]^{\mathcal{C}_1}$ continues while $[C_2]^{\mathcal{C}_2}$ is totally disconnected.

The listening set also determines how much information each agent needs to know about the rest of the system. That is, an agent $T$ can be: \rom{1} fully informed if it listens to all channels $Y$ of the centralised system in every state; \rom{2} uninformed if it is in a state $s$ and only listens to channels that $T$ initiates on, i.e., it can only initiate transitions from $s$ which are autonomous by construction; or \rom{3} partially informed otherwise. 

Clearly, case\rom{1} is hardly possible unless the centralised system is nondeterministically initiating transitions on all channels in every state;  case\rom{2} can be found in every agent in Fig.~\ref{fig:sm}. For instance, consider state ($1$) of the server in subfigure (a). Indeed, the server cannot tell if any other agent is going to react/receive $\msf{f}$. However, after the autonomous initiate on $\msf{f}$ the server reaches state ($2$), and becomes aware that some other agents might be (or will be) able to initiate on $\msf{q1}$ or $\msf{q2}$, namely the server is in case\rom{3}, i.e., is partially informed.

\end{toappendix}

\end{document}